\documentclass[sigplan,noacm]{acmart}
\usepackage{graphicx}
\usepackage{subcaption}
\usepackage{listings}
\usepackage{xcolor}
\usepackage{float}
\usepackage[linesnumbered,algo2e,ruled,vlined]{algorithm2e}
\usepackage{algorithm}
\usepackage{algpseudocode}
\usepackage{amsmath}
\usepackage{pifont}
\usepackage{booktabs}
\usepackage{wasysym}
\usepackage{soul}
\usepackage{booktabs}
\usepackage{tabularx}
\usepackage{geometry}

\definecolor{deepblue}{rgb}{0,0,0.5}
\definecolor{deepred}{rgb}{0.6,0,0}
\definecolor{deepgreen}{rgb}{0,0.5,0}

\definecolor{color1}{rgb}{0.1,0.7,0.8} 
\definecolor{color2}{rgb}{0.9,0.1,0.1} 
\definecolor{color3}{rgb}{0.7,0.3,0.7} 
\definecolor{color4}{rgb}{0.3,0.3,0.7} 
\definecolor{color5}{RGB}{8, 102, 3} 
\definecolor{color6}{rgb}{0.53, 0.66, 0.42} 

\lstdefinestyle{customstyle}{
    language=Python, 
    basicstyle=\ttfamily\footnotesize,
    keywordstyle=\color{blue},
    commentstyle=\color{gray},
    stringstyle=\color{red},
    breaklines=true,
    frame=single,
    morekeywords={const, let, function, if, else, while, return, true, false} 
}

\renewcommand\footnotetextcopyrightpermission[1]{}

\author{Zhixiang Wei}
\affiliation{%
\institution{Shanghai Jiao Tong University}
\country{China}
}
\email{tonywei_sjtu@sjtu.edu.cn}

\author{James Yen}
\affiliation{%
\institution{Shanghai Jiao Tong University}
\country{China}
}
\email{jamesyen2202002@gmail.com}

\author{Jingyi Chen}
\affiliation{%
\institution{Shanghai Jiao Tong University}
\country{China}
}
\email{2744234012@qq.com}

\author{Ziyang Zhang}
\affiliation{%
\institution{Shanghai Jiao Tong University}
\country{China}
}
\email{functioner@sjtu.edu.cn}

\author{Zhibai Huang}
\affiliation{%
\institution{Shanghai Jiao Tong University}
\country{China}
}
\email{paynqueller@sjtu.edu.cn}

\author{Chen Chen}
\affiliation{%
\institution{Shanghai Jiao Tong University}
\country{China}
}
\email{chenchen825@sjtu.edu.cn}

\author{Xingzi Yu}
\affiliation{%
\institution{Shanghai Jiao Tong University}
\country{China}
}
\email{edittriendl@sjtu.edu.cn}

\author{Yicheng Gu}
\affiliation{%
\institution{Shanghai Jiao Tong University}
\country{China}
}
\email{guyicheng98@sjtu.edu.cn}

\author{Chenggang Wu}
\affiliation{%
\institution{Shanghai Jiao Tong University}
\country{China}
}
\email{wuchenggang@sjtu.edu.cn}

\author{Yun Wang}
\affiliation{%
\institution{Shanghai Jiao Tong University}
\country{China}
}
\email{yunwang94@sjtu.edu.cn}

\author{Mingyuan Xia}
\affiliation{%
\institution{UltraRISC Shanghai}
\country{China}
}
\email{xiamy@ultrarisc.com}

\author{Jie Wu}
\affiliation{%
\institution{Cloud Computing Research Institute, China Telecom}
\country{China}
}
\email{wujie@chinatelecom.cn}

\author{Hao Wang}
\affiliation{%
\institution{Stevens Institute of Technology}
\country{United State}
}
\email{hwang9@stevens.edu}

\author{Zhengwei Qi}
\affiliation{%
\institution{Shanghai Jiao Tong University}
\country{China}
}
\email{qizhwei@sjtu.edu.cn}

\begin{document}

\title{Equinox: Holistic Fair Scheduling in Serving Large Language Models}

\begin{abstract}

Large Language Model serving faces unprecedented demand, yet existing schedulers struggle to allocate resources fairly across diverse workloads. Current deployment strategies like First-Come-First-Served offer no client isolation against monopolization, while Requests Per Minute policy wastes resources during off-peak hours. The Virtual Token Counter attempts fair-sharing by tracking token consumption but inadequately captures LLM resource dynamics. The Transformer architecture's prefill-decode bifurcation creates conflicting resource patterns: memory-bound decode operations dominate latency, compute-bound prefill operations determine throughput, and GPU utilization follows independent batch-refresh patterns, making single-metric fairness unachievable. 

We address these limitations with a dual-counter framework separating user and operator perspectives. The User Fairness Counter measures quality of service via weighted tokens and latency; the Resource Fairness Counter measures operational efficiency through throughput and GPU utilization. Since these metrics are only available post-execution, creating a scheduling paradox, we introduce a deterministic Mixture of Prediction Experts (MoPE) framework to predict user-perceived latency, output tokens, throughput, and GPU utilization. These predictions enable calculation of a unified Holistic Fairness score that balances both counters through tunable parameters for proactive fairness-aware scheduling. We implement this in Equinox, an open-source system with other optimizations like adaptive batching, and stall-free scheduling. Evaluations on production traces (ShareGPT, LMSYS) and synthetic workloads demonstrate Equinox achieves up to $1.3\times$ higher throughput, 60\% lower time-to-first-token latency, and 13\% higher fairness versus VTC while maintaining 94\% GPU utilization, proving fairness under bounded discrepancy across heterogeneous platforms.

\end{abstract}

\maketitle

\pagestyle{plain}

\section{Introduction}

\begin{figure}[t]
    \centering
    \setlength{\abovecaptionskip}{0.1cm}
    \setlength{\belowcaptionskip}{-0.1cm}
    \includegraphics[width=0.48\textwidth]{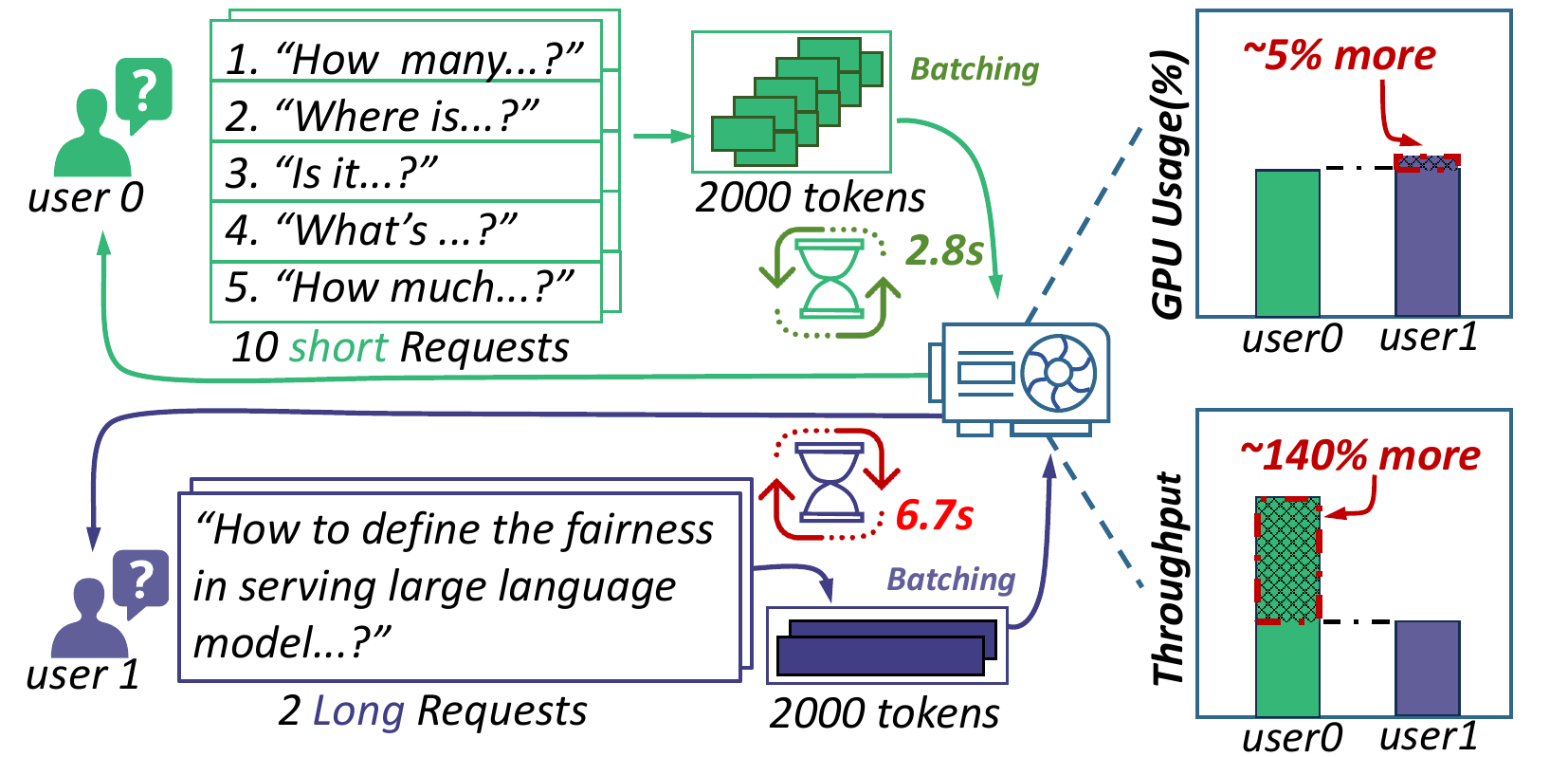}
    \caption{Token-count scheduling is unfair for independent, single-turn requests: with equal total tokens, many short vs.\ few long requests yield very different user latency, GPU utilization, and throughput. Batching mainly \emph{shifts} the advantage—short identical requests benefit in-batch; without batching, queued short requests can lose. Setup: S-LoRA~\cite{slora} on LMSYS Chat-1M~\cite{zheng2024lmsyschat1mlargescalerealworldllm}, A100-80GB, matching VTC~\cite{VTC}.}
    \label{fig:Problem_Intuition}
    \vspace{-1.0em}
\end{figure}

\begin{figure*}[htbp]
    \centering
    \begin{subfigure}[b]{0.32\linewidth}
        \centering
        \includegraphics[width=\linewidth]{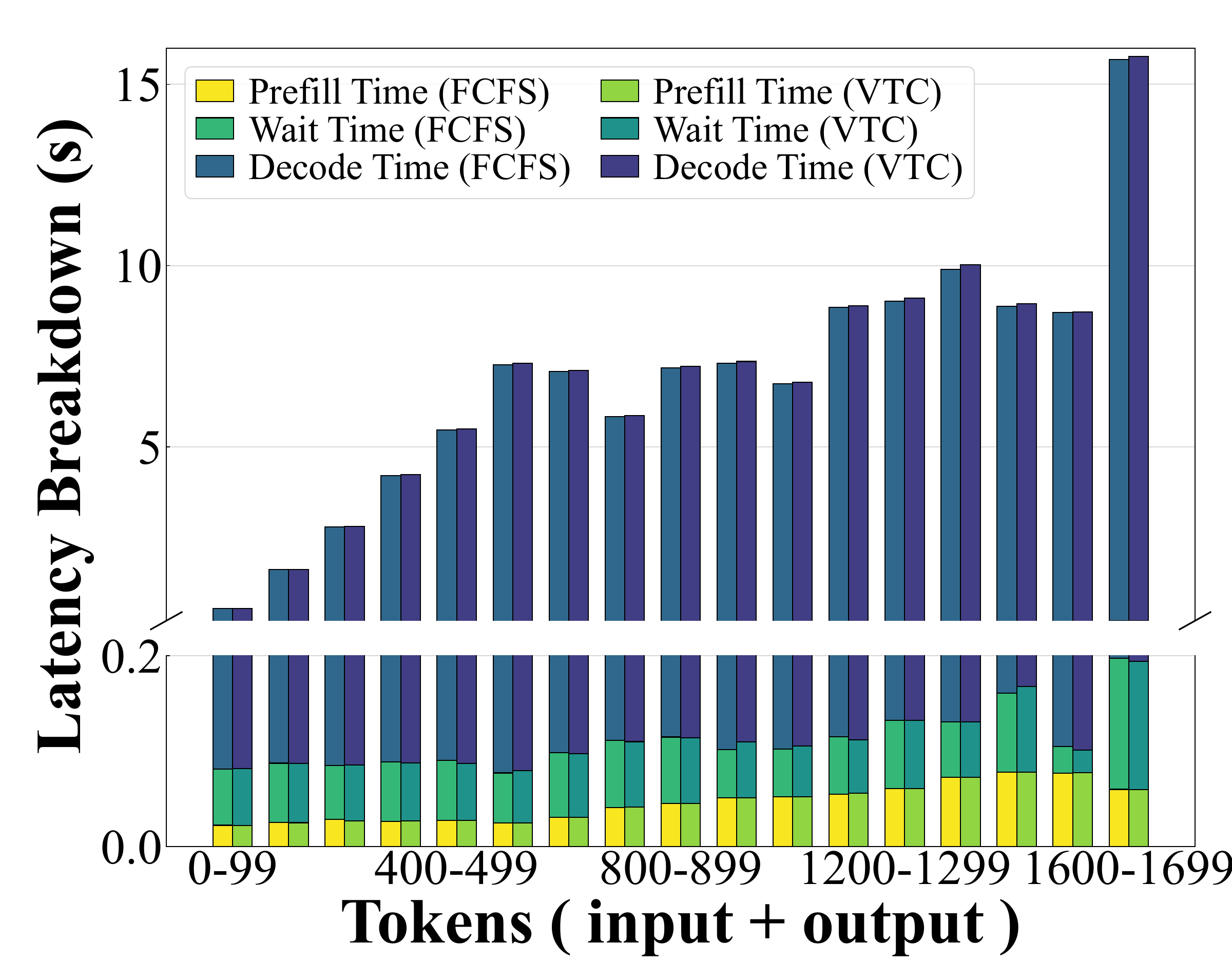}
        \caption{FCFS/VTC shows \textbf{non-linear} latency variation across token ranges.}
        \label{fig:motivation1}
    \end{subfigure}
    \hfill
    \begin{subfigure}[b]{0.32\linewidth}
        \centering
        \includegraphics[width=\linewidth]{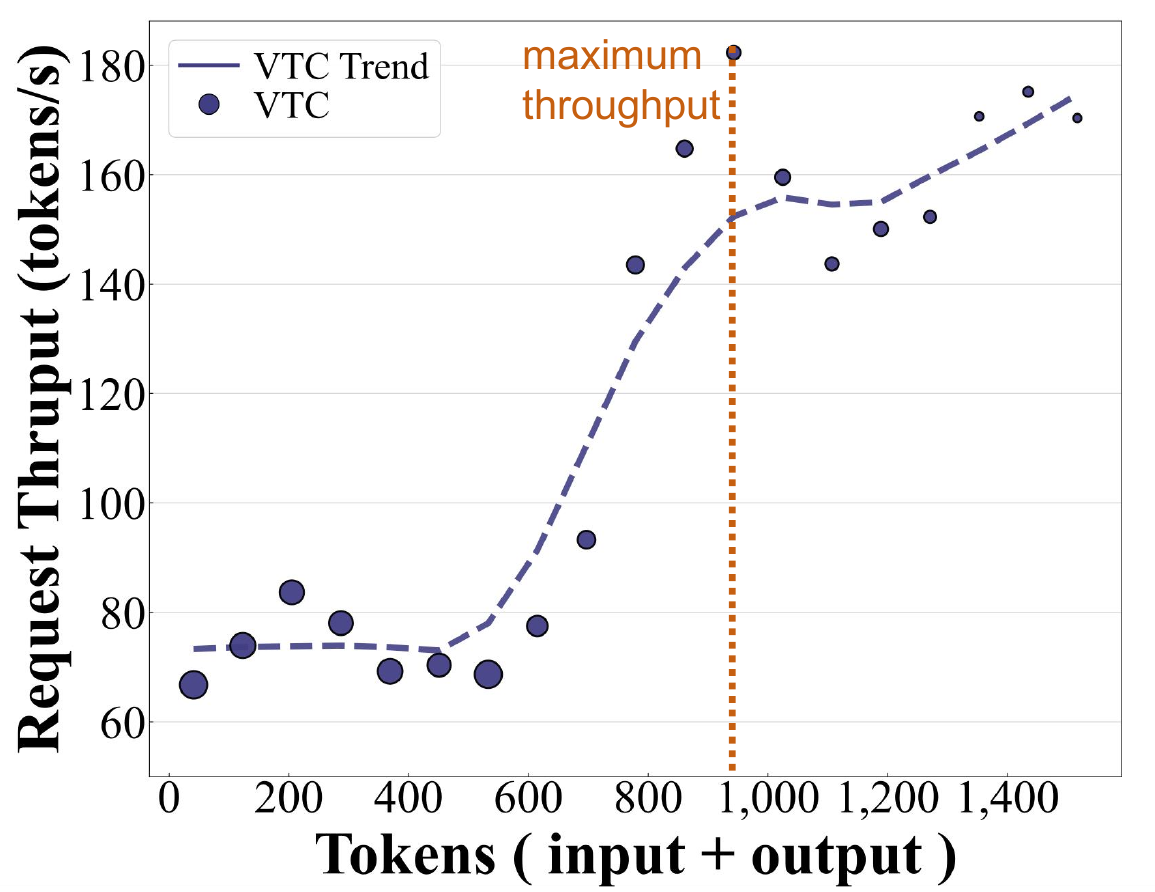}
        \caption{VTC exhibits \textbf{non-linear} throughput patterns across token ranges.}
        \label{fig:motivation2}
    \end{subfigure}
    \hfill
    \begin{subfigure}[b]{0.32\linewidth}
        \centering
        \includegraphics[width=\linewidth]{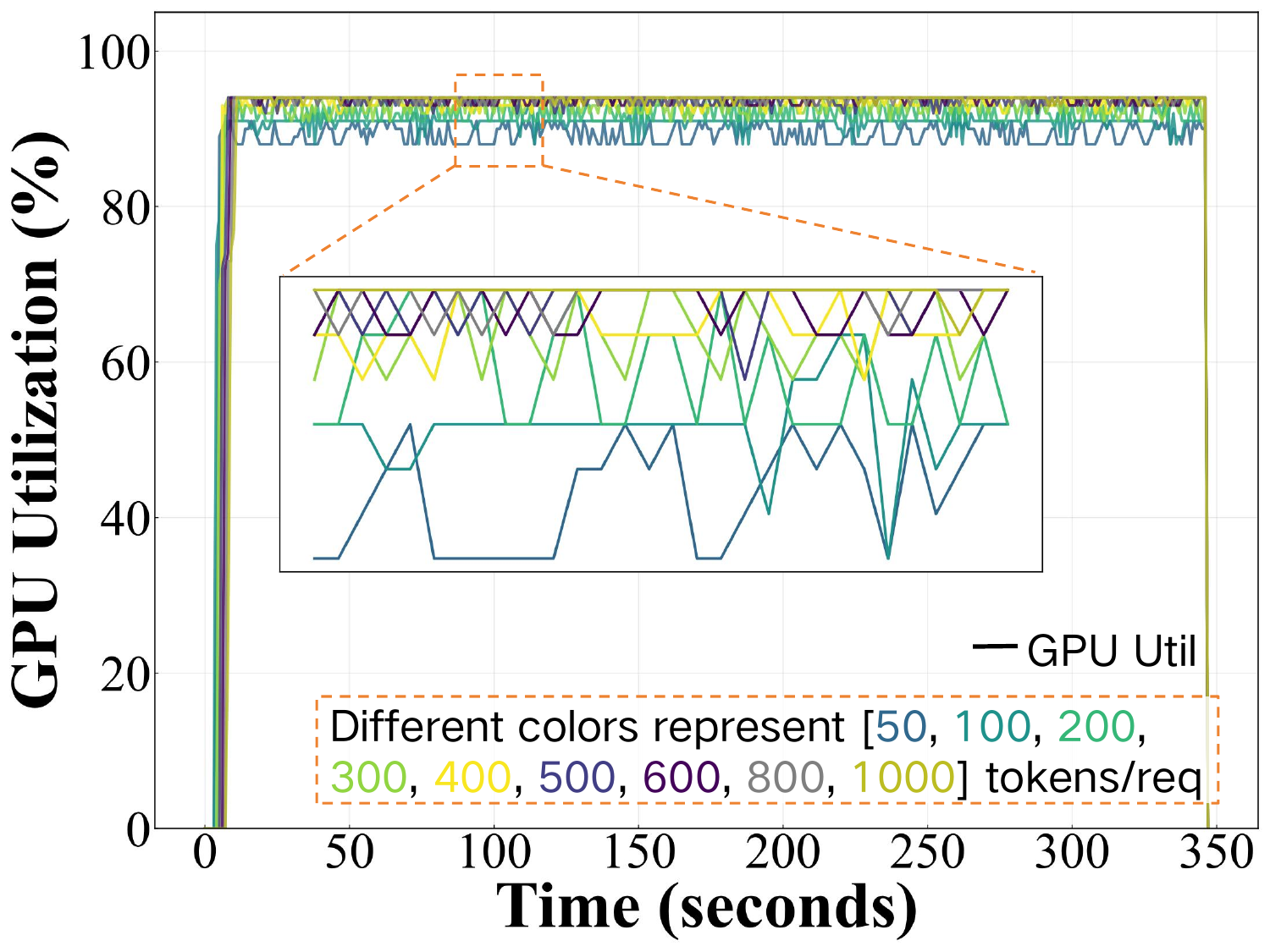}
        \caption{VTC demonstrates step-like GPU resource usage across token intervals.}
        \label{fig:motivation3}
    \end{subfigure}

    \caption{Experiments on NVIDIA A100-80GB use real-world trace (lmsys-chat-1m) for (a,b) and synthetic workload for (c) with fixed total tokens (RPS $\times$ tokens/request) and 1:1 input:output ratio (prefill:decode). Real-world traces across vLLM and SGLang (\autoref{fig:motivation-overall}) confirm these patterns stem from architectural properties, not implementation choices.}

    \label{fig:motivation-fcfs-vtc}
\end{figure*}

Large Language Models (LLMs) have reshaped artificial intelligence by combining modern neural networks with scalable computing and vast datasets. Their versatility spans tasks from content generation to software engineering, driving rapid adoption~\cite{GPT-3,BERT,GPT-2,Attention-is-all-you-need,LSTM}. For example, ChatGPT gained 1 million users in five days—and 100 million in two months—while DeepSeek topped the U.S. iOS chart shortly after its January 2025 release~\cite{chatgpt-popular,deepseek-popular}. Yet, most prior research has primarily targeted improvements in model quality, inference speed~\cite{llm_prediction_2,10.1145/3694715.3695948}, and GPU utilization~\cite{llm_inference_5,10.1145/3676641.3716011}, leaving fair multi‑tenant serving largely unexplored. This gap prompts a central question: \emph{How can we allocate resources both efficiently and equitably as LLM demand and workload diversity grow?}

Most production deployments~\cite{llm_serving_1,llm_serving_2,llm_serving_3,llm_serving_4, 298764, wang2025fairmoefairnessorientedmixtureexperts} rely on First Come First Served (FCFS) scheduling, which processes requests strictly in arrival order, thereby offering no client isolation and allowing compute-intensive requests to monopolize resources and degrade co-tenants. To mitigate this, providers often impose static Requests Per Minute (RPM) quotas~\cite{RPM}, a rate-limiting mechanism that caps the number of requests a client can send within a one-minute interval; however, these waste capacity during off-peak periods as the fixed allowance goes unused when demand is low.

The Virtual Token Counter (VTC)~\cite{VTC} addresses these scheduling limitations by adapting fair-share scheduling from network queuing and operating systems~\cite{fair_queueing_networking,fair_queueing_operating_systems,fair_queueing_1,fair_queueing_2,fair_queueing_3,fair_queueing_4}. In traditional domains, resource consumption patterns allow a single metric like CPU seconds or transmitted bytes to effectively represent overall usage, as these metrics naturally correlate with actual resource costs. VTC attempts to replicate this approach in LLM serving by adopting token count as its scheduling metric. The scheduler tracks cumulative input and output tokens per user and implements a work-conserving policy~\cite{work-conserving} that prioritizes users with lower token totals. This approach solves both prior failures: it prevents FCFS starvation by ensuring proportional service allocation while avoiding RPM's capacity waste through dynamic resource allocation instead of static quotas. 

\smallskip
\noindent \textbf{Key Insight.} Although VTC improves upon FCFS and RPM, it inadequately adapts fair-share scheduling principles to LLM serving environments. Unlike traditional domains where one metric comprehensively captures resource usage, LLM serving exhibits metrics that actively conflict rather than correlate. The Transformer architecture's bifurcated nature drives this conflict because \emph{prefill} processes prompts in parallel with superlinear compute-bound costs, while \emph{decode} generates tokens sequentially with cache-dominated memory-bound costs~\cite{llm_prediction_2}. Equal token counts thus yield vastly different resource consumption. As \autoref{fig:Problem_Intuition} demonstrates, identical \textbf{token counts} produce divergent outcomes across \textbf{latency}, \textbf{throughput}, and \textbf{GPU utilization} due to opposing computational constraints. Latency grows monotonically as a memory-bound operation, with decode consuming over 90\% of end-to-end time (\autoref{fig:motivation1}). Throughput exhibits compute-bound non-monotonic patterns, initially increasing then declining after 1,000 tokens (\autoref{fig:motivation2}). GPU utilization shows stepwise plateaus because shorter requests trigger frequent batch refreshes, introducing idle intervals during CPU-bound operations (\autoref{fig:motivation3}). These observations show that token count cannot serve as a universal fairness metric and motivate a shift from single metric fairness to an explicitly multi objective formulation.

\smallskip
\noindent \textbf{Key Idea.} Motivated by this conflict, we propose a dual counter framework comprising the User Fairness Counter (UFC) and the Resource Fairness Counter (RFC) that acknowledges distinct stakeholder needs. The UFC ensures \emph{equal} tenant Quality of Service (QoS) by combining weighted tokens with user perceived latency, with weighting justified because decode dominates both pricing and latency. The RFC ensures equitable operator allocation using GPU utilization and throughput, without weighting since throughput bottlenecks arise in prefill. We define holistic fairness (HF) as the joint consideration of UFC and RFC, reframing scheduling as the co-optimization of user experience and system efficiency rather than the division of a single resource. Implementing the dual-counter framework requires solving two key challenges: a paradox where computing UFC and RFC needs metrics that schedulers cannot predict, and the resulting tension between minimizing tenant latency and maximizing operator utilization.

\smallskip
\noindent \textbf{Our Approach.} We address both challenges through our open-source system Equinox, pairing a predictive framework with novel scheduling policy. Our Mixture of Prediction Experts (MoPE) provides low-overhead estimates for fairness components: user-perceived latency and output tokens for UFC; GPU utilization, and throughput for RFC. These predictions inform a fairness-aware scheduler jointly optimizing our composite metric, enabling the holistic balance that token-centric methods cannot provide.

\smallskip
\noindent \textbf{Contributions.} Our key contributions are:
\begin{itemize}
    \item 
    \textbf{Formalizing Holistic Fairness in LLM Serving.} We model LLM serving as a max-min fairness problem using UFC for per-client latency and weighted tokens, and RFC for GPU utilization and throughput, jointly optimizing metrics that transcend token-level fairness.

    \item 
    \textbf{Deterministic Prediction Framework for Scheduling.} We present the first deterministic MoPE framework that, unlike single proxy approaches~\cite{predictor_proxy} struggling with diverse output lengths, routes requests to specialized experts, reducing L1 prediction error for output tokens \textbf{from 80 to 33} versus prior methods.
    
    \item 
    \textbf{Robust System Architecture for Fairness Enforcement.} Equinox, our open-source implementation combining MoPE with our fair scheduling algorithm, cuts worst-case service gaps by \textbf{42\%} and average gaps by \textbf{86\%} versus VTC, achieving fairness comparable to an \texttt{Oracle} predictor with only a 17\% gap.
\end{itemize}

\section{Background and Motivation}

\subsection{Large Language Models Serving}

\begin{figure}[t]
    \includegraphics[width=\linewidth]{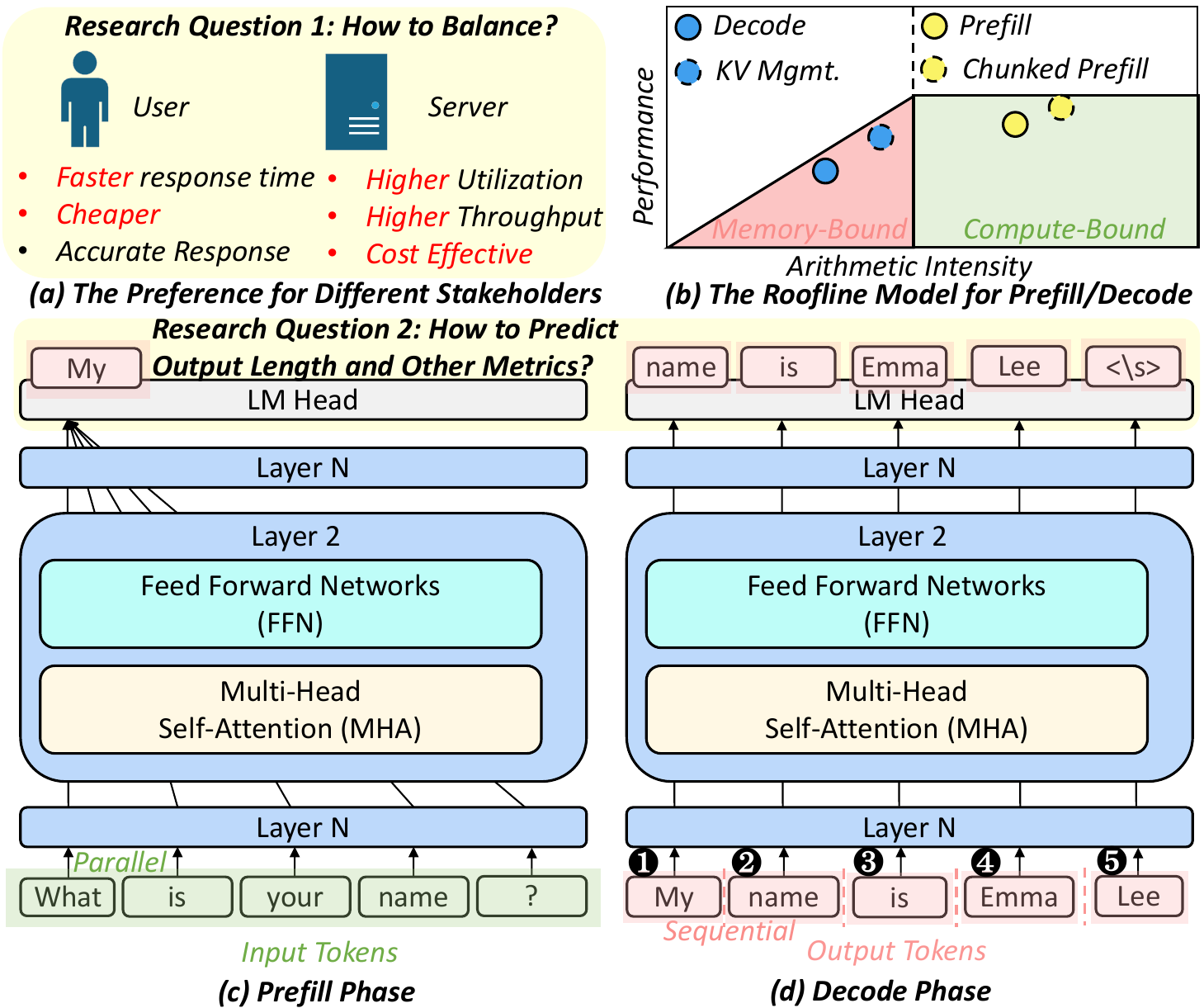}
    \caption{The core challenge of serving Large Language Models lies in balancing its two distinct inference phases: the parallel, compute-bound Prefill phase for processing prompts, and the sequential, memory-bound Decode phase for generating responses. Efficiently managing these conflicting workloads is essential to simultaneously achieve low latency for users and high throughput for service providers.}
    \label{fig:background}
    \vspace{-1em}
\end{figure}

\noindent \textbf{The LLM Serving Challenge.} While the introduction established the dual-counter framework for balancing tenant experience (UFC) and operator efficiency (RFC), implementing this framework reveals fundamental challenges in LLM serving mechanics. Production deployments must handle concurrent multi-tenant requests while managing transformer inference's unique computational patterns, illustrated in \autoref{fig:background}.

The prefill-decode bifurcation creates compounding non-linearity beyond the basic distinction noted earlier~\cite{llm_inference_1,llm_inference_2,llm_inference_3}. As the KV cache expands during sequential token generation, both memory consumption and per-token latency increase disproportionately~\cite{llm_inference_4}. Although continuous batching improves GPU utilization by interleaving requests~\cite{llm_inference_5}, the combination of workload heterogeneity and non-linear scaling properties demonstrates why token counts fundamentally fail as cost proxies.

\vspace{\baselineskip} \noindent \textbf{From Token-Level to Holistic Fairness.} Achieving holistic fairness requires co-optimizing four interconnected metrics: tenant-facing latency and token usage (UFC) alongside operator-facing GPU utilization and throughput (RFC). VTC's reliance on accumulated token counts cannot capture this multi-dimensional optimization space, manifesting in two critical failures. First, requests with identical token costs experience vastly different latencies, violating the principle that similar payments yield similar service. Second, prioritizing purely by token accumulation ignores actual GPU cycles consumed, memory bandwidth utilized, and throughput achieved (\autoref{fig:intuition_1}). This disconnection between the scheduling metric and actual objectives necessitates fundamental framework redesign.

\medskip
\noindent
\fbox{%
   \parbox{0.45\textwidth}{
       \textbf{Research Question 1:} How can we design a scheduling framework that moves beyond token-counting to co-optimize the four metrics of holistic fairness: tenant-facing latency and token usage, and operator-facing GPU utilization and throughput?
   }
}
\medskip

\subsection{LLM Token-Length Prediction}

\begin{figure*}[htbp]
    \centering
    \begin{subfigure}[b]{0.49\linewidth}
        \centering
        \includegraphics[width=\linewidth]{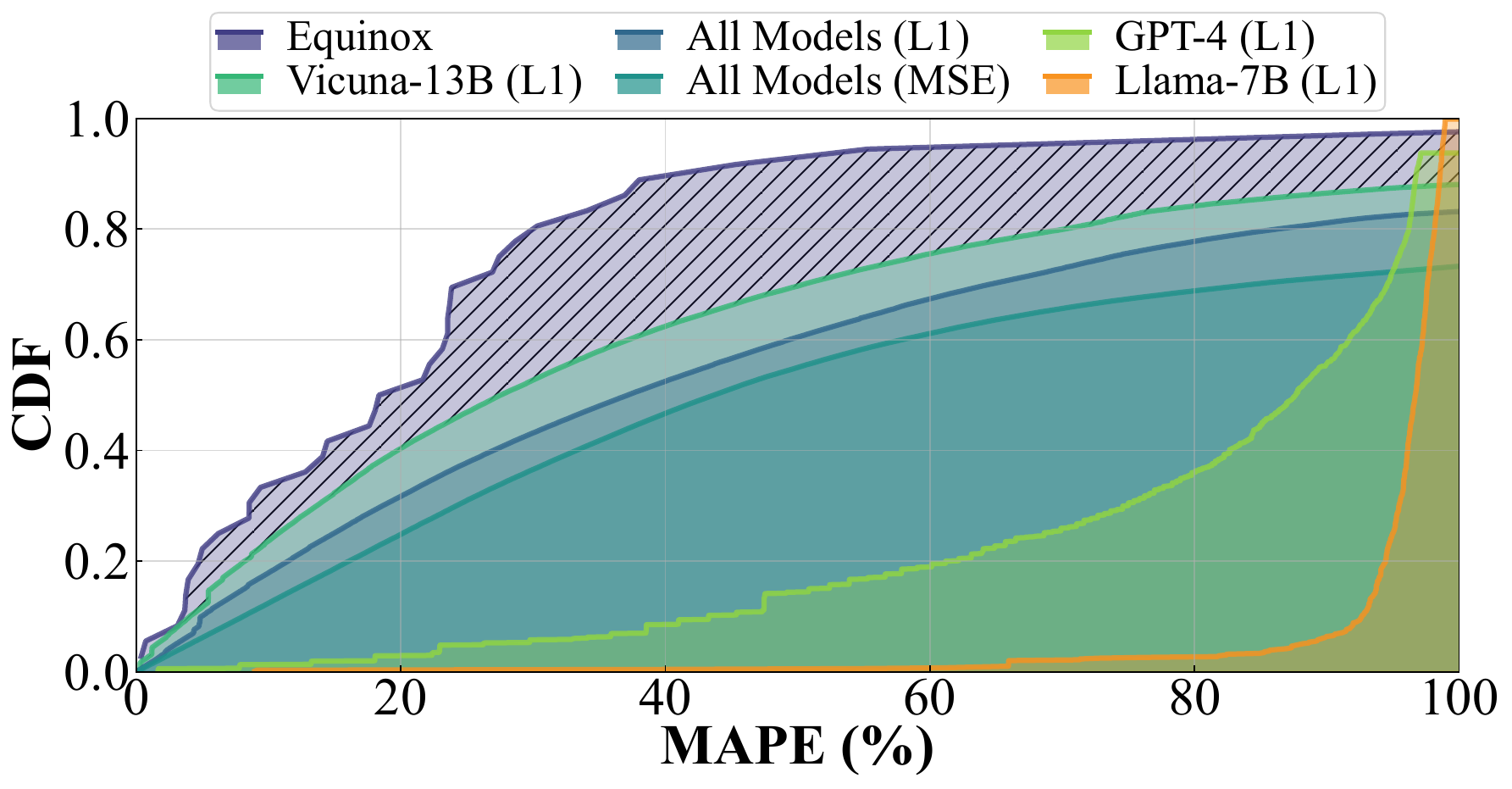}
        \caption{CDF of prediction error (MAPE) comparing single proxy models, unified models, and Equinox.}
        \label{fig:mope1}
    \end{subfigure}
    \hfill
    \begin{subfigure}[b]{0.49\linewidth}
        \centering
        \includegraphics[width=\linewidth]{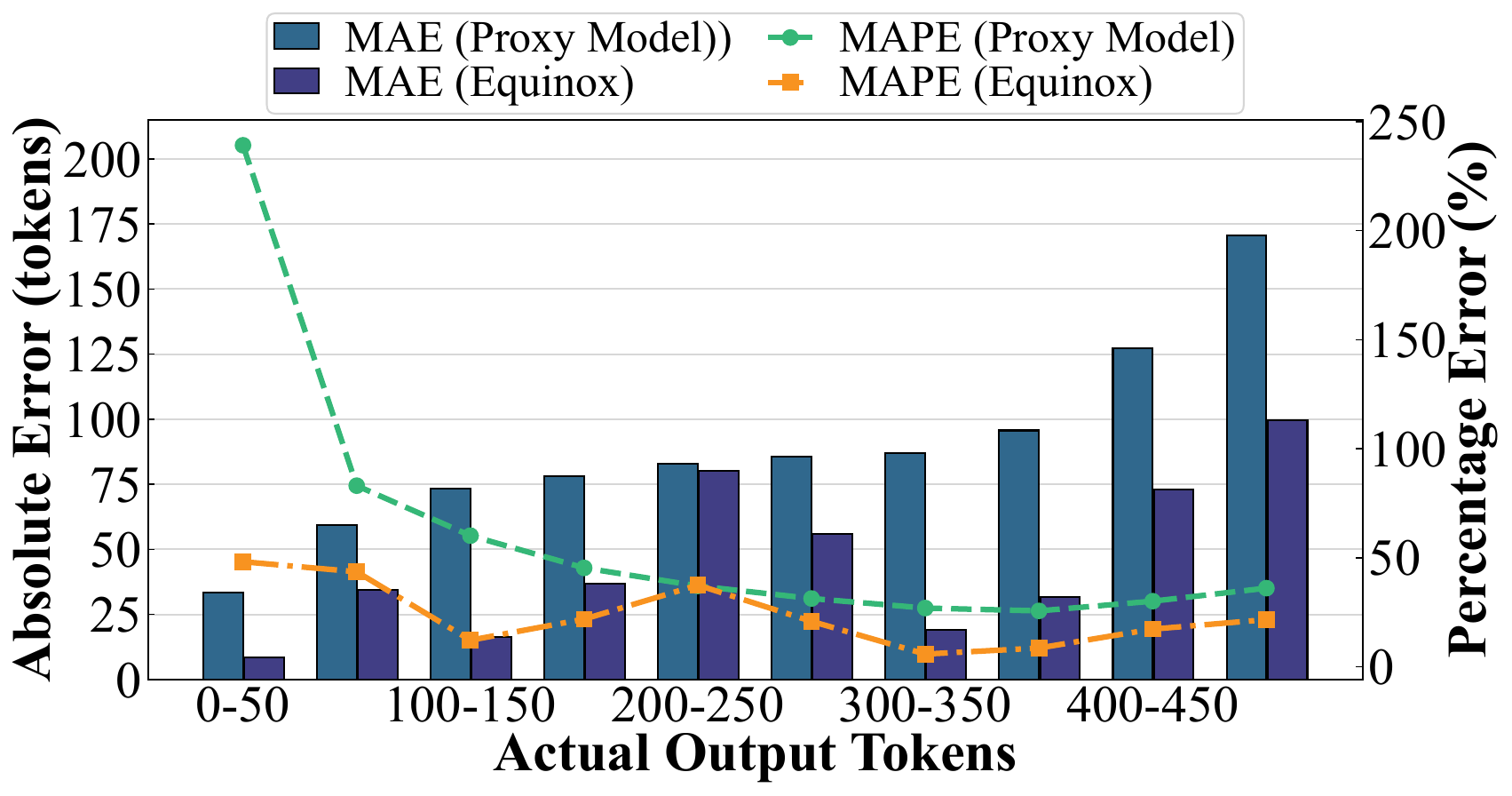}
        \caption{MAE and MAPE breakdown by actual output tokens for proxy models versus Equinox.}
        \label{fig:mope2}
    \end{subfigure}
    \caption{Error analysis of LLM prediction approaches demonstrates (a) limitations of current proxy models in capturing complex behaviors and (b) challenges in managing diverse token ranges, both addressed by Equinox through dynamic routing and expert specialization (detailed in \autoref{sec:mope}).}
    \label{fig:error-analysis}
    \vspace{-1em}
\end{figure*}

\noindent \textbf{The Predictability Paradox in LLM Inference.} Implementing Research Question 1's framework requires accurate predictions of all four fairness metrics before GPU pipeline entry. Paradoxically, while output generation remains inherently unpredictable due to dynamic token production, LLM inference's computational characteristics prove surprisingly amenable to prediction. The rapid, parallel prefill phase enables comprehensive input characteristic gathering before scheduling decisions, while the sequential decode phase exhibits predictable memory-bound behavior through systematic KV cache traversal.

Current approaches fail to exploit this opportunity, ranging from coarse proxy model categorizations~\cite{predictor_proxy,298496} to regression models and lightweight heuristics~\cite{llm_prediction_4,llm_prediction_5,llm_prediction_6,llm_prediction_7,llm_prediction_8,10946802,10.1145/3676641.3716011}. \autoref{fig:mope1} reveals that single proxy models trained on specific chatting datasets (Llama-7B, GPT-4, Vicuna-13B) and unified models trained across all data (All Models) using different loss function (L1/MSE) struggle to capture LLM complexity, resulting in high Mean Average Percentage Error (MAPE) for a large fraction of predictions. \autoref{fig:mope2} further demonstrates the challenges facing state-of-the-art proxy models by breaking down Mean Average Error (MAE) and MAPE by actual output tokens. While absolute error increases with length for both approaches, a single proxy model show significantly higher absolute error compared to Equinox, particularly struggling with long sequences where prediction errors compound dramatically. Training these models employs the same methodology~\cite{298496} and we evaluate them on lmsys-chat-1m dataset~\cite{zheng2024lmsyschat1mlargescalerealworldllm}.

\vspace{\baselineskip} \noindent \textbf{Beyond Token Prediction: The Multi-Metric Challenge.} The fundamental limitation extends beyond accuracy. Existing predictors estimate only token counts, providing merely one-quarter of holistic fairness's required metrics. Computing UFC demands predicting both user-perceived latency and tokens, as latency determines service quality. Computing RFC requires GPU utilization and throughput predictions that capture true resource consumption. Since token-to-metric relationships prove neither linear nor monotonic, schedulers equipped solely with token-length predictions remain blind to actual costs imposed on tenants and operators. While Research Question 1 defines the necessary scheduling framework, implementation requires solving a more fundamental prediction challenge—schedulers cannot optimize unmeasurable metrics, and current predictors provide insufficient information.

\medskip 
\noindent
\fbox{
   \parbox{0.45\textwidth}{
       \textbf{Research Question 2:} How can we design a prediction framework that accurately estimates not only token length but also the latency, GPU utilization, and throughput required to enable a holistically fair scheduler?
   }
}
\medskip
\vspace{-1em}

\section{Defining Holistic Fairness}

As we argued, the interdependent nature of batched GPU execution necessitates a fairness model that moves beyond the traditional "fair share" paradigm to consider the distinct objectives of both principals in the service agreement. We therefore formalize this concept of holistic fairness by evaluating it through two complementary lenses: the \textbf{user} (representing the tenant or service consumer) and the \textbf{system} (representing the service operator).

User fairness guarantees equitable access by equalizing response latency and token allocation, thereby consistently satisfying user expectations. System fairness, in turn, emphasizes efficient resource allocation to maximize throughput and maintain full GPU utilization, which directly benefits the operator by lowering operational costs. To formalize and co-optimize these competing goals, we introduce two distinct fairness counters, which represent four metrics include token counts, latency, GPU utilization and throughput.

\subsection{User-Fairness Counter}
\label{sec:ufc}

\begin{figure}[htbp]
    \includegraphics[width=\linewidth]{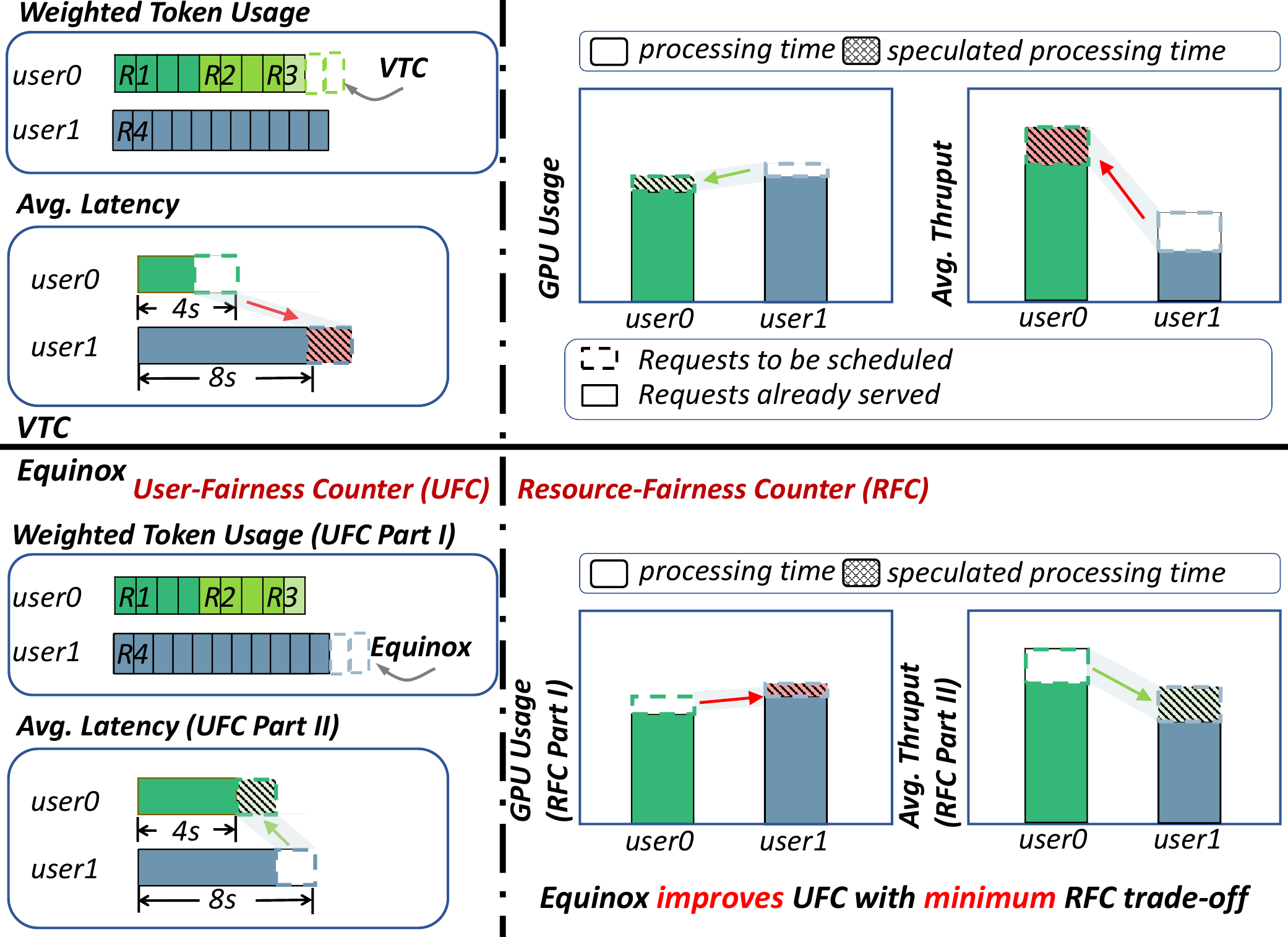}
    \caption{\textbf{Holistic fairness in LLM scheduling.} VTC picks \emph{user0} for having fewer tokens yet fairness spans users and resources. \emph{Equinox} serves the request with the smallest Holistic Fairness score computed from two counters, a User Fairness Counter that blends token usage and latency and a Resource Fairness Counter that reflects GPU utilization and throughput, achieving max min fairness; weights \(\alpha\) and \(\beta\) tune priority with \(\alpha\) greater than \(\beta\) to favor user experience.}
    \label{fig:intuition_1}
    \vspace{-1em}
\end{figure}

\begin{figure*}[t]
    \centering
    \includegraphics[width=1\textwidth]{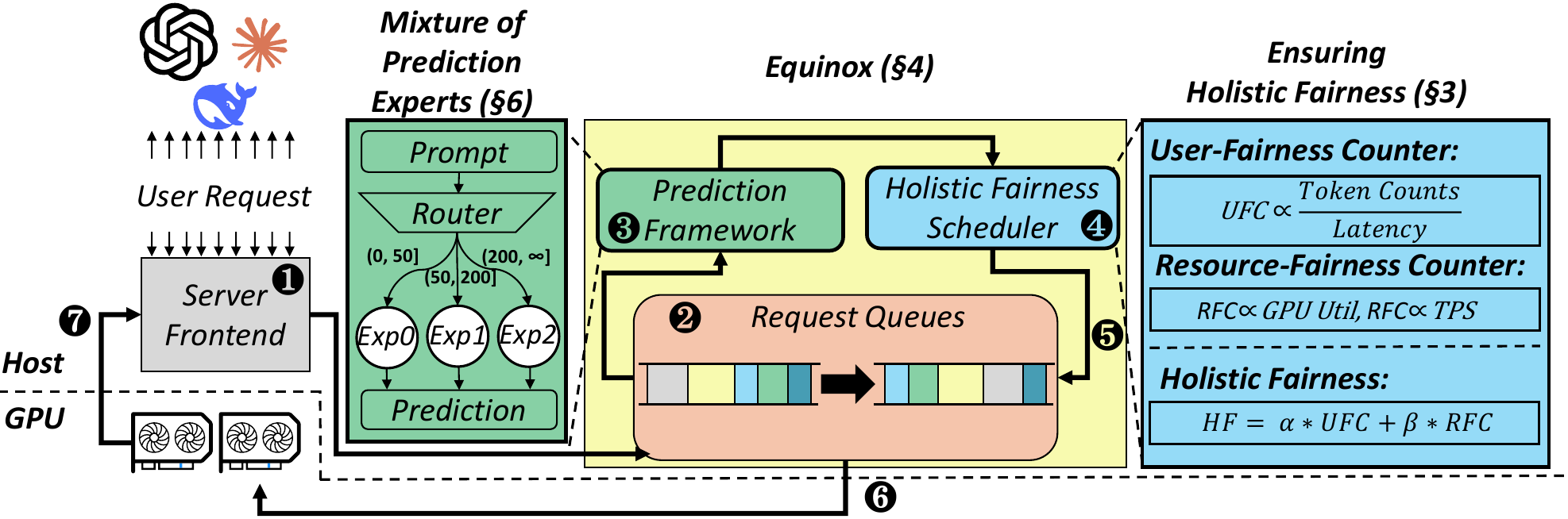} 
    \caption{System Design Overview}
    \label{fig:system_design}
    \vspace{-1em}
\end{figure*}

The User-Fairness Counter (UFC) quantifies service quality from the user's perspective by combining two essential metrics: token consumption and perceived latency. When the scheduler selects a request for GPU execution, it updates the corresponding client's UFC to reflect their accumulated resource usage. The token component follows established industry pricing models~\cite{openai_pricing_2025,google_ai_pricing_2025,deepseek_pricing_2025,anthropic_pricing_2025} and VTC conventions, weighting predicted output tokens ($\mathrm{Tokens_{req}^{\text{out}}}$) four times more heavily than input tokens ($\mathrm{Tokens_{req}^{\text{in}}}$). This \textbf{4x} multiplier reflects the computational asymmetry between prefill and decode phases, where generating output tokens consumes substantially more resources than processing input prompts. The latency component captures the complete user experience through two measurements. First, $\mathrm{WaitTime_{req}}$ tracks the elapsed time since request submission to the server queue. Second, $\mathrm{PredictTime_{req}}$ estimates the expected GPU inference duration once execution begins. Together, these metrics represent the total service time users experience from submission to completion.

For each user $f$, the UFC updates when selecting a new request (req) according to the following calculation:
\[
UFC \leftarrow UFC + \omega_f \frac{ \mathrm{Tokens}_{req}^{\mathrm{in}} + 4 \cdot \mathrm{Tokens}_{req}^{\mathrm{out}} }{1 + \delta \cdot (\mathrm{WaitTime}_{req} + \mathrm{PredictTime}_{req})}
\], 
where $\omega_f$ is the priority weight of user $f$, and $\delta$ configures the latency compensation factor, which is tested and set as 0.1 in our system. This factor prioritizes backlogged users by scaling their UFC with accumulated latency.

\subsection{Resource-Fairness Counter}

The Resource-Fairness Counter (RFC) maintains server-side fairness by dynamically updating after processing each request batch. RFC integrates 2 per-batch critical performance metrics: system throughput in tokens per second, GPU utilization percentage. For every client, the update operation of RFC per request follows the implementation exactly:
\[
RFC \leftarrow RFC + \omega_f \cdot \text{TPS} \cdot \text{Util}_{GPU}, 
\]
where $\text{TPS}$ is the estimated request throughput (tokens per second in GPU), $\text{Util}_{GPU}$ is the estimated GPU utilization percentage, scaled from 0 to 1.

\subsection{Putting it Together}
\label{sec:hfs}
With the UFC and RFC established, we can now define holistic fairness. The core challenge lies in co-optimizing these two, often competing, sets of objectives. A scheduler that exclusively minimizes UFC might select latency-sensitive requests at the cost of poor batching, harming system throughput. Conversely, a scheduler that strictly maximizes RFC could prioritize high-throughput batches, potentially starving users with less computationally convenient requests.

Our approach is to unify these counters into a single, composite score that guides scheduling decisions. We formally define the Holistic Fairness (HF) for a user $f$ as a weighted linear combination of their normalized UFC and RFC values:
\[
\label{eq:hfs}
\text{HF}_f = \alpha \cdot \text{UFC}_f + \beta \cdot \text{RFC}_f, 
\]
where $\alpha$ and $\beta$ are weighting parameters that represent the system's strategic priorities, with $\alpha + \beta = 1$. 

The scheduler's objective is to maintain max-min fairness by always prioritizing requests from the user with the lowest current HF. This ensures that over time, all users receive a balanced QoS according to our holistic definition. Crucially, the weights $\alpha$ and $\beta$ allow operators to tune the fairness policy. In our framework, we assert that the end-user experience is the ultimate goal of the service. Therefore, we assign a higher weight to the UFC, setting $\alpha > \beta$, reflecting a strong but not absolute preference for user-centric metrics over system-centric ones. The sensitivity and impact of $\alpha$ and $\beta$ are evaluated in ~\autoref{sec:ufc_rfc}. As an illustration, \autoref{fig:intuition_1} applies Eq.~\eqref{eq:hfs} and our policy choice \(\alpha>\beta\): a traditional token-aware scheduler (VTC) would select \emph{user0} solely because it has accumulated fewer tokens. However, Equinox recognizes that \emph{user0} already enjoys lower latency. Because our policy weights the latency-sensitive UFC more heavily, the composite HF score correctly identifies \emph{user1} as more underserved, demonstrating a more nuanced approach to fairness.
\vspace{-1em}

\section{System Design}

We introduce Equinox, a multi-stage system designed to handle the demanding task of serving Large Language Model (LLM) inference requests by enforcing holistic fairness. Holistic fairness represents our core architectural principle, balancing user-centric metrics (token allocation and latency) with system-centric objectives (GPU utilization and throughput) to achieve equitable service delivery. The architecture comprises four components that collectively implement this holistic fairness paradigm:
(i) The Server Frontend ingests requests, performs authentication and semantic validation, and applies rate limiting.  
(ii) Request Queues buffer validated tasks, decouple ingestion from execution and support dynamic reordering based on holistic fairness scores.  
(iii) A Mixture of Prediction Experts predicts output lengths by routing queries to specialized proxy models, supplying accurate scheduling estimates essential for holistic fairness calculations.  
(iv) The Holistic Fairness Scheduler reprioritizes requests by computing and balancing UFC and RFC scores through the holistic fairness equation, ensuring equitable resource allocation while maximizing system efficiency.

\noindent \textbf{Workflow Overview.} ~\autoref{fig:system_design} illustrates the end to end pipeline implementing holistic fairness. Clients submit requests to the \textbf{Server Frontend} (\ding{182}); invalid inputs are dropped, whereas valid ones enter the distributed \textbf{Request Queues} (\ding{183}). The \textbf{Prediction Framework} estimates response token counts and maps historical performance metrics for each queued request (\ding{184}), providing critical input for holistic fairness computation. The \textbf{Holistic Fairness Scheduler} employs these predictions and mappings to calculate accurate UFC and RFC scores. The scheduler then applies the holistic fairness equation to rank clients and batches requests for GPU execution, selecting those from clients with the lowest holistic fairness scores (\ding{185}). After reprioritization based on holistic fairness (\ding{186}), selected requests execute on GPU accelerators (\ding{187}), where the system continuously monitors key performance indicators like latency, compute utilization, and throughput, leveraging insights derived from offline profiling on representative workloads such as the lmsys-chat-1m dataset. These metrics feed back into the holistic fairness calculation, creating a continuous optimization loop. Upon completion, generated tokens stream back through the Server Frontend for response assembly and delivery (\ding{188}), completing the holistic fairness driven service cycle.

\begin{algorithm2e}[t!]
\SetAlgoLined
\caption{Holistic Fair Scheduling Algorithm}
\label{alg:holistic}
\KwIn{$\alpha, \beta, \delta$, batch size $L_b$, GPU memory $M$} \KwOut{$B$}
Initialize Predictor $P$, and PriorityQueue $Q$\;

\While{continuously}{
    \ForEach{incoming $req$ from client $c$}{
        $\mathrm{req.Tokens_{out}} \leftarrow P.\mathrm{predict}(req)$\;
        $(\mathrm{req.Latency}, \mathrm{req.GPU_{util}}, \mathrm{req.TPS}) \leftarrow P.\mathrm{map}(\mathrm{Tokens_{out}})$\;
        record $req.\mathrm{arrival}$\;
        $Q.\mathrm{enqueue}(req, \mathrm{HF}_c)$\;
    }
    
    $B \leftarrow \emptyset$\;
    \While{$Q$ not empty}{
        $c^* \leftarrow \arg\min_{c \in Q} \mathrm{HF}_c$\;
        $req \leftarrow Q.\mathrm{dequeue}(c^*)$\;
        
        \If{$\mathrm{canSchedule}(req, B, M, L_b)$}{
            $B \leftarrow B \cup \{req\}$\;
            $\mathrm{updateCounter}(req, c^*)$\;
        }
    }
    
    Execute $B$ on GPU\;
    \ForEach{completed request from client $c$}{
        Update $\mathrm{HF}_c$ and $P.map()$ with actual metrics\;
    }
}
\end{algorithm2e}
\section{Algorithm Analysis}

The core logic of Equinox’s fair scheduling algorithm is presented in \autoref{alg:holistic}. The algorithm maintains a per-client holistic score $\mathrm{HF}_c$. It initializes the predictor $P$ and a priority queue $Q$ keyed by $\mathrm{HF}_c$. The loop runs continuously to be consistent with continuous batching. For each incoming request $req$ from client $c$, the scheduler invokes $P.\mathrm{predict}(req)$ to estimate $\mathrm{Tokens_{out}}$ and uses $P.\mathrm{map}(\cdot)$ to attach predicted $(\mathrm{Latency},\mathrm{GPU_{util}},\mathrm{TPS})$ to the request. The request’s arrival time is recorded, and the request is enqueued keyed by the current $\mathrm{HF}_c$.

The scheduler creates an empty batch $B$ and, while $Q$ is non-empty, repeatedly selects the client $c^*$ with the minimum $\mathrm{HF}_{c^*}$, dequeues its head request, and check feasibility via $\mathrm{canSchedule}(req,B,M,L_b)$. $\mathrm{canSchedule}(req,B,M,L_b)$ checks whether the new request satisfies the current batch-size and GPU-memory resource constraints. If feasible, the request is loaded into $B$ and the scheduler calls $\mathrm{updateCounter}(req,c^*)$ to update $\mathrm{HF}_{c^*}$ using the request’s predicted $(\mathrm{Tokens_{out}},$ $\mathrm{Latency}, \mathrm{GPU_{util}}, \mathrm{TPS})$ together with the observed wait time $(\mathrm{now}-req.\mathrm{arrival})$. Because selection always picks the smallest $\mathrm{HF}$, clients with slower-growing UFC tend to remain with lower $\mathrm{HF}$ and are preferentially admitted next, while the RFC nudges the batch toward efficient GPU utilization.

After building $B$, the batch executes on the GPU. Upon completion, the scheduler refreshes $\mathrm{HF}_c$ using the actual metrics and updates $P.\mathrm{map}(\cdot)$ accordingly. This closes the feedback loop and continuously calibrates predictions to the observed hardware behavior.
\vspace{-1em}

\begin{figure}[t]
    \centering
    \begin{subfigure}[b]{0.48\linewidth}
        \centering
        \includegraphics[width=\linewidth]{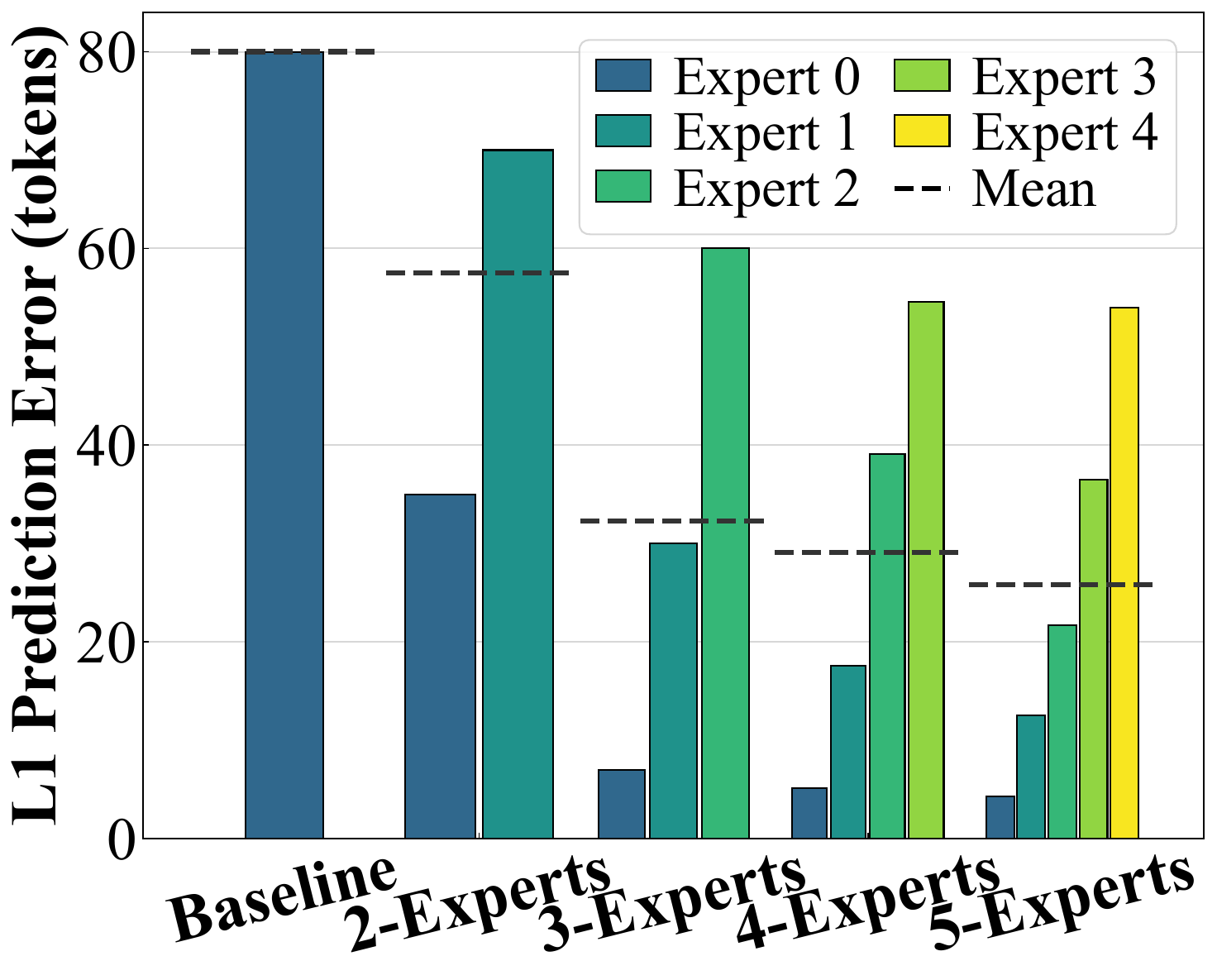}
        \caption{Expert specialization.}
        \label{fig:mope3}
    \end{subfigure}
    \hfill
    \begin{subfigure}[b]{0.48\linewidth}
        \centering
        \includegraphics[width=\linewidth]{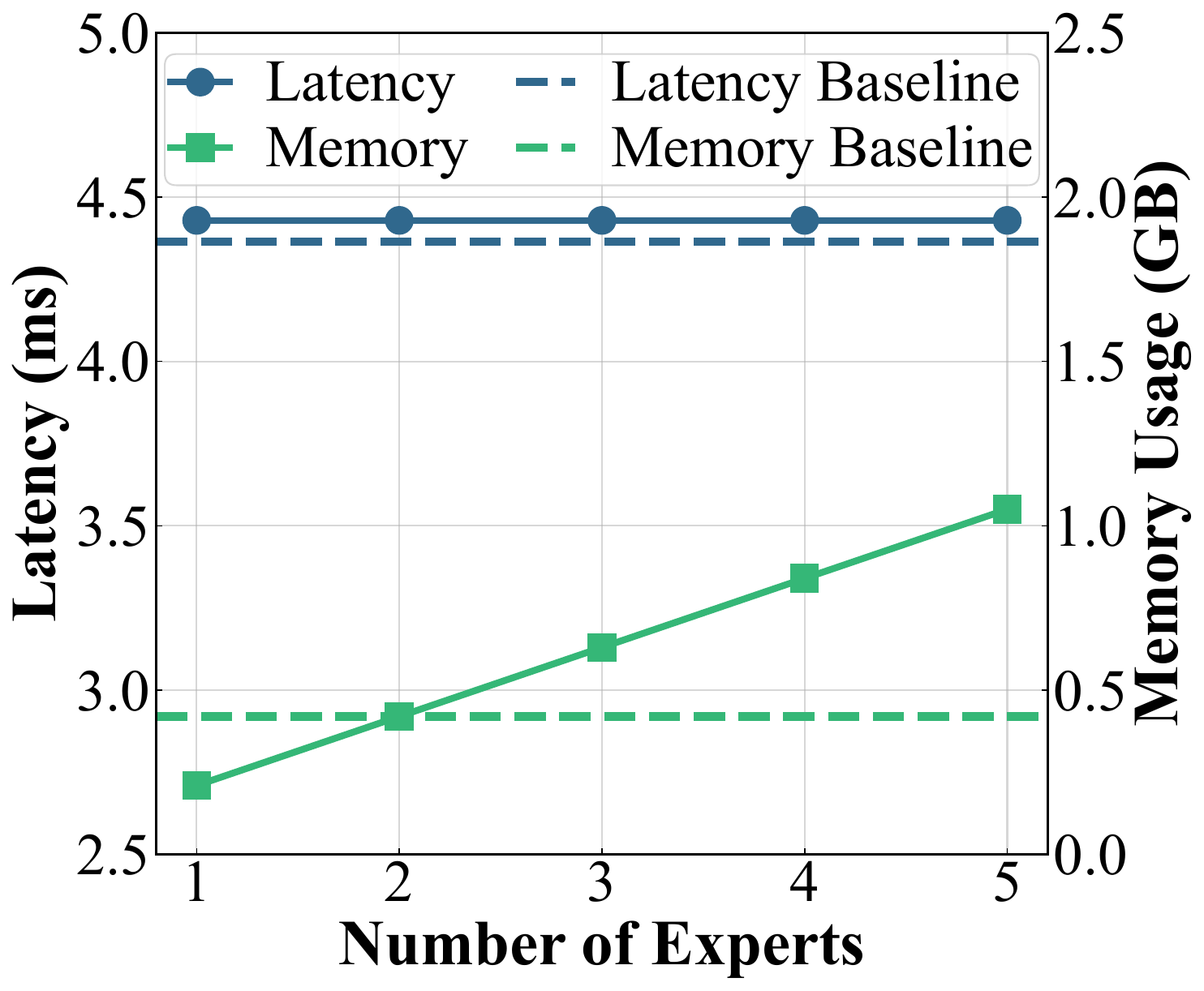}
        \caption{System trade‑offs.}
        \label{fig:mope4}
    \end{subfigure}
    \hfill
    \begin{subfigure}[b]{0.48\linewidth}
        \centering
        \includegraphics[width=\linewidth]{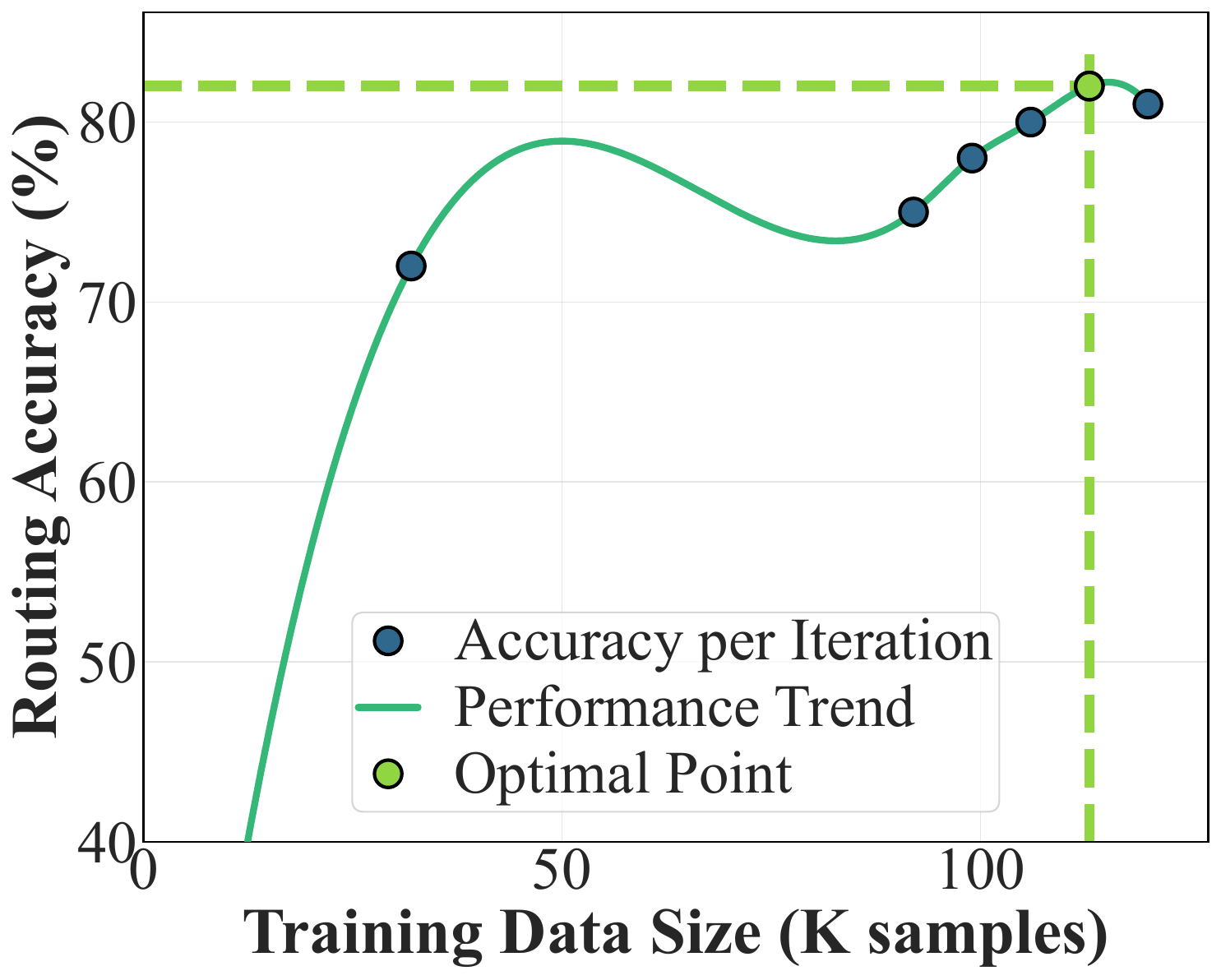}
        \caption{Router learning patterns.}
        \label{fig:mope5}
    \end{subfigure}
    \hfill
    \begin{subfigure}[b]{0.48\linewidth}
        \centering
        \includegraphics[width=\linewidth]{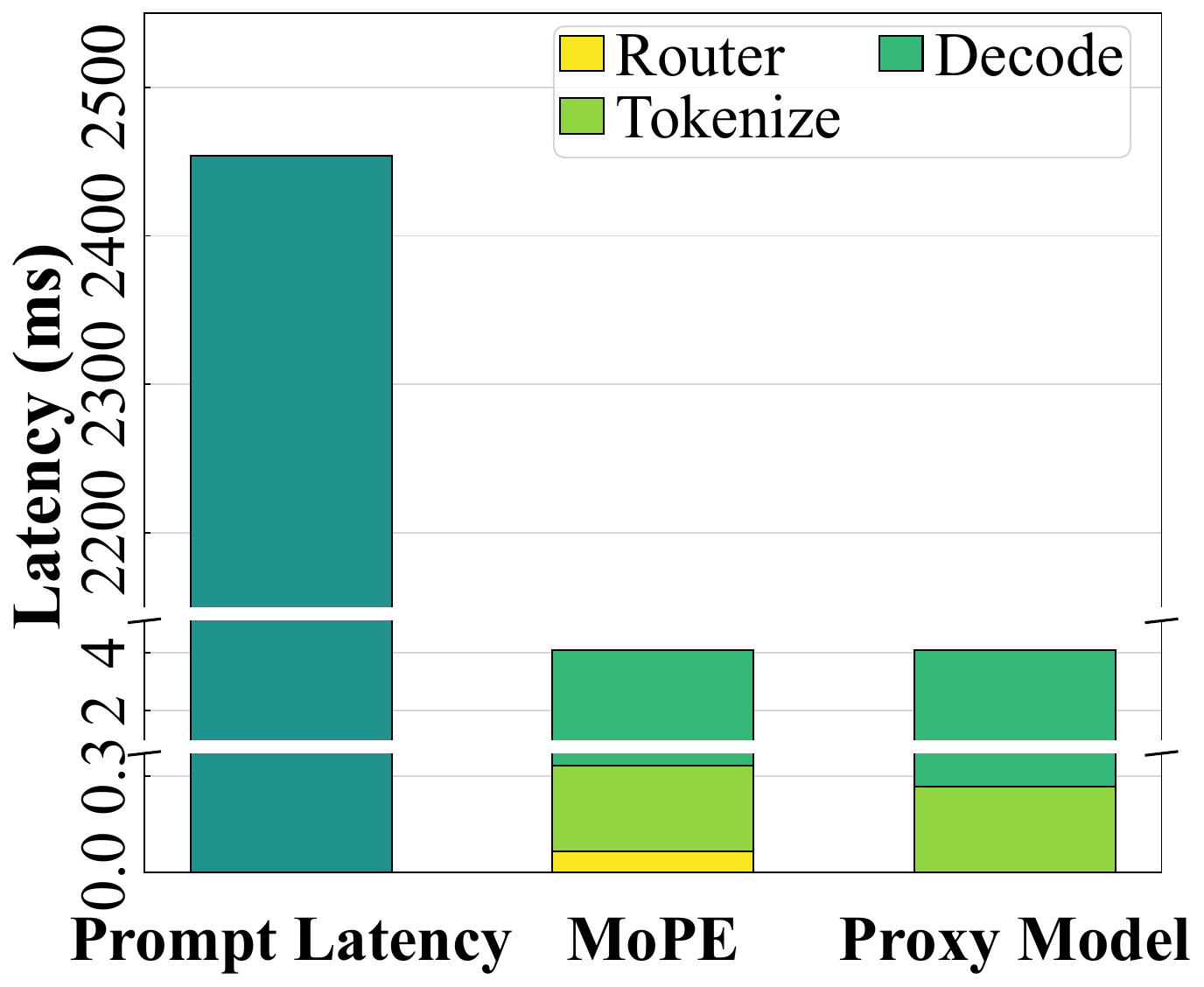}
        \caption{Latency breakdown.}
        \label{fig:mope6}
    \end{subfigure}
    \caption{MoPE insights and data analysis.}
    \label{fig:mope-insight}
    \vspace{-2em}
\end{figure}

\section{Prediction Framework for Scheduling}
\label{sec:mope}

To address the limitations of relying on a single, generic proxy model for token count prediction, we introduce \textit{MoPE}, a framework comprising multiple specialized predictors (\emph{experts}) coordinated by a routing layer. MoPE dynamically assigns incoming requests to the most suitable expert, effectively handling diverse prompt types and output token lengths characteristic of modern LLM workloads. Unlike conventional mixture-of-experts architectures focused on generation, MoPE specifically targets accurate token count prediction within a lightweight structure suitable for production LLM serving.

\begin{figure}[t]
    \centering
    \setlength{\abovecaptionskip}{0.1cm}
    \setlength{\belowcaptionskip}{-0.1cm}
    \includegraphics[width=0.5\textwidth]{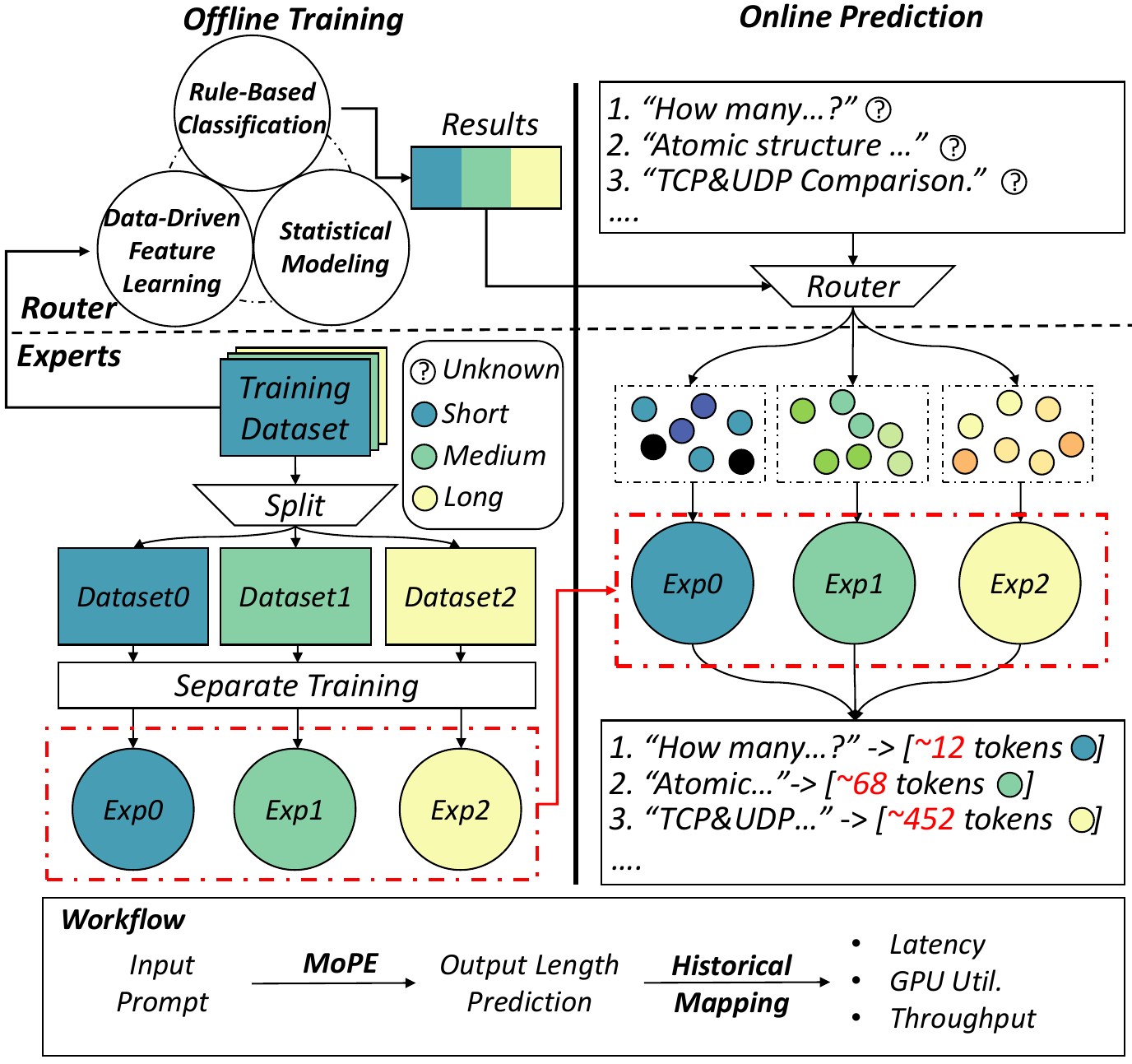} 
    \caption{Overview and workflow of MoPE architecture}
    \label{fig:moe_system_design}
    \vspace{-1em}
\end{figure}

The offline training pipeline (\autoref{fig:moe_system_design}, left) first trains the router. Using the true output lengths from the training corpus, the router learns to classify prompts based on input length thresholds and automatically identified keywords indicative of output length classes. This process employs feature embedding and similarity lookups, balancing different signals via a mixing weight, and iteratively selects data to maximize classification accuracy.

The expert training phase then partitions the corpus according to the router's learned classifications (e.g., short, medium, long regimes). For each partition, a specialized regression BERT-base model is fine-tuned, typically by adapting a pretrained encoder with a regression head~\cite{predictor_proxy}. Training focuses on prediction accuracy (e.g., minimizing L1 error) via standard techniques like stratified splits and early stopping.

As detailed in ~\autoref{fig:mope-insight}, empirical evaluations show that a configuration of three experts strikes the optimal balance. ~\autoref{fig:mope3} Specifically, the L1 prediction error is 80 for one expert, 33 for three experts, and 25 for five experts. We reduce the model weights from FP32 to BF16 precision to decrease memory consumption for multiple experts, therefore the resource usage (memory, latency) increases substantially beyond three experts (\autoref{fig:mope4}). For router training, We present the router accuracy performance across training data sizes up to 120k from LMSYS dataset~\cite{zheng2024lmsyschat1mlargescalerealworldllm}. Accuracy gradually increases before 50k samples, reaching a local optimum. It then achieves peak accuracy of 80\% at approximately 110k samples in \autoref{fig:mope5}. We report the end-to-end latency breakdown of MoPE and prompt inference in~\autoref{fig:mope6}. MoPE only introduces router overhead (0.02 ms) compared to the original proxy model, with total MoPE inference time being 4.5 ms. This represents less than 1\% of the average prompt latency (2,400 ms), demonstrating that the overall latency added by MoPE remains negligible compared to total prompt processing time. We also compare the end-to-end prediction errors between MoPE and the stae-of-the-art proxy model, as detailed in~\autoref{fig:error-analysis}. These experiments collectively support the three-expert configuration as maximizing predictive accuracy relative to system cost.

During online prediction, the router rapidly classifies an incoming prompt and directs it to the relevant expert (\texttt{short}, \texttt{medium}, or \texttt{long}), incorporating the target LLM identity during preprocessing. The selected expert performs a forward pass to generate the token length prediction. Finally, we model token-latency-throughput-GPU utilization relationships using historical offline data in ~\autoref{fig:motivation-fcfs-vtc}. For inputs processed by MoPE, we use this historical mapping to predict all metrics required for computing Holistic Fairness (HF). The offline modeling method we use has been validated for latency prediction~\cite{predictor_roofline}, we extend it to estimate GPU utilization and throughput.

\section{Evaluation}

\begin{figure*}[htbp]
    \centering
    \begin{subfigure}[b]{0.24\linewidth}
        \includegraphics[width=\linewidth]{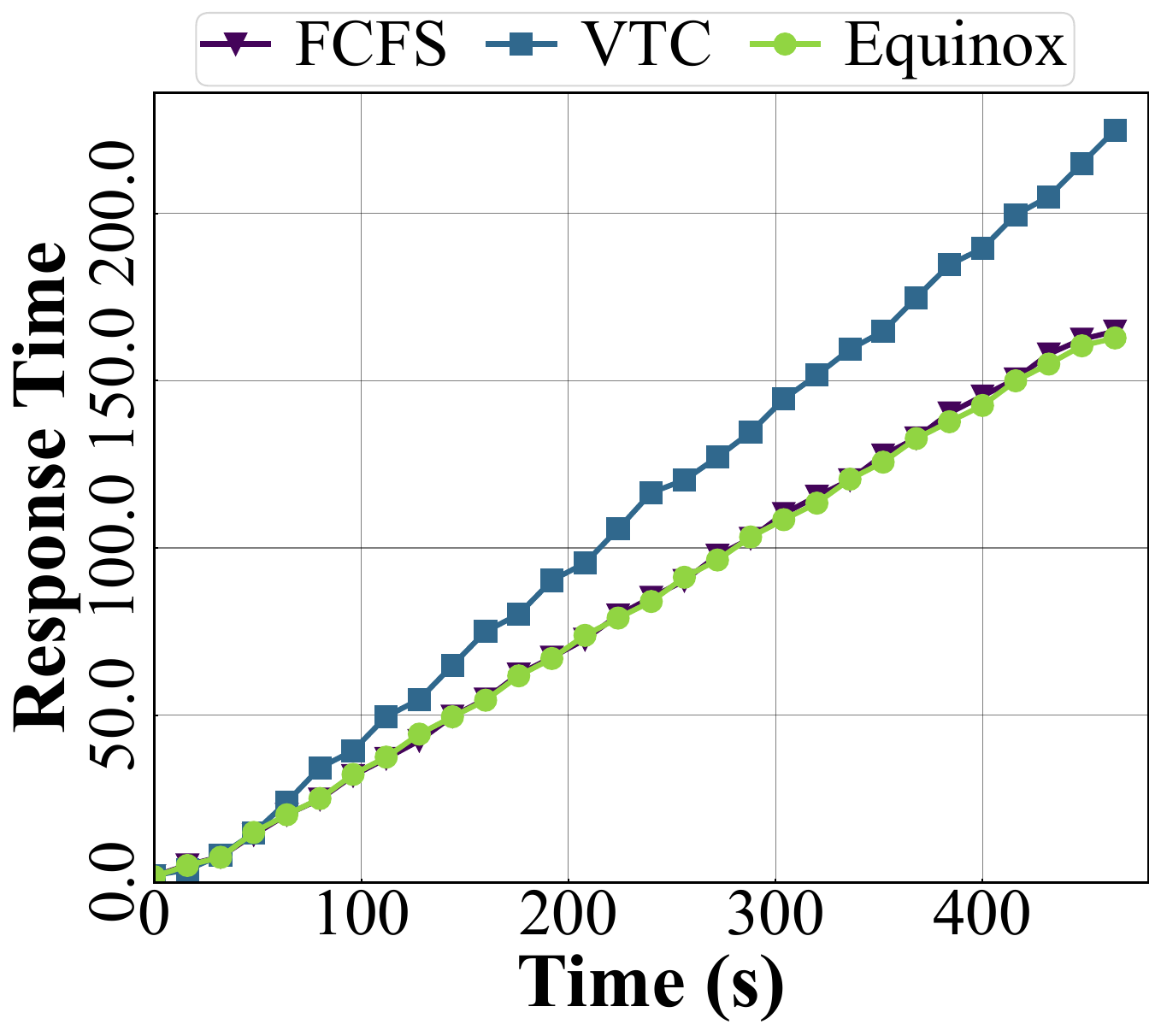}
        \caption{Total Response Time}
        \label{fig:balanced_resp_time}
    \end{subfigure}
    \hfill
    \begin{subfigure}[b]{0.24\linewidth}
        \includegraphics[width=\linewidth]{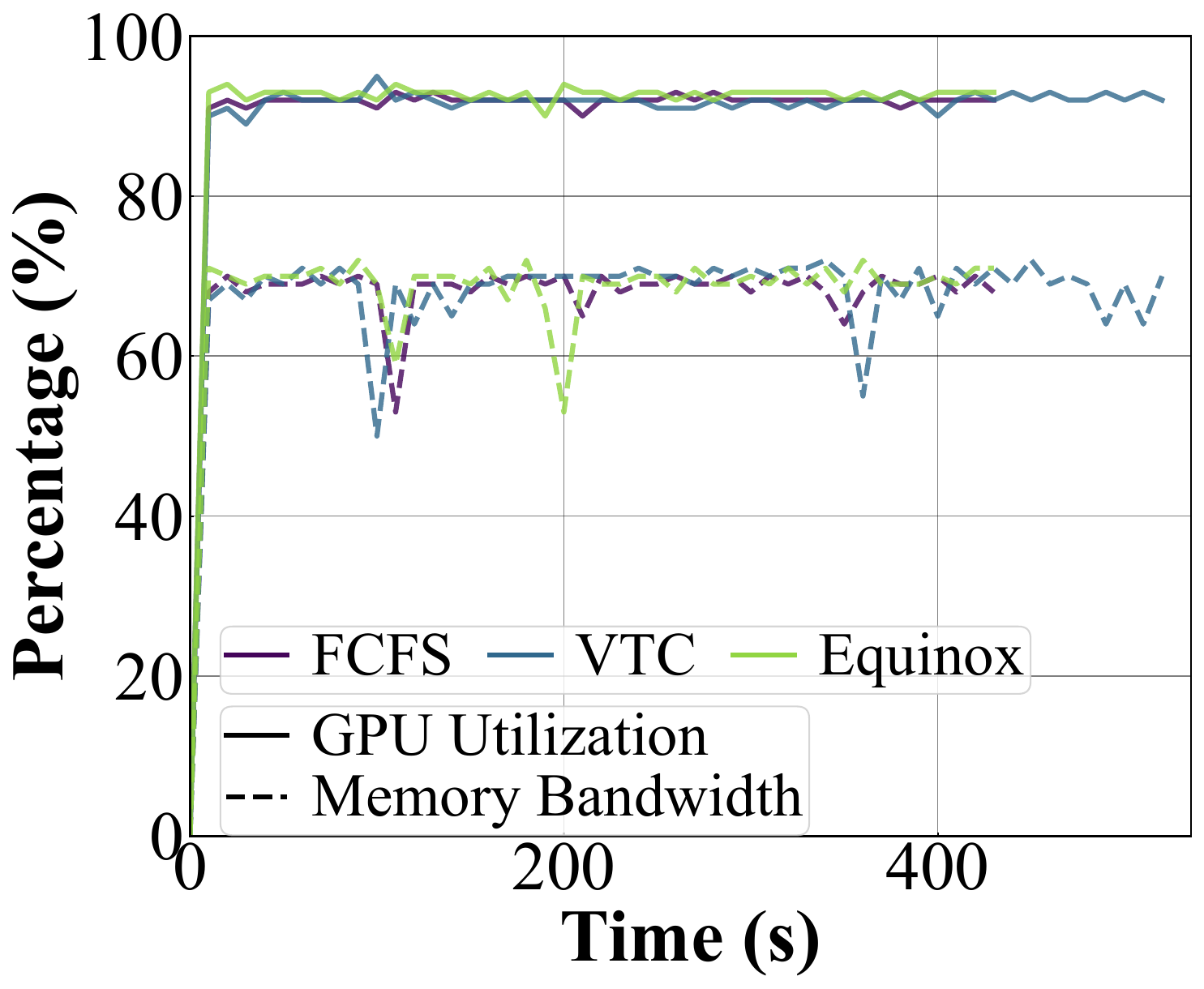}
        \caption{GPU Util. \& Bandwidth}
        \label{fig:balanced_gpu}
    \end{subfigure}
    \hfill
    \begin{subfigure}[b]{0.24\linewidth}
        \includegraphics[width=\linewidth]{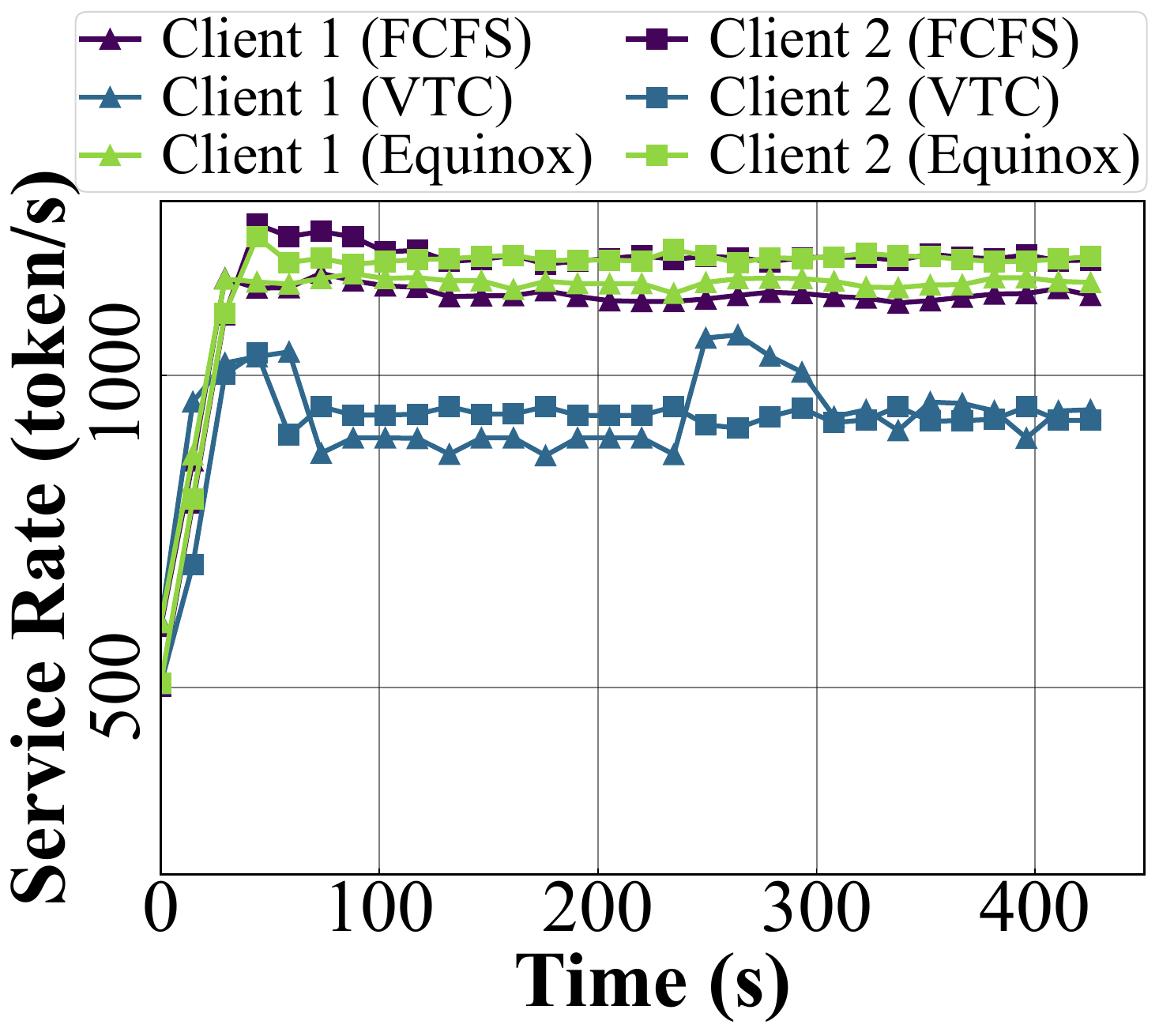}
        \caption{Per-client Service Rate}
        \label{fig:balanced_service_rate}
    \end{subfigure}
    \hfil
    \begin{subfigure}[b]{0.24\linewidth}
        \includegraphics[width=\linewidth]{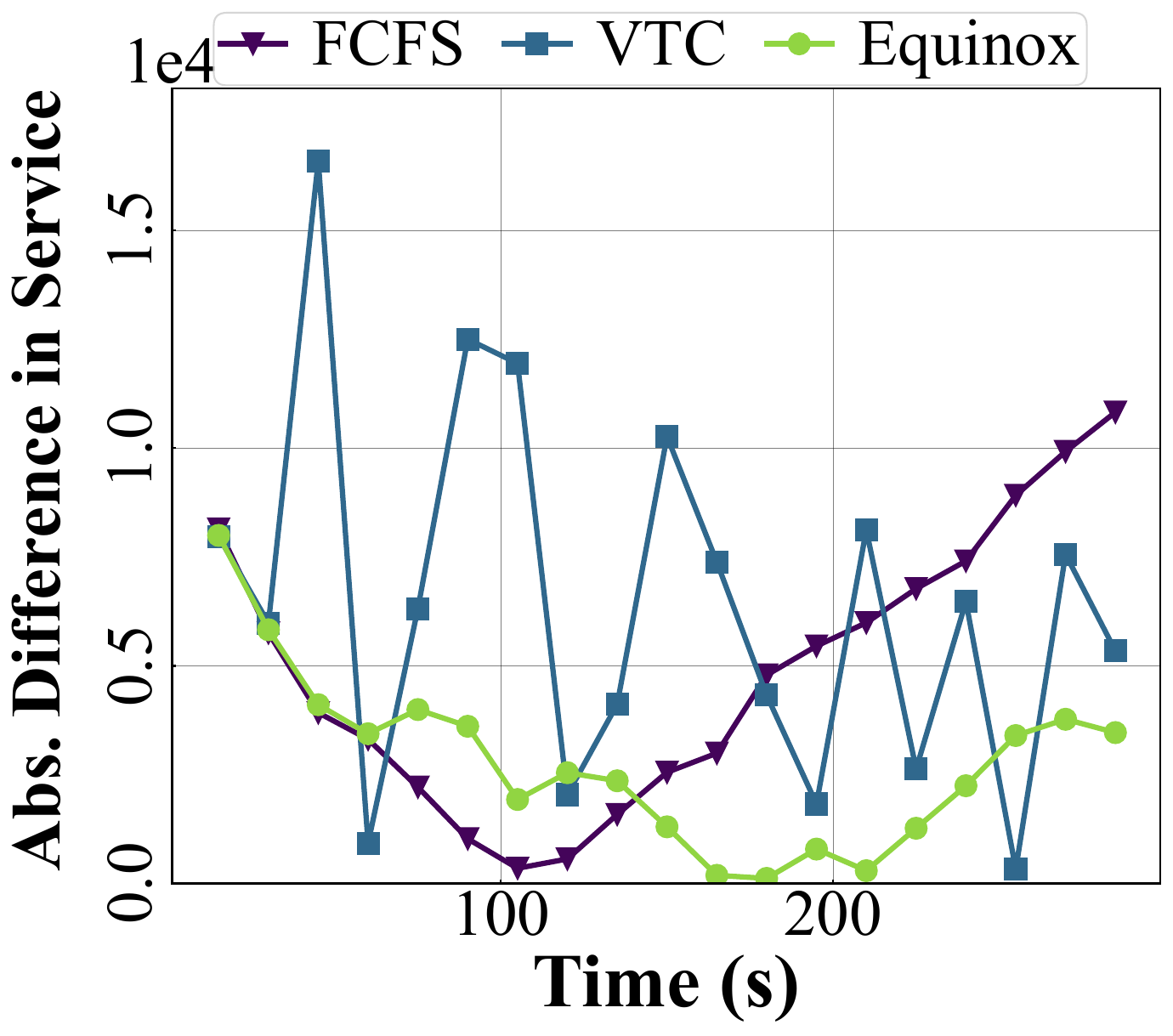}
        \caption{Service Difference}
        \label{fig:balanced_service_diff}
    \end{subfigure}
    \caption{Balanced load scenario. Equinox maintains fairness despite different request rates and lengths.}
    \label{fig:synthetic_balanced}
    \vspace{-1em}
\end{figure*}

\begin{figure*}[htbp]
    \centering
     \begin{subfigure}[b]{0.24\linewidth}
        \includegraphics[width=\linewidth]{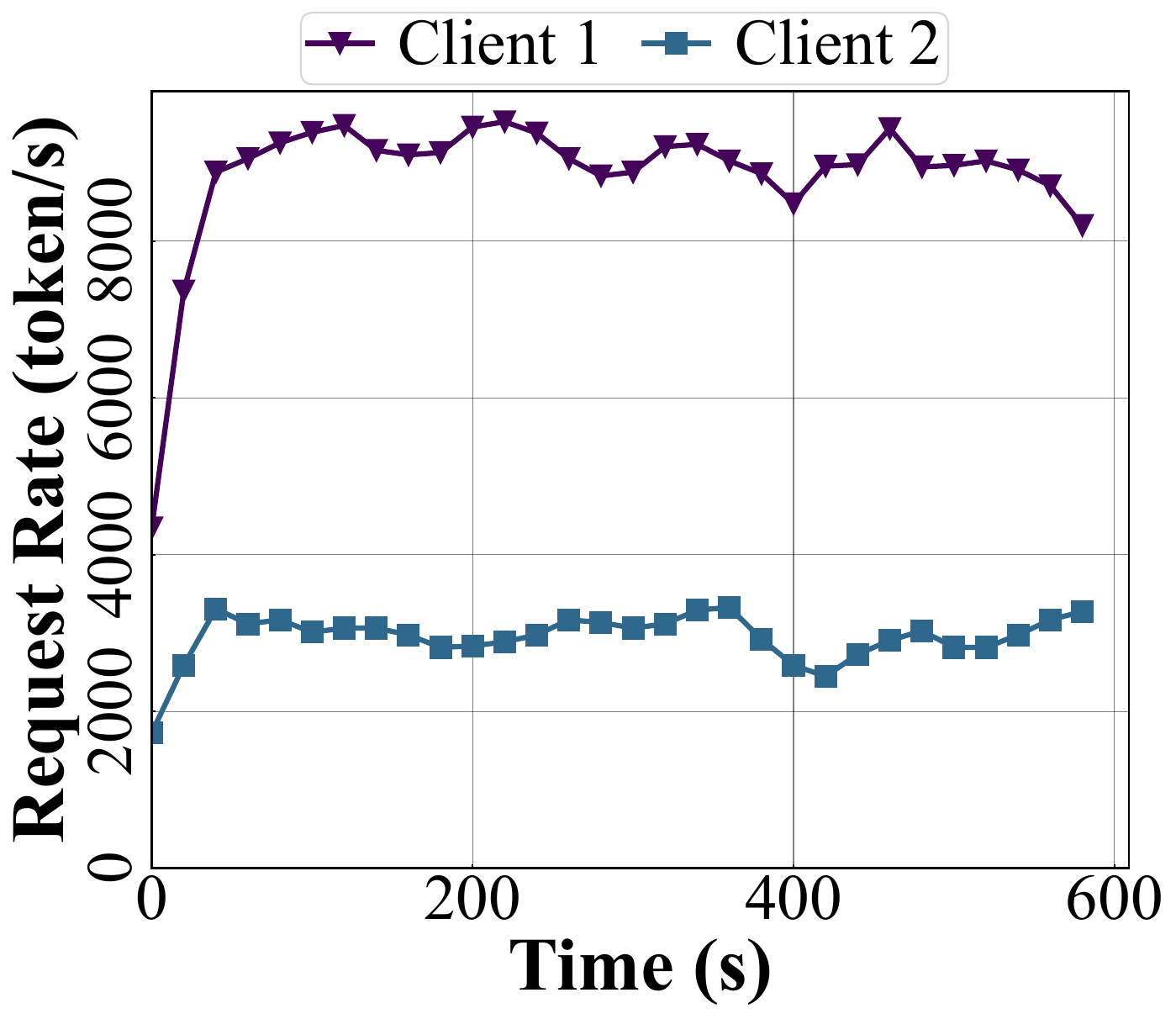}
        \caption{Client Request Rate}
        \label{fig:poisson_req_rate}
    \end{subfigure}
    \hfill
    \begin{subfigure}[b]{0.24\linewidth}
        \includegraphics[width=\linewidth]{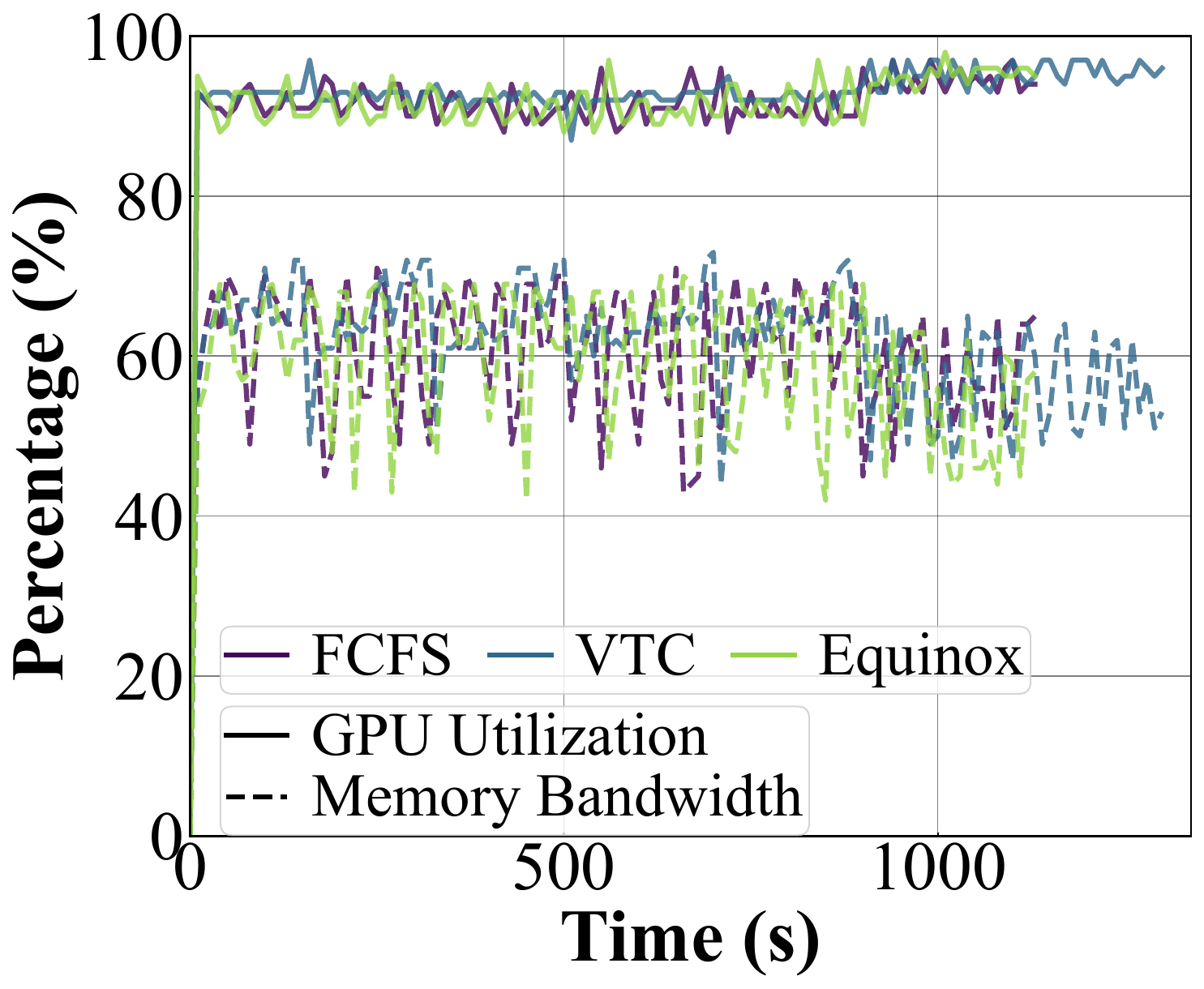}
        \caption{GPU Util. \& Bandwidth}
        \label{fig:poisson_gpu}
    \end{subfigure}
    \hfill    
    \begin{subfigure}[b]{0.24\linewidth}
        \includegraphics[width=\linewidth]{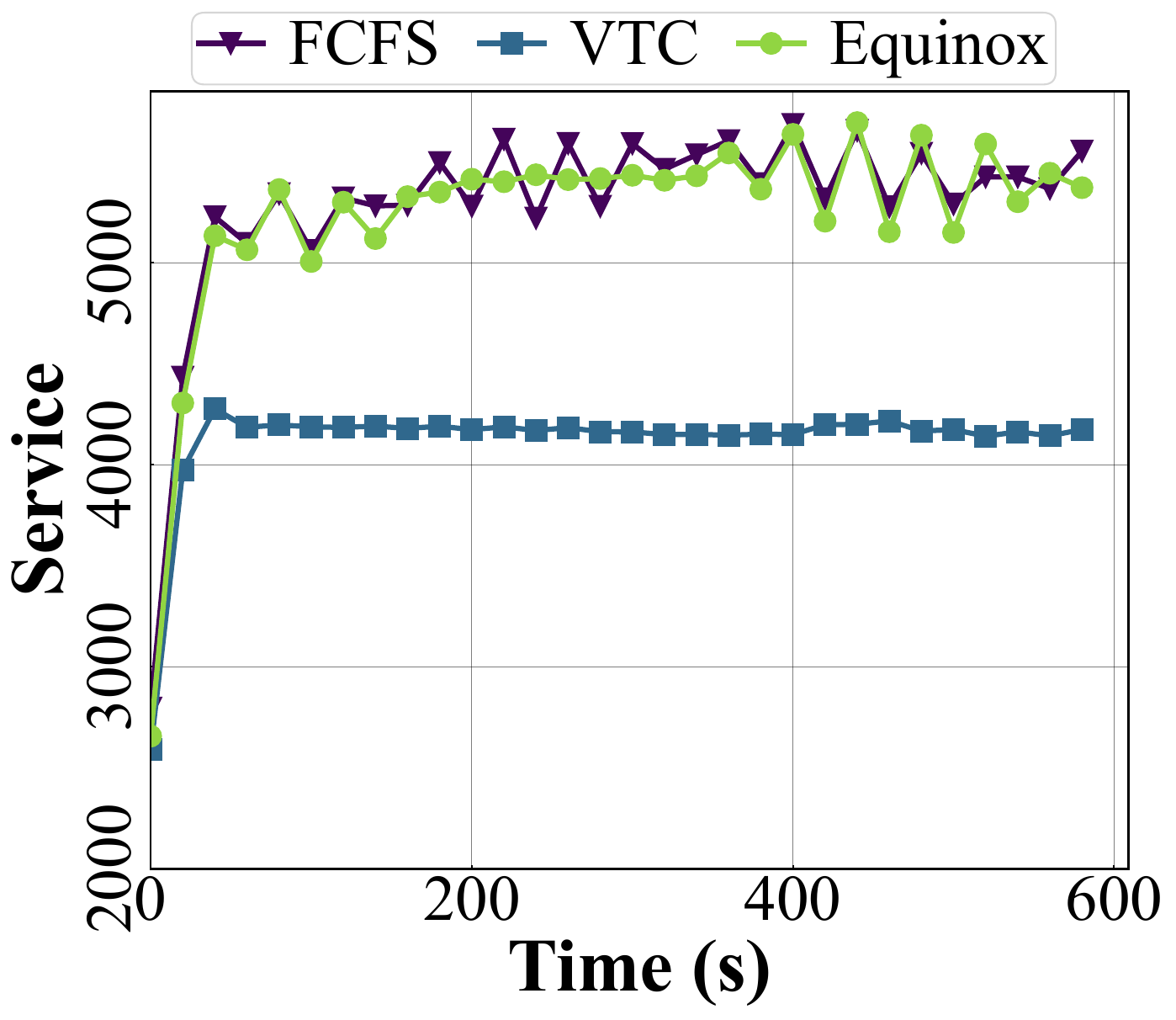}
        \caption{Total Service Rate}
        \label{fig:poisson_service_rate}
    \end{subfigure}
    \hfill
    \begin{subfigure}[b]{0.24\linewidth}
        \includegraphics[width=\linewidth]{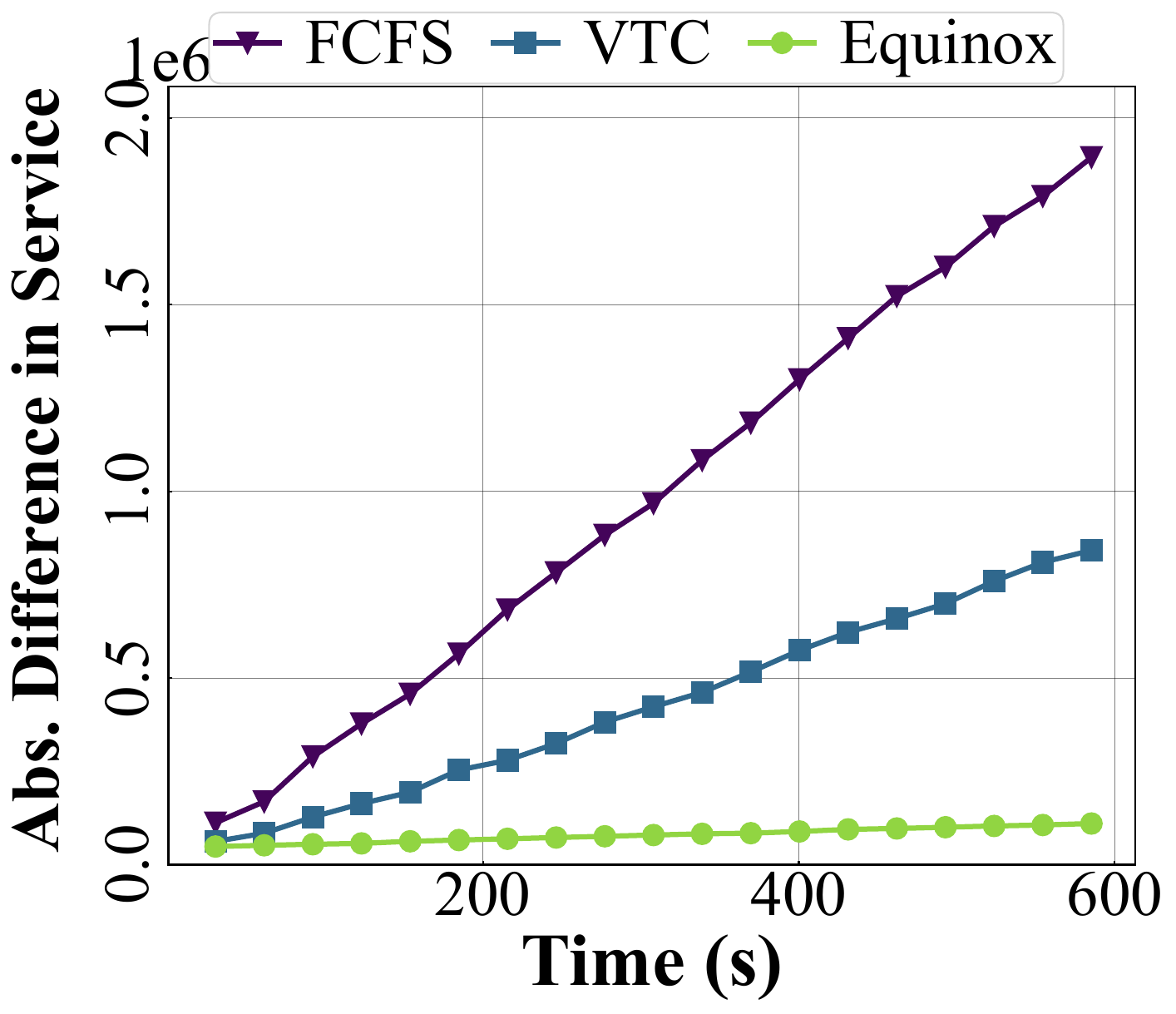}
        \caption{Service Difference}
        \label{fig:poisson_service_diff}
    \end{subfigure}
    \caption{Poisson arrivals with heterogeneous requests. Equinox maintains fairness under stochastic load.}
    \label{fig:synthetic_poisson}
    \vspace{-1em}
\end{figure*}

We implemented Equinox atop existing systems (S-LoRA ~\cite{slora}, vLLM ~\cite{10.1145/3600006.3613165}, SGLang~\cite{sglang}) using $\sim$1000 lines of Python. Its lightweight layer coordinates optimizations like continuous batching, chunked prefill ~\cite{llm_prediction_1}, PagedAttention ~\cite{10.1145/3600006.3613165}, and KV-cache reuse, and supports adapter-level fairness (e.g., per-LoRA).

\subsection{Evaluation Methodology}
\label{para:methodology}

\noindent \textbf{Workload and Testbed.} We use synthetic workloads and real traces from LMSYS Chatbot Arena ~\cite{zheng2024lmsyschat1mlargescalerealworldllm} and ShareGPT ~\cite{ShareGPT}. During runs, MoPE predicts token lengths and other resource demands, and the scheduler interleaves prefill/decode to balance UFC and RFC. For synthetic traces, experiments run on an A100 GPU (80GB) with Llama-2-7b, mirroring VTC's setup. For real-world traces, experiments run on an 8-A100 GPU (40GB) cluster with Llama-2-70b. The CPU is Intel(R) Xeon(R) Gold 5218@2.30GHz with 256 GB DRAM. For vLLM and SGLang, we use Tensor Parallelism (TP) = 8.

\noindent \textbf{Baselines.} For scheduling algorithm, we compare Equinox against FCFS and VTC. For prediction, we choose single proxy model~\cite{predictor_proxy} as the baseline. We further conduct an ablation study removing MoPE to isolates its contribution to fairness.

\noindent\textbf{MoPE Configuration.} Following Section~\ref{sec:mope}, we use 3 experts and a fariness metirc mapping trained on LMSYS Chatbot Arena dataset. Boundaries partition the training data into $33th, 66th$ and $99th$ points based on output length (<53, 53-210, >210 tokens). Generalizability is tested on the unseen ShareGPT dataset (Section~\ref{sec:real_world_evaluation}).

\noindent \textbf{Metrics.} Key metrics include:
\begin{itemize}
    \item \textbf{Service Rate:} per-client weighted tokens processed per second.
    \item \textbf{Service Difference:} Absolute difference in accumulated weighted service between clients.
    \item \textbf{latency:} We use response time (time to first token) and end-to-end latency to measure latency in different cases.
    \item \textbf{Jain's Fairness Index} is a widely-used metric for evaluating the fairness of resource allocation in networked systems. In our context, we compute Jain's Fairness Index by letting $x_i$ denote the \textbf{HF} of client $i$, which is defined in ~\autoref{sec:hfs}. The index is mathematically defined as:
\begin{equation}
J(x_1, x_2, \dots, x_n) = \frac{\left( \sum_{i=1}^n x_i \right)^2}{n \cdot \sum_{i=1}^n x_i^2}
\end{equation}
where $x_i$ represents the allocation for the $i^{th}$ client, and $n$ is the total number of clients. The value of $J$ ranges from $\frac{1}{n}$ (minimum fairness, when one client monopolizes all resources) to 1 (maximum fairness, when resources are equally distributed).

\end{itemize}

\subsection{Evaluation with Synthetic Workloads}
\label{sec:synthetic_evaluation}

We design experiments using synthetic workloads to rigorously evaluate the fairness properties of Equinox under controlled conditions. These scenarios allow for clear visualization and analysis of the scheduler's behavior. This section focuses on two key scenarios: a balanced load with heterogeneous client demands and a stochastic arrival pattern mimicking real-world unpredictability. Further analyses under overload, underload, and dynamic load conditions are detailed in Appendix~\ref{sec:appendix_synthetic}.





\begin{figure*}[htbp]
    \centering
    \begin{subfigure}[b]{0.33\linewidth}
        \includegraphics[width=\linewidth]{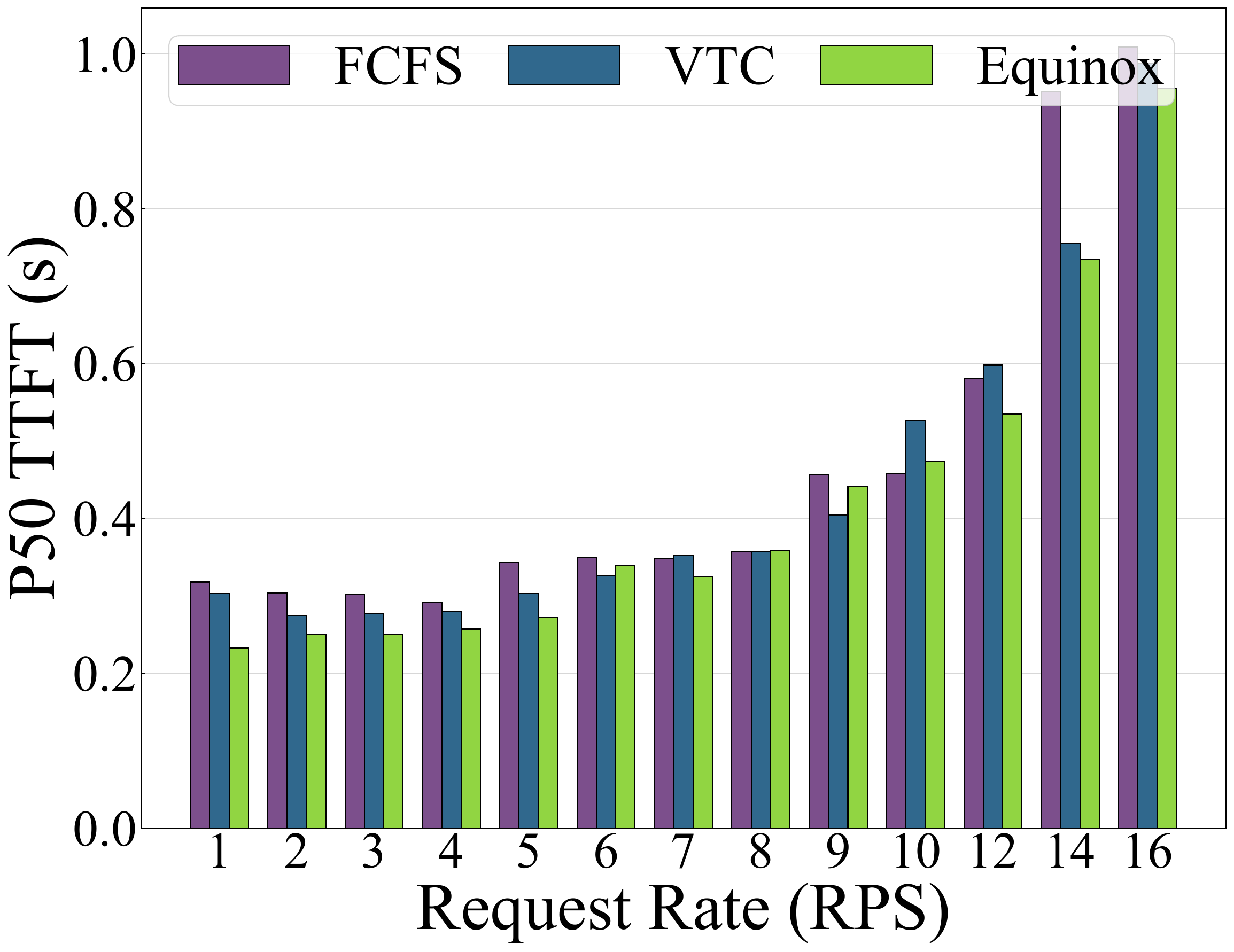}
        \caption{P50 TTFT} 
        \label{fig:SGLang_1}
    \end{subfigure}
    \hfill
    \begin{subfigure}[b]{0.33\linewidth}
        \includegraphics[width=\linewidth]{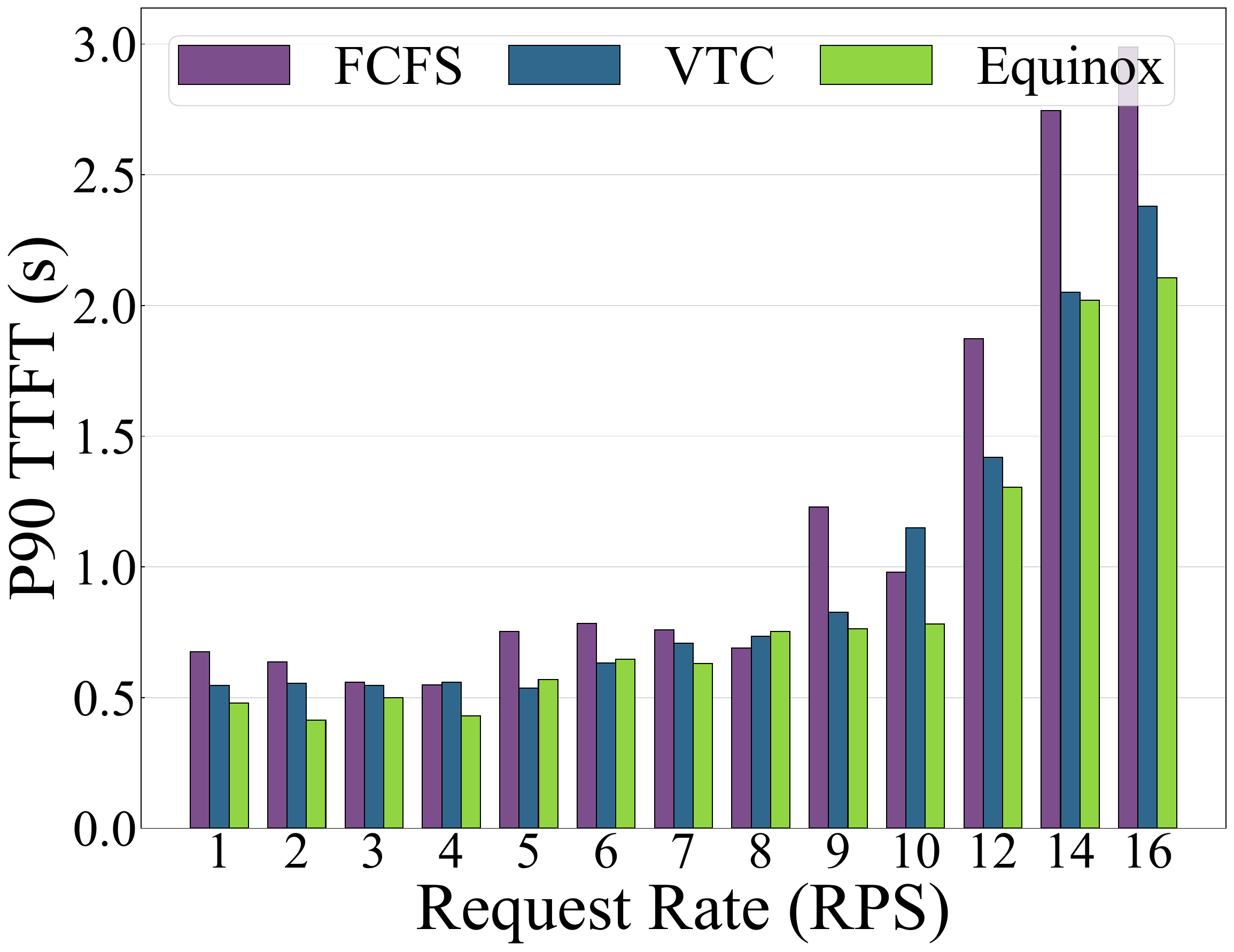}
        \caption{P90 TTFT} 
        \label{fig:SGLang_2}
    \end{subfigure}
    \hfill
    \begin{subfigure}[b]{0.33\linewidth}
        \includegraphics[width=\linewidth]{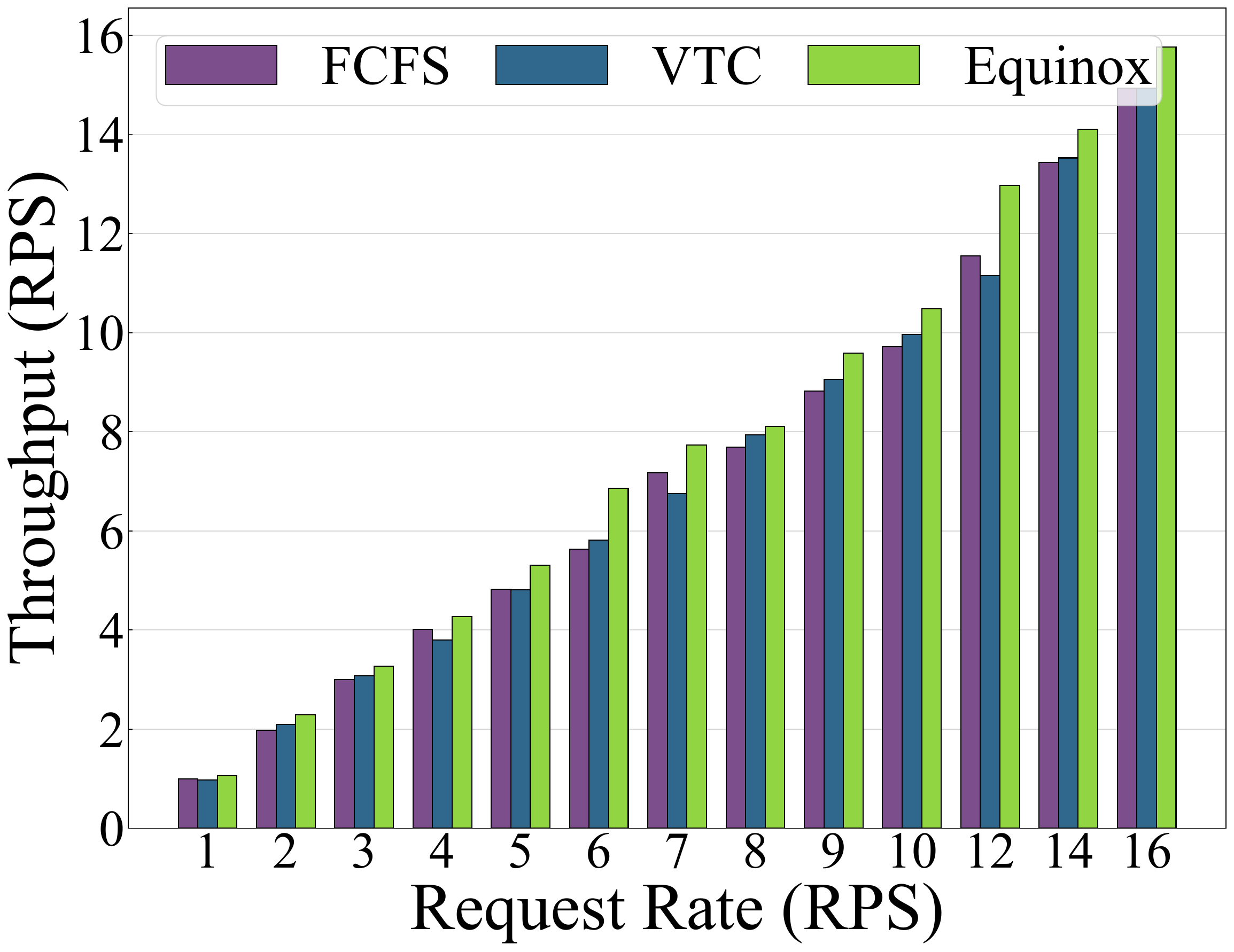}
        \caption{Throughput} 
        \label{fig:SGLang_3}
    \end{subfigure}
    \vspace{-0.15in}
    \caption{Performance comparison using ShareGPT trace in SGLang.}
    \vspace{-0.1in}
    \label{fig:SGLang}
\end{figure*}

\subsubsection{Balanced Load Scenario}
\label{sec:synthetic_balanced_scenario}

We configure a scenario where the system operates under a balanced load but serves clients with distinct demands. Client 1 sends requests at a constant rate of 2 req/s for input lengths of 100 and output lengths of 400, while Client 2 sends requests at a lower rate of 1 req/s but for longer output lengths of 900. This setup assesses fairness when request rates and per-request service needs differ under a manageable load.

\noindent {\bf Overall performance.} Equinox successfully maintains fairness between clients with differing request characteristics while achieving high efficiency, as illustrated in~\autoref{fig:synthetic_balanced}. Appendix \autoref{fig:appendix_overload_service_rate} show that, Equinox’s fairness algorithm prioritizes throughput and delivers a higher total service rate than FCFS or VTC (1.3x). Because the balanced workload alternates between light and heavy contention, Equinox is able to capture both advantages: its service distance and aggregate service rate are simultaneously better than those of FCFS and VTC.

\noindent {\bf Breakdown.} The fairness achieved by Equinox is evident in the per‑client service rate (\autoref{fig:balanced_service_rate}), showing comparable service allocation over time and modestly higher absolute throughput than FCFS or VTC (1.3x). Crucially, the accumulated service difference remains tightly bounded (\autoref{fig:balanced_service_diff}), demonstrating consistent fairness enforcement. Mirroring the underload study, Equinox also maintains up to 60 \% lower response times than VTC (\autoref{fig:balanced_resp_time}) while sustaining high GPU utilization (\autoref{fig:balanced_gpu}), confirming that fairness need not come at the cost of efficiency.

\subsubsection{Stochastic Arrivals Scenario}
\label{sec:synthetic_poisson_scenario}
To simulate more realistic, unpredictable traffic, we employ a Poisson process for request arrivals. Client 1 issues requests with an average rate of 16 req/s, primarily requiring prefill computation (input lengths of 512 and output lengths of 32). Client 2 issues requests at a lower average rate of 3 req/s, focusing on decoding (input lengths of 32 and output lengths of 512). Detailed request pattern is shown in \autoref{fig:poisson_req_rate}. This configuration tests Equinox's robustness under stochastic arrivals and mixed computational demands (prefill-heavy vs. decoding-heavy).

As depicted in ~\autoref{fig:synthetic_poisson}, Equinox demonstrates holistic fairness even when subjected to stochastic arrivals and heterogeneous request types. It effectively keeps the total service rate virtually identical to FCFS in ~\autoref{fig:poisson_service_rate}, confirming that its fairness logic does not sacrifice throughput even when arrivals are highly variable. Simultaneously, it significantly reduces the accumulated service difference compared to both baseline methods (\autoref{fig:poisson_service_diff}). VTC’s token‑level metric fails here because it does not have the predictor mechanism thus undervaluing the long‑decode requests of Client 2 while Equinox’s MoPE design corrects this bias.

\subsection{Evaluation with Real-world Workloads}
\label{sec:real_world_evaluation}

This section focuses on Equinox's performance within the SGLang and vLLM platform and presents a direct comparison of fairness across multiple setups and metrics. Analyses using traces within S-LoRA are detailed in ~\autoref{sec:appendix_real_world}. 

\subsubsection{SGLang with ShareGPT Dataset}
\label{sec:real_sglang_scenario}

\begin{figure*}[htbp]
    \centering
    \begin{subfigure}[b]{0.24\linewidth}
        \includegraphics[width=\linewidth]{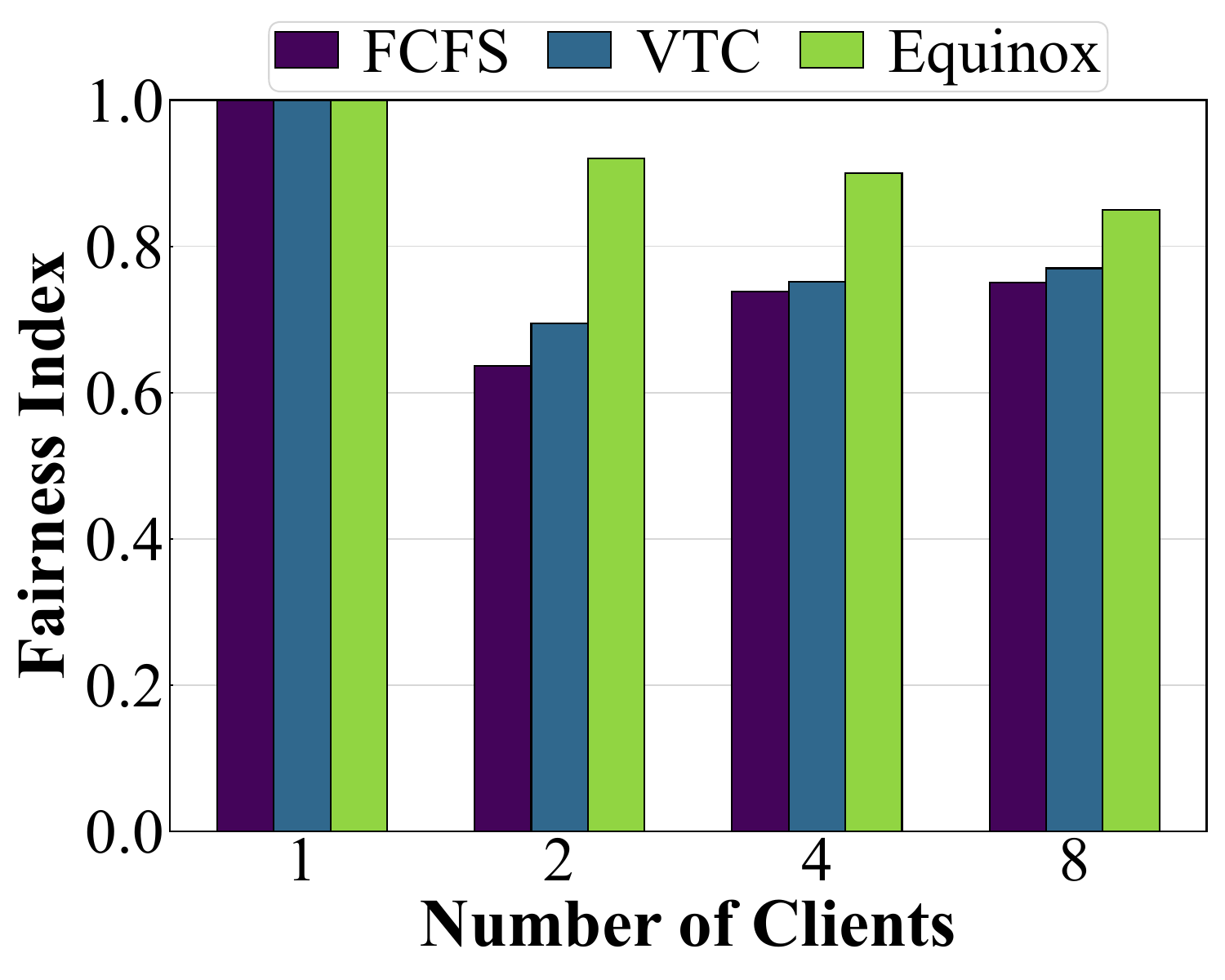} 
        \caption{Jain's Fairness Index} 
        \label{fig:vllm_1}
    \end{subfigure}
    \hfill
    \begin{subfigure}[b]{0.24\linewidth}
        \includegraphics[width=\linewidth]{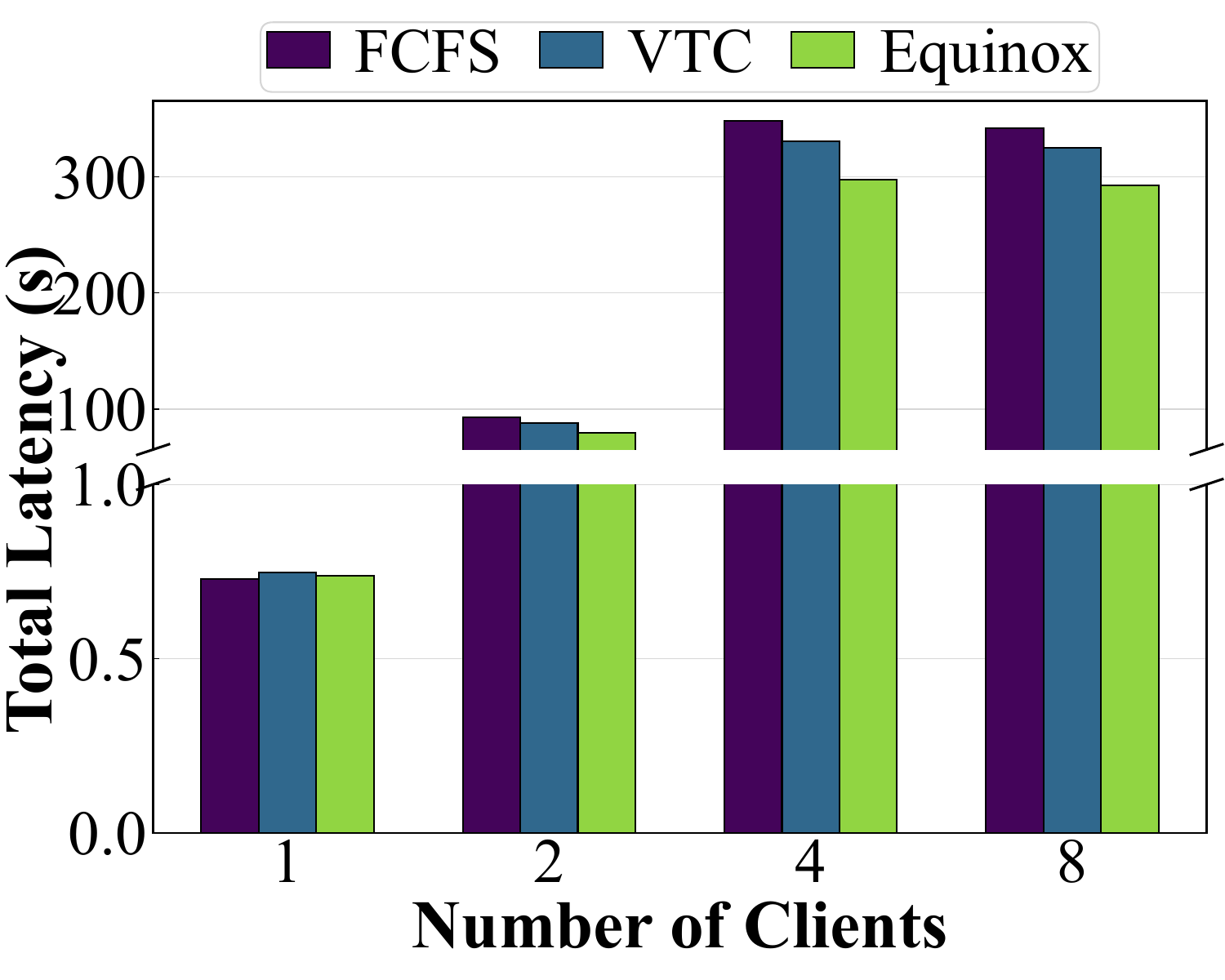}
        \caption{Per-client Response Time}
        \label{fig:vllm_2}
    \end{subfigure}
    \hfill
    \begin{subfigure}[b]{0.24\linewidth}
        \includegraphics[width=\linewidth]{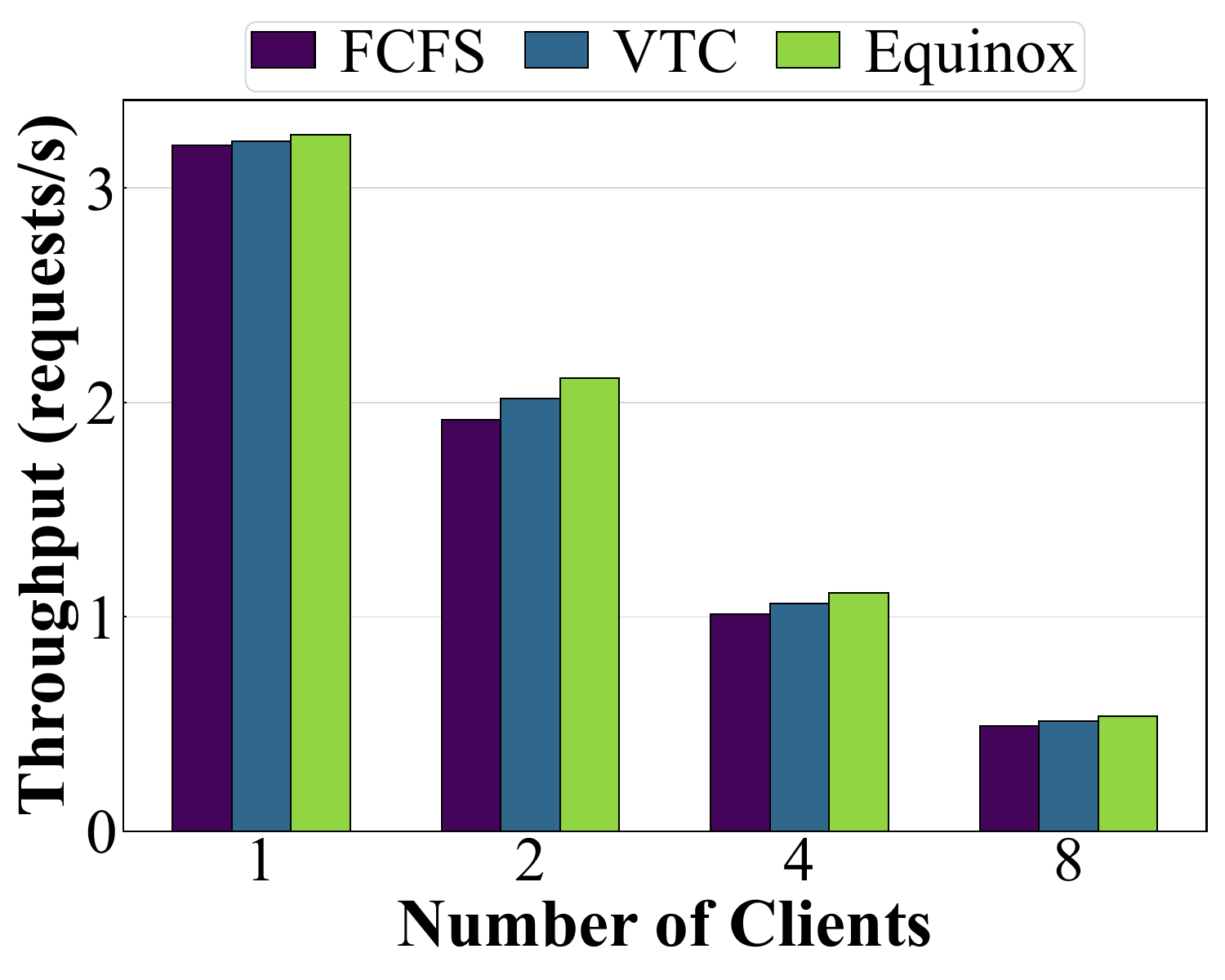}
        \caption{Per-client Recv. Service Rate}
        \label{fig:vllm_3}
    \end{subfigure}
    \hfill
    \begin{subfigure}[b]{0.24\linewidth}
        \includegraphics[width=\linewidth]{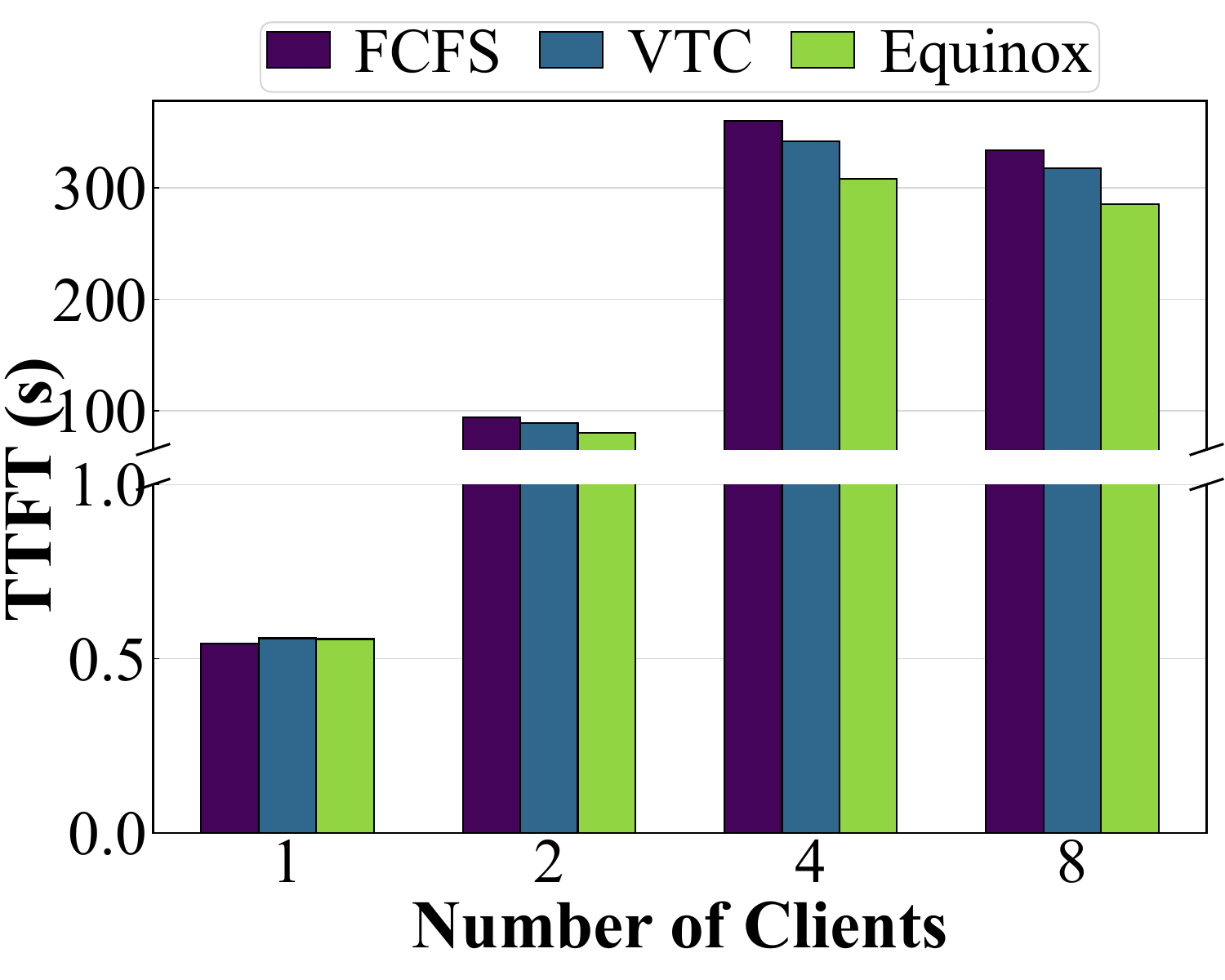} 
        \caption{Per-client TTFT} 
        \label{fig:vllm_4}
    \end{subfigure}
    \caption{Performance comparison using ShareGPT datasets in vLLM.} 
    \label{fig:vllm}
    \vspace{-1em}
\end{figure*}

\begin{figure}[t]
    \centering
    \setlength{\abovecaptionskip}{0.1cm}
    \setlength{\belowcaptionskip}{-0.1cm}
    \includegraphics[width=0.45\textwidth]{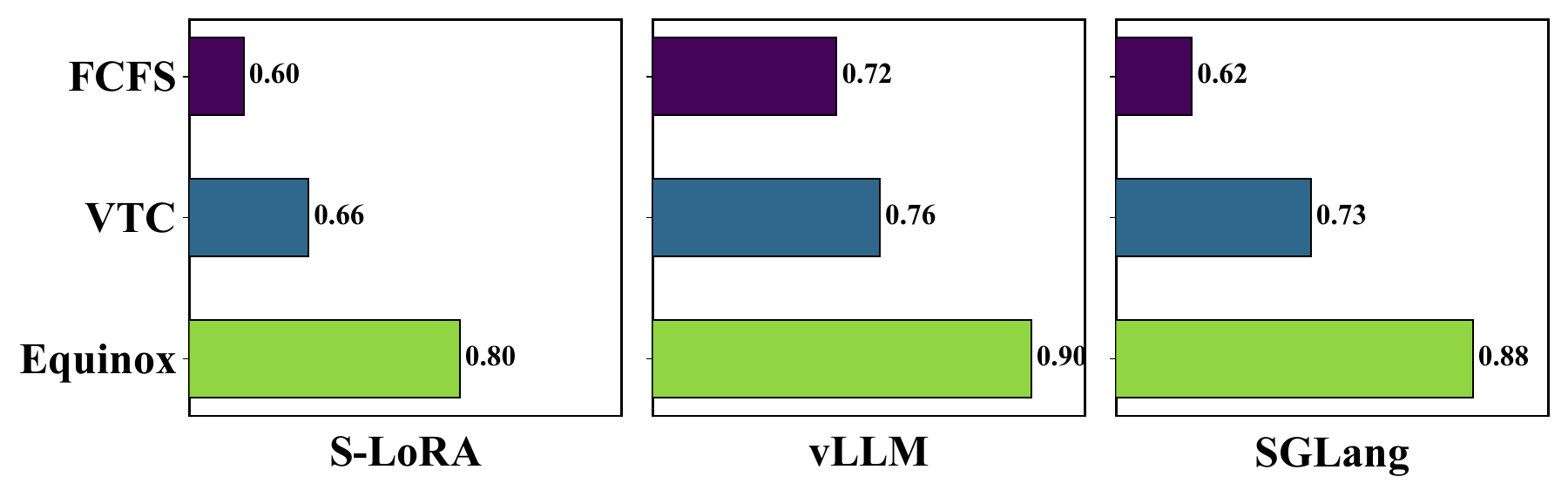}
    \caption{Fairness comparison across S-LoRA, vLLM and SGLang.} 
    \label{fig:fairness_comparison}
    \vspace{-1em} 
\end{figure}

We configure an experiment using the SGLang serving system~\cite{sglang} processing requests derived from the ShareGPT dataset. The workload comprises 256 simulated clients, with the aggregate request arrival rate dynamically varying between 1 and 16 requests per second (RPS) and the total number of prompts as 1280, which is a standard benchmark integrated into SGLang Benchmark\footnote{\url{https://github.com/sgl-project/sglang/tree/main/benchmark/hicache}}. This setup rigorously tests Equinox's performance, scalability, and responsiveness under fluctuating demand within a state-of-the-art serving platform designed for complex interaction patterns.

Equinox demonstrates significant performance advantages when integrated into the SGLang environment and subjected to the ShareGPT dataset, as illustrated in ~\autoref{fig:SGLang}. Compared to FCFS and VTC, Equinox achieves improvements in P50/P90 time to first token (TTFT) (up to 30\%) and mild system throughput (up to 25\%) when RPS is high, highlighting its efficiency in realistic settings. Although designed for fairness, Equinox monitors UFC/RFC and reorder and prioritize requests from clients are vulnerable to starvation. This mechanism, akin to priority scheduling based on aging or urgency, partially mitigates head-of-line (HOL) blocking, thereby optimizing TTFT and system throughput. 

\subsubsection{vLLM with ShareGPT Dataset}
We also evaluate performance using vLLM, again leveraging the ShareGPT workload. In this specific experimental setup, the number of concurrent clients varies from 1 to 8, and the per-client request rate is kept constant at 3.5 requests per second in a poisson distribution. Each client sends 1000 requests in total.

The ~\autoref{fig:vllm} illustrates both the performance and fairness gains of Equinox across multiple metrics. Equinox achieves higher and more stable Jain’s fairness index compared to FCFS and VTC (up to 33\% in \autoref{fig:vllm_1}) and slightly lower average TTFT and end-to-end latency ($\sim$5\% in \autoref{fig:vllm_2} and ~\autoref{fig:vllm_4}), ensuring better quality of service for each client. It maintains a slightly higher per-client service rate (\autoref{fig:vllm_3}), reflecting efficient token-level scheduling. Together, these improvements validate Equinox’s robustness and fairness under different setups of clients in vLLM.





\subsubsection{Cross-Serving System Fairness Comparison}
\label{sec:real_fairness_comparison}

To provide a consolidated assessment of fairness across different LLM serving system, we use Jain's Fairness Index to compare the fairness achieved by Equinox against FCFS and VTC. This comparison spans multiple popular serving systems: S-LoRA, vLLM, and SGLang, the detailed workload and setup are detailed in ~\autoref{para:methodology}. As depicted in ~\autoref{fig:fairness_comparison}, Equinox consistently delivers 13\% superior fairness compared to both FCFS and VTC. Counterintuitively, VTC’s fairness index on HF is no better than FCFS. This demonstrates that scheduling policies relying on multipled metrics improves and ensures equitable service opportunities for clients.

\begin{table}[t]
    \centering
    \caption{Ablation of fairness (Max/Avg/Var of Service Difference) across schedulers and predictors under synthetic load.}

    \label{tab:ablation_results}
    \vspace{-1em}
    \begin{tabular}{l|rrr}
        \toprule
        Scheduler Variant & Max Diff & Avg Diff & Diff Var \\
        \midrule
        FCFS & 1864.42 & 1400.40 & 44868.58 \\
        VTC & 1505.13 & 1106.31 & 45151.63 \\ 
        VTC + Single & 3344.00 & 2992.92 & 44588.27 \\ 
        \textbf{VTC + MoPE} & \textbf{1390.00} & \textbf{1003.82} & \textbf{35400.60} \\
        \textbf{VTC + Oracle} & \textbf{1375.80} & \textbf{999.12}  & \textbf{33859.38} \\
        \midrule
        Equinox + Single & 1385.23 & 1005.47 & 34210.75 \\ 
        \textbf{Equinox + MoPE}   & \textbf{865.62}  & \textbf{150.64}  & \textbf{28578.68} \\
        \textbf{Equinox + Oracle} & \textbf{715.12}  & \textbf{99.80}   & \textbf{22156.91} \\
        \bottomrule
    \end{tabular}
    \vspace{-1em}
\end{table}

\subsection{Ablation Study}
\label{sec:ablation_study}

We isolate the impact of MoPE within Equinox by comparing its fairness performance against baseline schedulers (FCFS, VTC) under a challenging synthetic workload (details in ~\autoref{sec:synthetic_poisson_scenario}). We use three predictors: MoPE, a single proxy model (Single)~\cite{298496}, and an Oracle (perfect prediction) as an ideal benchmark. Fairness is measured by the maximum, average, and variance of the accumulated absolute service difference between clients over the experiment (lower is better), presented in \autoref{tab:ablation_results}.

\noindent {\bf Analysis.} \autoref{tab:ablation_results} shows FCFS provides poor fairness. Baseline VTC, lacking predictive capabilities, also performs poorly as it cannot account for varying request costs. Incorporating token length prediction significantly boosts the fairness of VTC. VTC + MoPE approaches the Oracle performance, demonstrating the necessity of accurate prediction for token-level fairness. However, Equinox's core scheduling logic yields far greater fairness improvements. Equinox variants consistently outperform their VTC counterparts using the same predictor. Notably, Equinox + MoPE dramatically reduces service differences compared to VTC + MoPE, confirming the advantage of Equinox's holistic fairness approach over VTC's token-centric method. Equinox + single proxy model offers little benefit over VTC with the same predictor, highlighting that advanced scheduling requires good prediction.  The close performance of Equinox + MoPE to Equinox + Oracle further validates MoPE's effectiveness.

\begin{figure}[t]
    \centering
    \begin{subfigure}[b]{0.49\linewidth}
        \includegraphics[width=\linewidth]{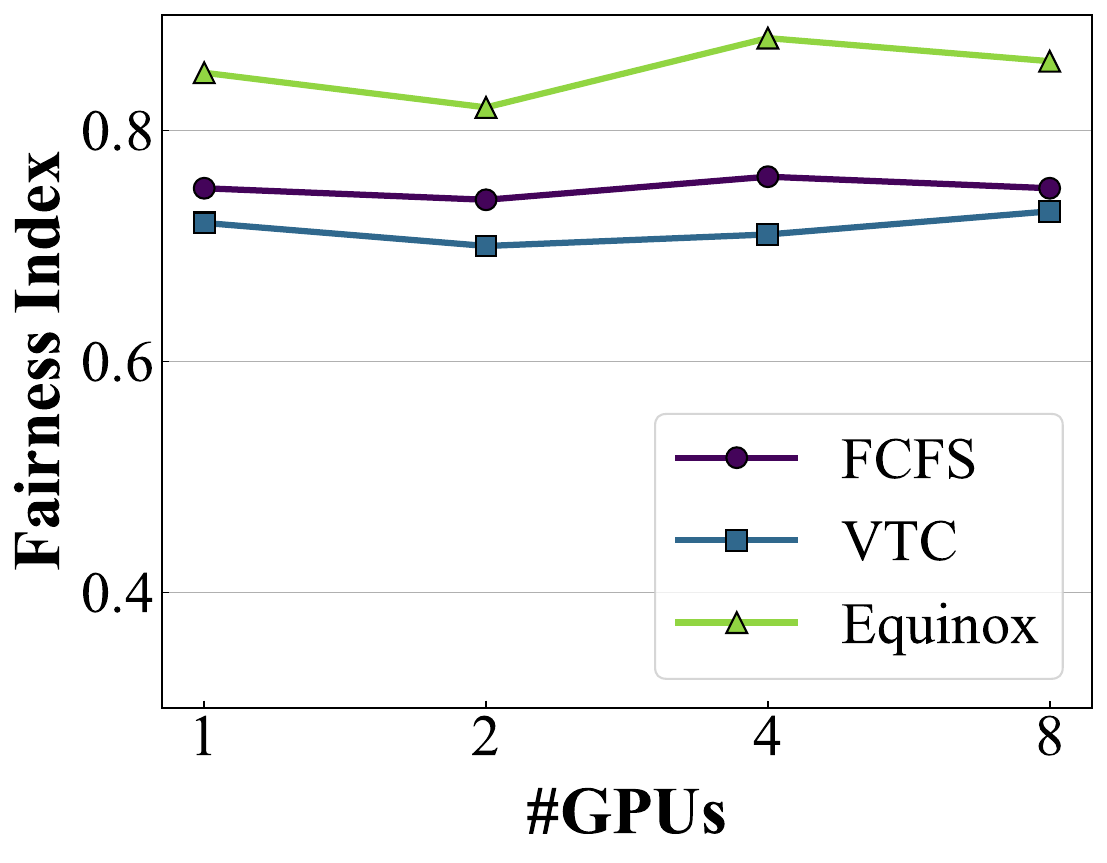} 
        \caption{vLLM Fairness Scalability} 
        \label{fig:scalability_1}
    \end{subfigure}
    \begin{subfigure}[b]{0.49\linewidth}
        \includegraphics[width=\linewidth]{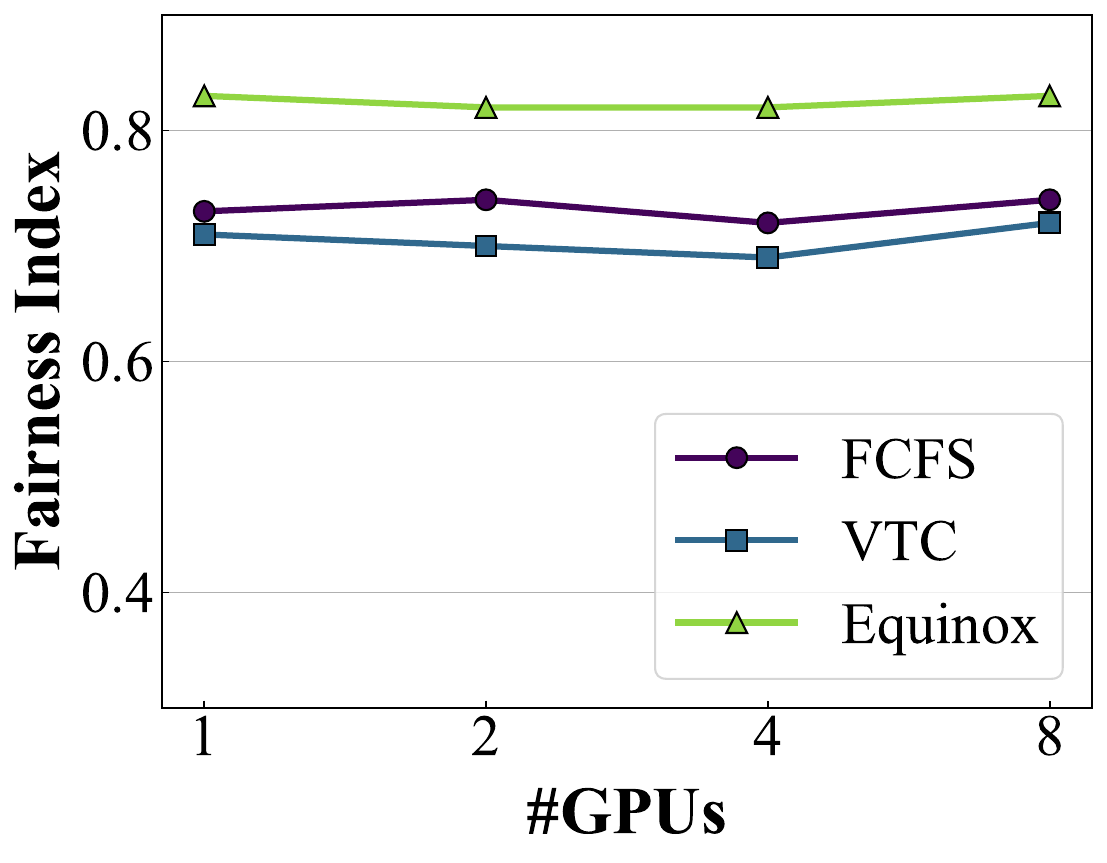}
        \caption{SGLang Fairness Scalability}
        \label{fig:scalability_2}
    \end{subfigure}
    \caption{Experiments to test Jain's Fairness on vLLM and SGLang scaling GPU counts from 1 to 8}
    \label{fig:scalability}
    \vspace{-1em}
\end{figure}

\subsection{Scalability Analysis}
We repeated experiments on vLLM and SGLang, scaling GPU counts from 1 to 8 with proportional tensor parallelism (TP) adjustments. Across all configurations, Equinox consistently outperformed VTC and FCFS in fairness. This demonstrates that Equinox's core innovations—UFC/RFC, MoPE, and scheduling algorithm—exhibit robust setup-agnostic behavior. While extending Equinox to multi-node deployment would require additional engineering efforts (e.g., distributed UFC/RFC updates), we maintain that its high-level design principles remain fundamentally sound.

\begin{figure}[t]
    \centering
    \setlength{\abovecaptionskip}{0.1cm}
    \setlength{\belowcaptionskip}{-0.1cm}
    \includegraphics[width=0.45\textwidth]{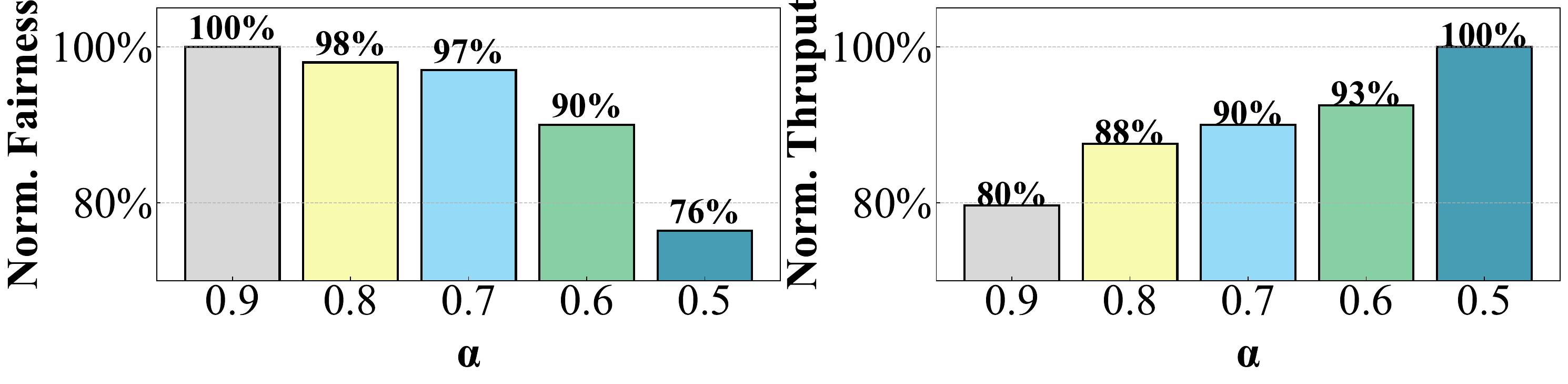}
    \caption{Impact of $\alpha$/$\beta$ weight ratios on client fairness and system throughput} 
    \label{fig:ufc_rfc}
    \vspace{-1em} 
\end{figure}

\subsection{Hyperparameter Analysis}
\label{sec:ufc_rfc}
We repeated experiments on SGLang at RPS=16, adjusting $\alpha$ from 0.5 to 0.9 while setting $\beta$ to 1 - $\alpha$. For each configuration, we recorded Jain's Fairness Index for client P90 TTFT and system throughput (requests per second), normalizing values to their maximums. As shown in \autoref{fig:ufc_rfc}, reducing $\alpha$ decreases latency fairness but increases throughput: At $\alpha = 0.9$, peak fairness occurs but throughput decreases by 20\% compared to $\alpha = 0.5$; conversely at $\alpha = 0.5$, maximum throughput is achieved but fairness drops by 23\%. We choose the optimal balance as $\alpha = 0.7$ and $\beta = 0.3$, which maintains 97\% peak fairness (only 3\% below maximum) while achieving 90\% of maximum throughput. This configuration was selected to prioritize user-perceived latency while sustaining high system performance.

\section{Related Work}


\noindent \textbf{LLM Serving Systems.} Recent systems have optimized latency and throughput, but they do not fundamentally resolve the resource allocation challenges posed by the prefill-decode bifurcation. For instance, \emph{Sarathi-Serve} introduces chunked-prefills to admit new requests without stalling ongoing decodes, while \emph{DistServe} disaggregates prefill and decode phases onto separate GPU pools to eliminate interference. Other innovations include dynamic early exits in \emph{Apparate} and neuron caching in \emph{PowerInfer}. Critically, while these methods mitigate interference, they don't alter the underlying per-iteration computation, leaving the pipeline bound by either prefill or decode. Consequently, they optimize around the architectural split without addressing its fairness implications, typically defaulting to simplistic FCFS or token-based schedulers.

\noindent \textbf{Fair Scheduling in LLM.} Token-based fairness mechanisms, used in systems like VTC, aim to prevent resource monopolization by prioritizing users with the fewest accumulated tokens. However, this approach is flawed because the prefill-decode bifurcation means identical token counts correspond to vastly different resource consumption; decode disproportionately affects latency, while prefill drives throughput bottlenecks. This mismatch makes token-counting an inaccurate proxy for the actual costs imposed on users or the system. Our dual-counter framework overcomes this limitation by explicitly separating user concerns (the UFC, combining weighted tokens and latency) from operator concerns (the RFC, tracking GPU utilization and throughput). This transforms fairness from a single-metric problem into a multi-objective optimization that reflects the true computational dynamics of LLM serving.

\noindent \textbf{Length Prediction in LLM.} Existing prediction models often focus on a single metric, such as using proxy models to estimate output token length for shortest-job-first scheduling. While techniques like range prediction or learning-to-rank offer some improvement, they are insufficient for holistic fairness, which requires estimating four distinct metrics: user latency and token counts, alongside operator utilization and throughput. Our Mixture of Prediction Experts (MoPE) reconceptualizes this task. Instead of predicting one value, MoPE routes prompts to specialized experts that estimate all four fairness components with minimal overhead, enabling the comprehensive scheduling decisions that single-metric predictors cannot support.
\vspace{-1em}

\section{Conclusion}

Equinox redefines fairness in multi-tenant LLM serving by recognizing that the prefill-decode bifurcation makes single-metric fairness fundamentally unachievable. Its dual-counter framework separates user concerns through the User Fairness Counter (combining weighted tokens with perceived latency) from operator concerns through the Resource Fairness Counter (tracking GPU utilization and throughput), transforming fairness from resource allocation to multi-objective optimization. The deterministic MoPE enables this framework by predicting required metrics before execution, resolving the scheduling paradox where fairness computation requires post-execution information. Experiments show that Equinox achieves up to 1.3$\times$ higher throughput, 60\% lower time-to-first-token latency, and 13\% higher fairness compared to VTC while maintaining 94\% GPU utilization on production traces. These results demonstrate that holistic fairness, which balances both user experience and operator efficiency, is achievable in heterogeneous LLM serving systems when scheduling decisions account for the computational dynamics of transformer .

\clearpage
\bibliographystyle{ACM-Reference-Format}
\bibliography{references}

\clearpage
\appendix

\begin{figure*}[htbp]
    \centering

    \begin{subfigure}[b]{0.32\linewidth}
        \centering
        \includegraphics[width=\linewidth]{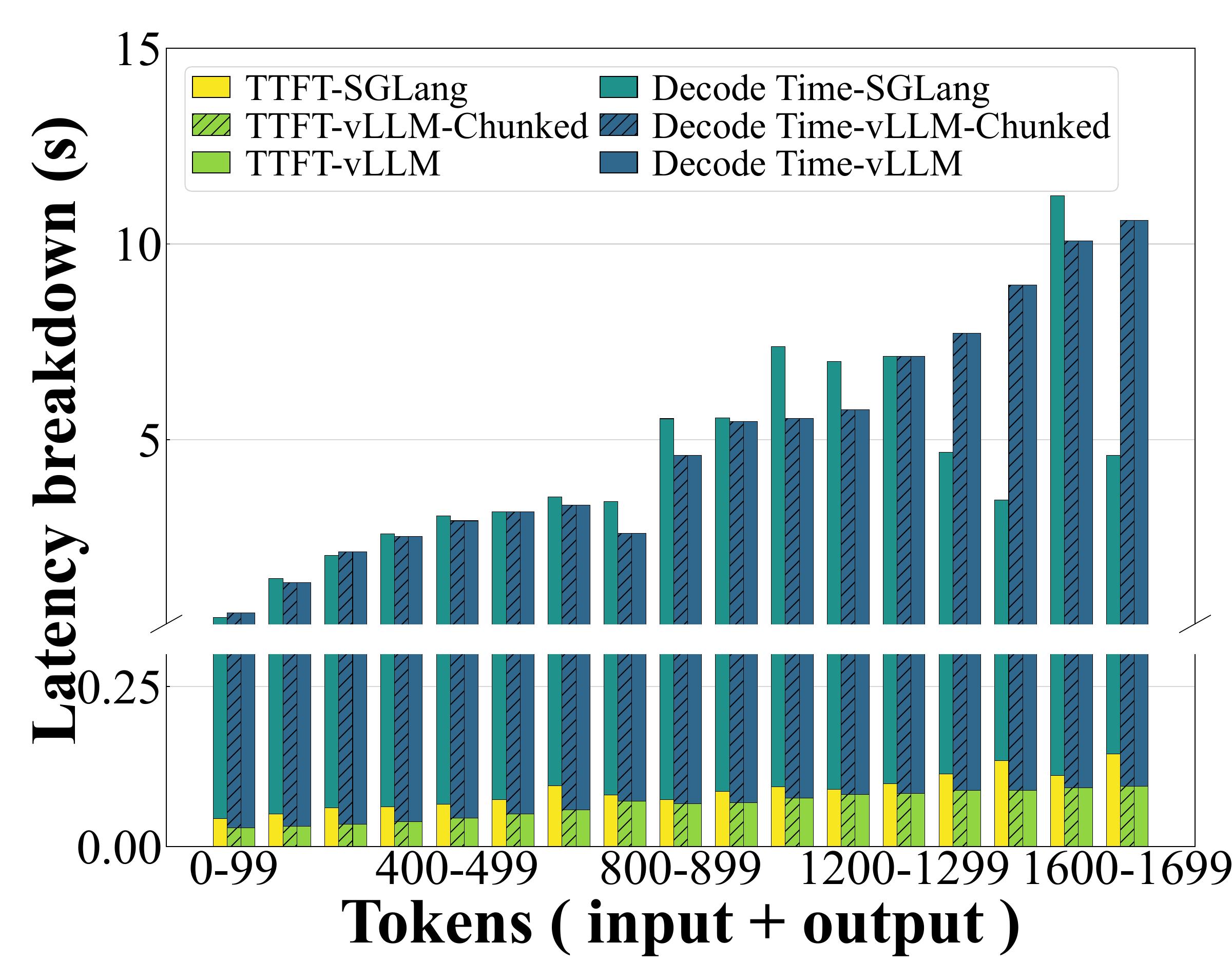}
        \caption{vLLM and SGLang exhibit similar latency patterns.}
        \label{fig:motivation4}
    \end{subfigure}
    \hfill
    \begin{subfigure}[b]{0.32\linewidth}
        \centering
        \includegraphics[width=\linewidth]{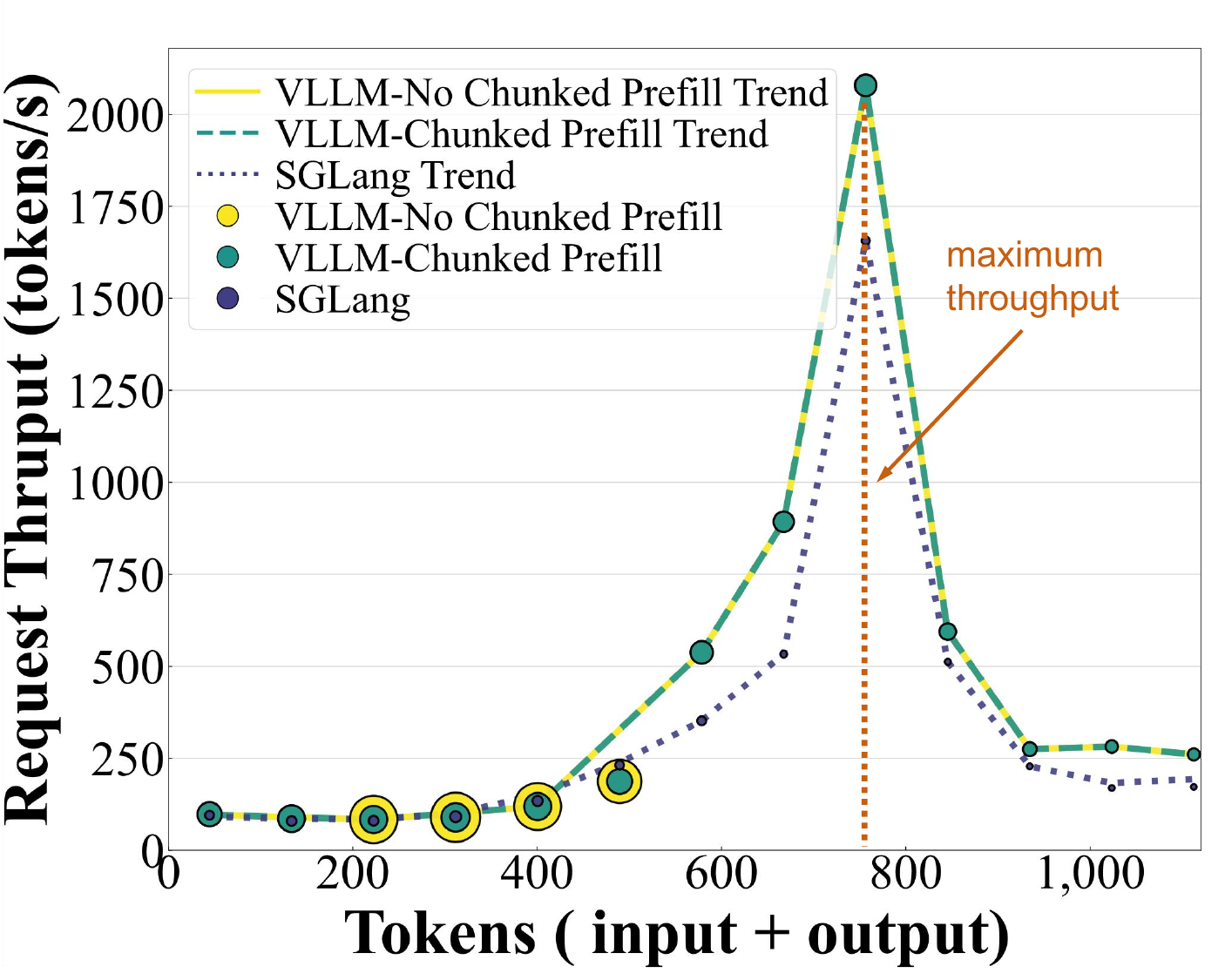}
        \caption{vLLM and SGLang show non-linear throughput gains across token ranges.}
        \label{fig:motivation5}
    \end{subfigure}
    \hfill
    \begin{subfigure}[b]{0.32\linewidth}
        \centering
        \includegraphics[width=\linewidth]{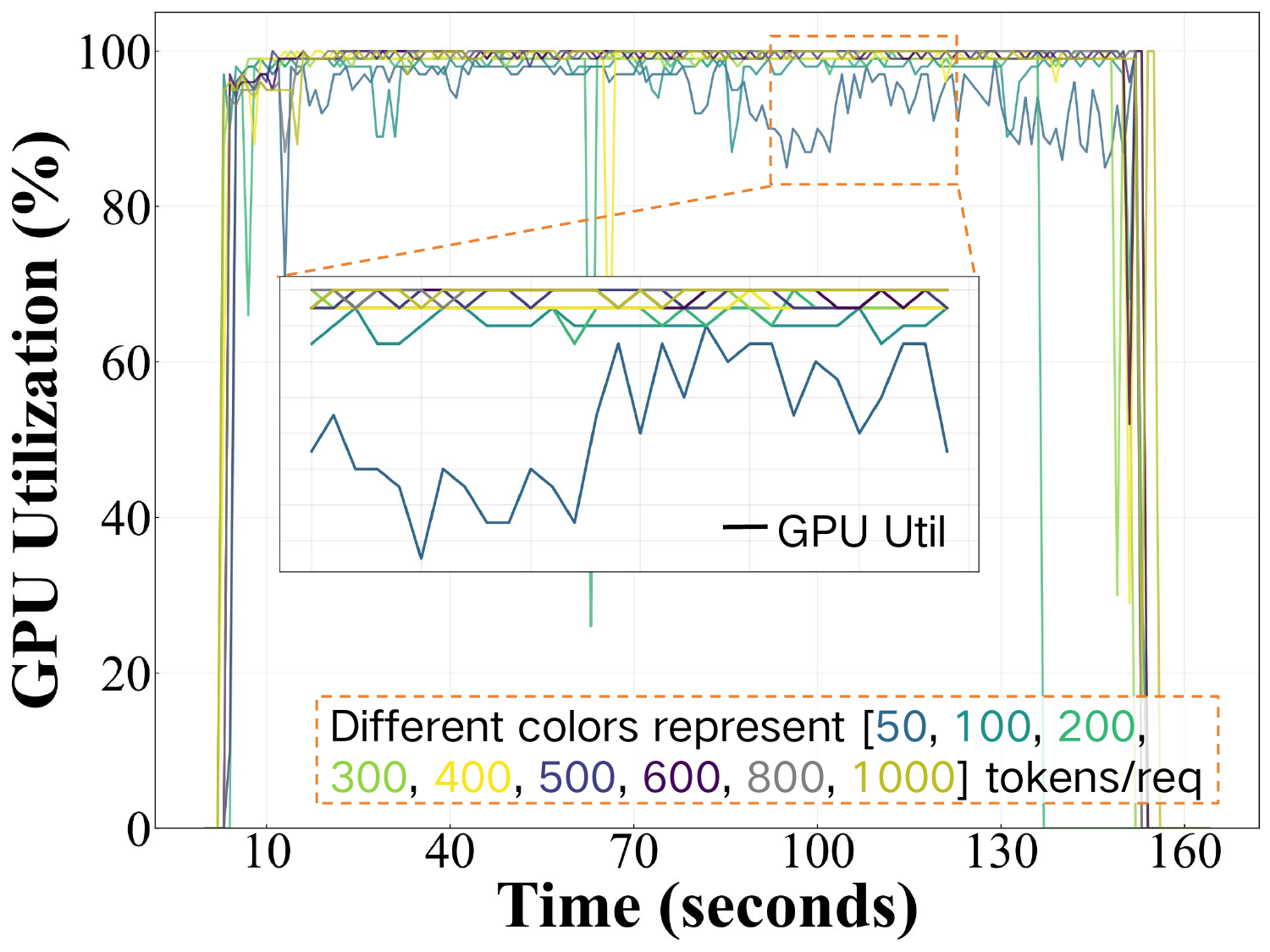}
        \caption{vLLM and SGLang display step-like GPU resource usage patterns.}
        \label{fig:motivation6}
    \end{subfigure}

    \caption{
        Results show metric independence from token count under varied scheduling and serving systems. Analysis reveals \textbf{non-linear}, inconsistent latency, throughput, and GPU utilization, contradicting token-level fairness assumptions. Crucially, even with optimizations like chunked prefill, per-token metrics unreliably predict performance across adjacent token ranges. Experiments utilized an NVIDIA A100 GPU with ShareGPT (a-b), and controlled synthetic workloads (c).
    }
    \label{fig:motivation-overall}
\end{figure*}

\begin{figure*}[h]
    \centering
     \begin{subfigure}[b]{0.24\linewidth}
        \includegraphics[width=\linewidth]{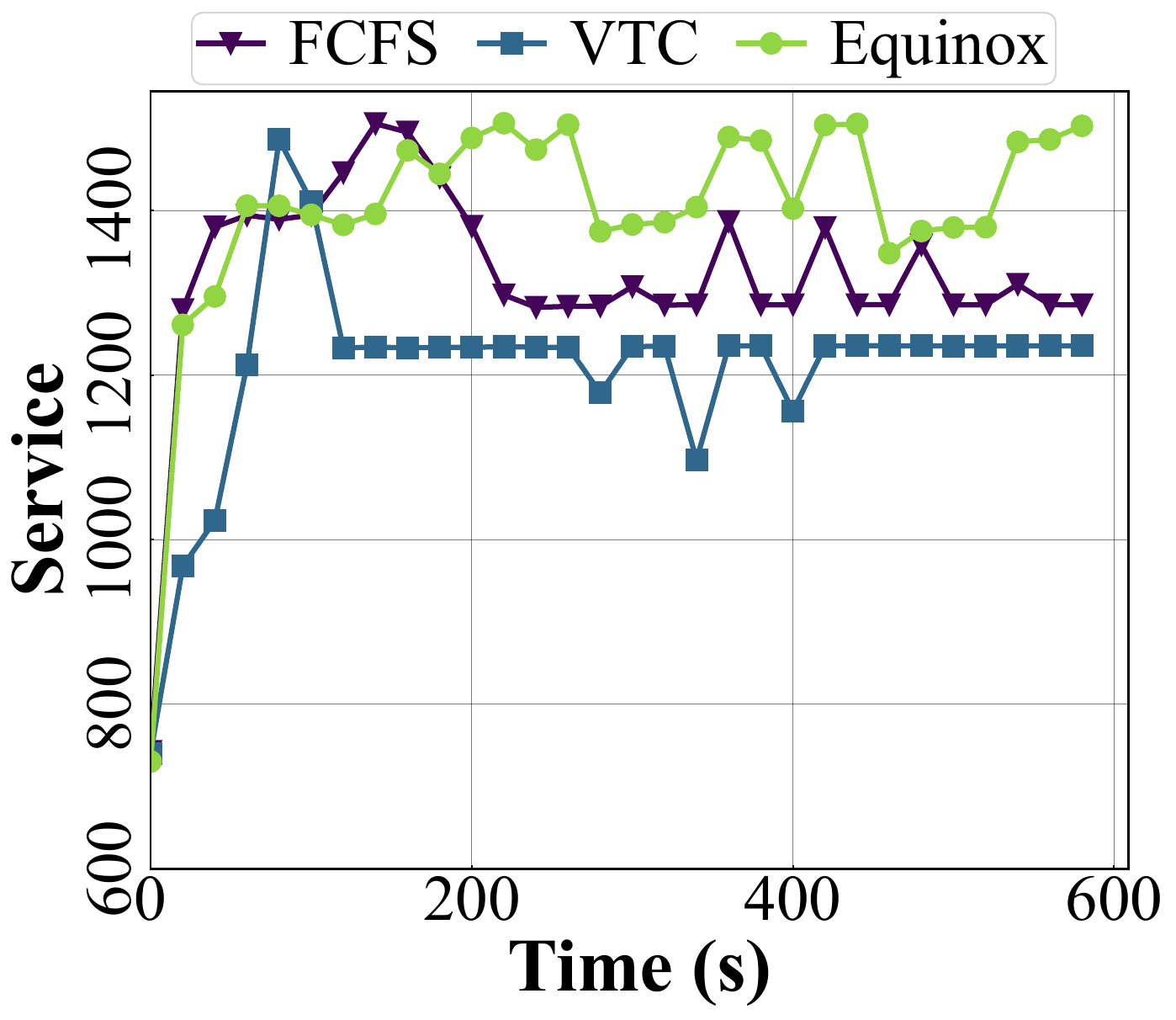}
        \caption{Total Service Rate}
        \label{fig:appendix_overload_service_rate}
    \end{subfigure}
    \hfill
    \begin{subfigure}[b]{0.24\linewidth}
        \includegraphics[width=\linewidth]{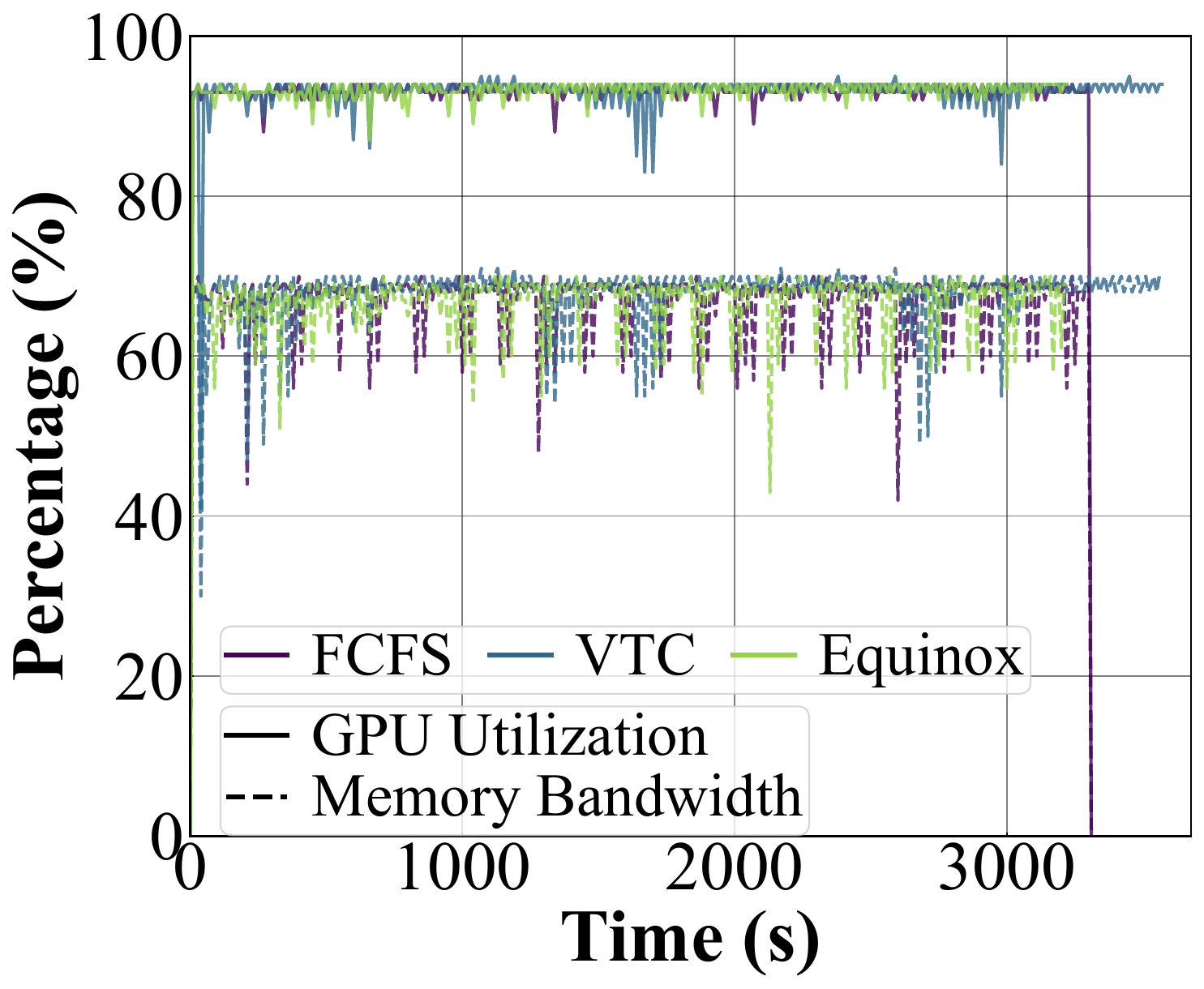}
        \caption{GPU Util. \& Bandwidth}
        \label{fig:appendix_overload_gpu}
    \end{subfigure}
    \hfill
    \begin{subfigure}[b]{0.24\linewidth}
        \includegraphics[width=\linewidth]{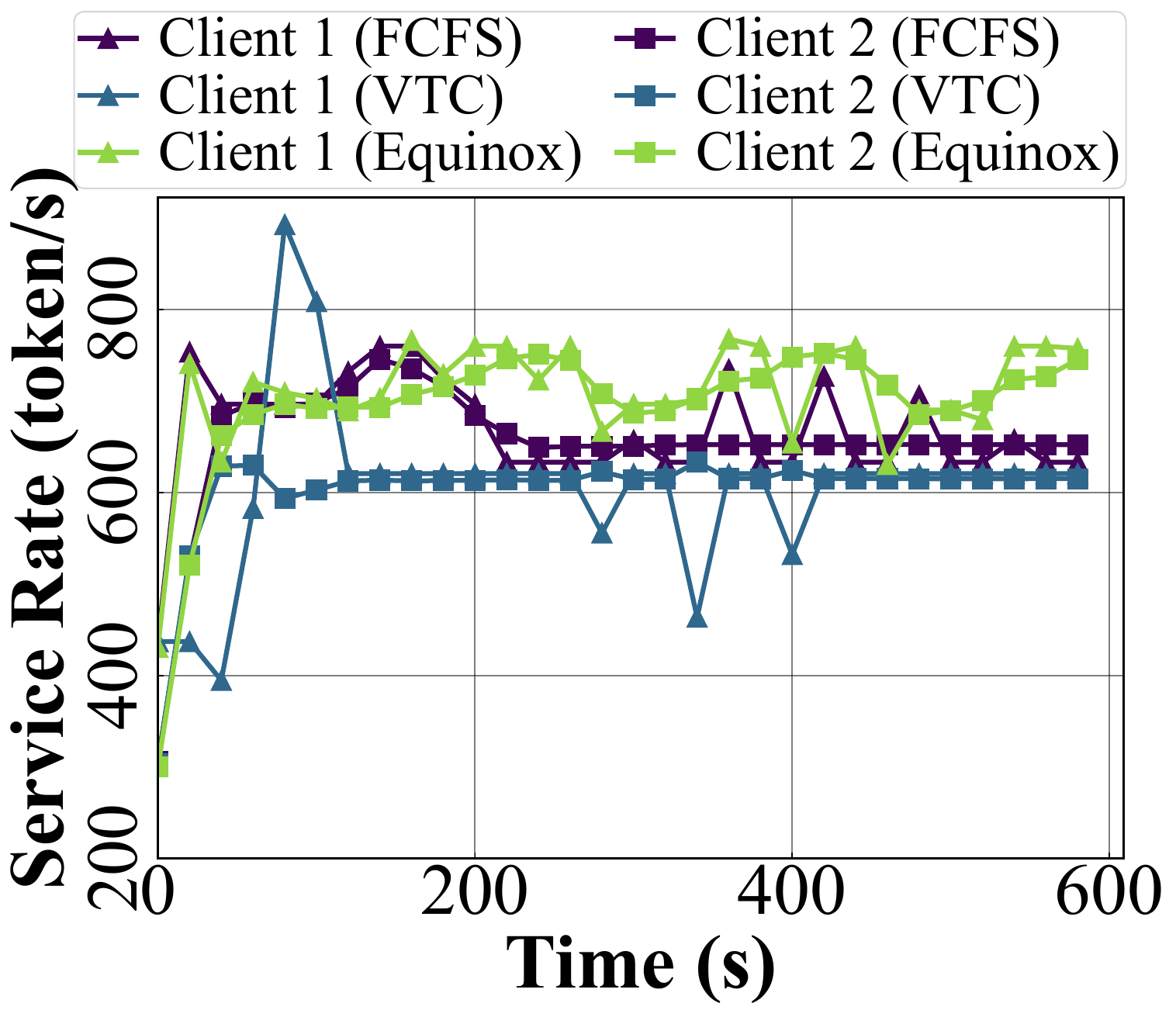}
        \caption{Per-client Service Rate}
        \label{fig:appendix_overload_per_client_service}
    \end{subfigure}
    \hfil
    \begin{subfigure}[b]{0.24\linewidth}
        \includegraphics[width=\linewidth]{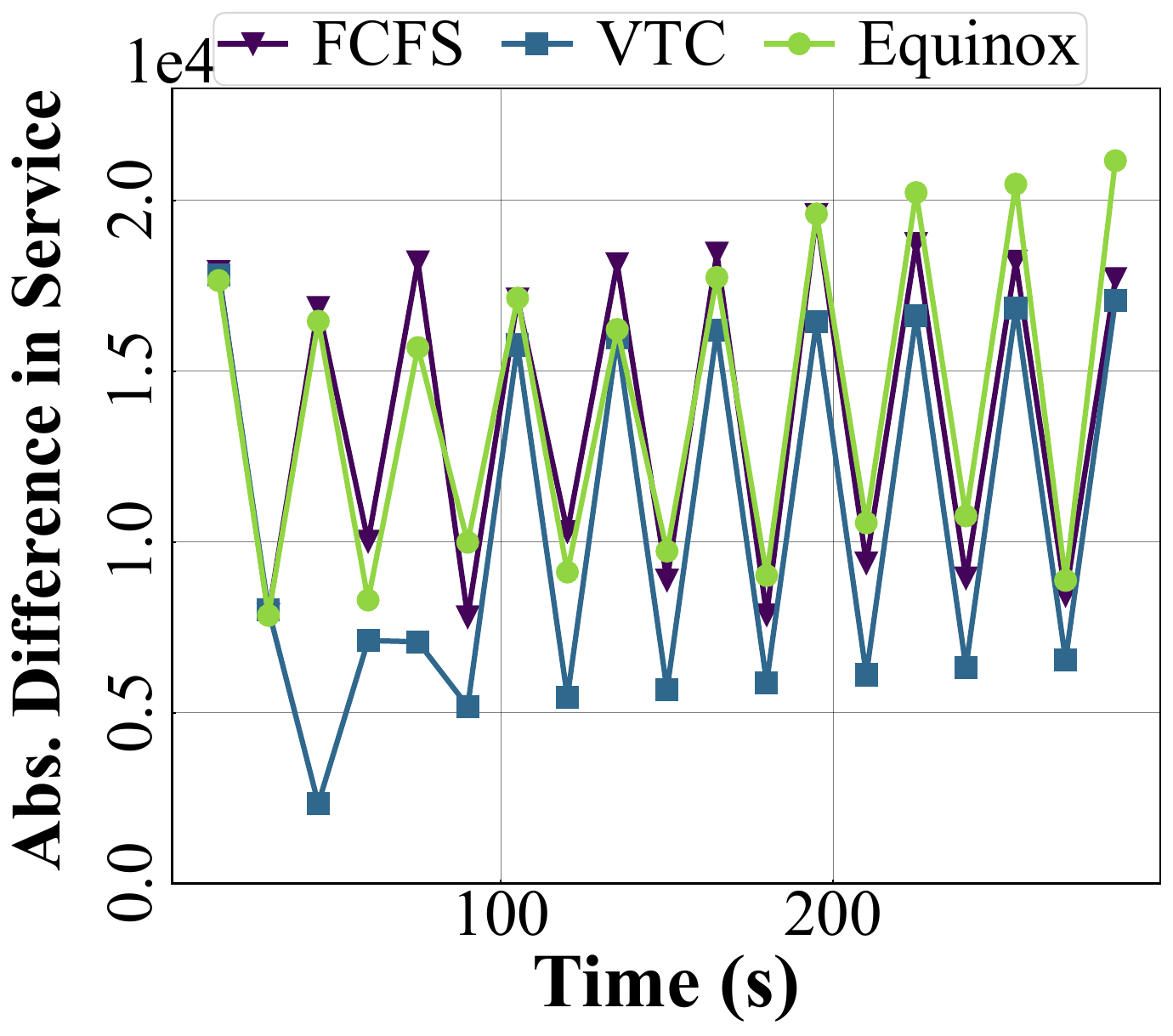}
        \caption{Service Difference}
        \label{fig:appendix_overload_service_diff}
    \end{subfigure}
    \caption{Constant overload scenario.}
    \label{fig:appendix_overload}
\end{figure*}


\begin{figure*}[h]
    \centering
    \begin{subfigure}[b]{0.24\linewidth}
        \includegraphics[width=\linewidth]{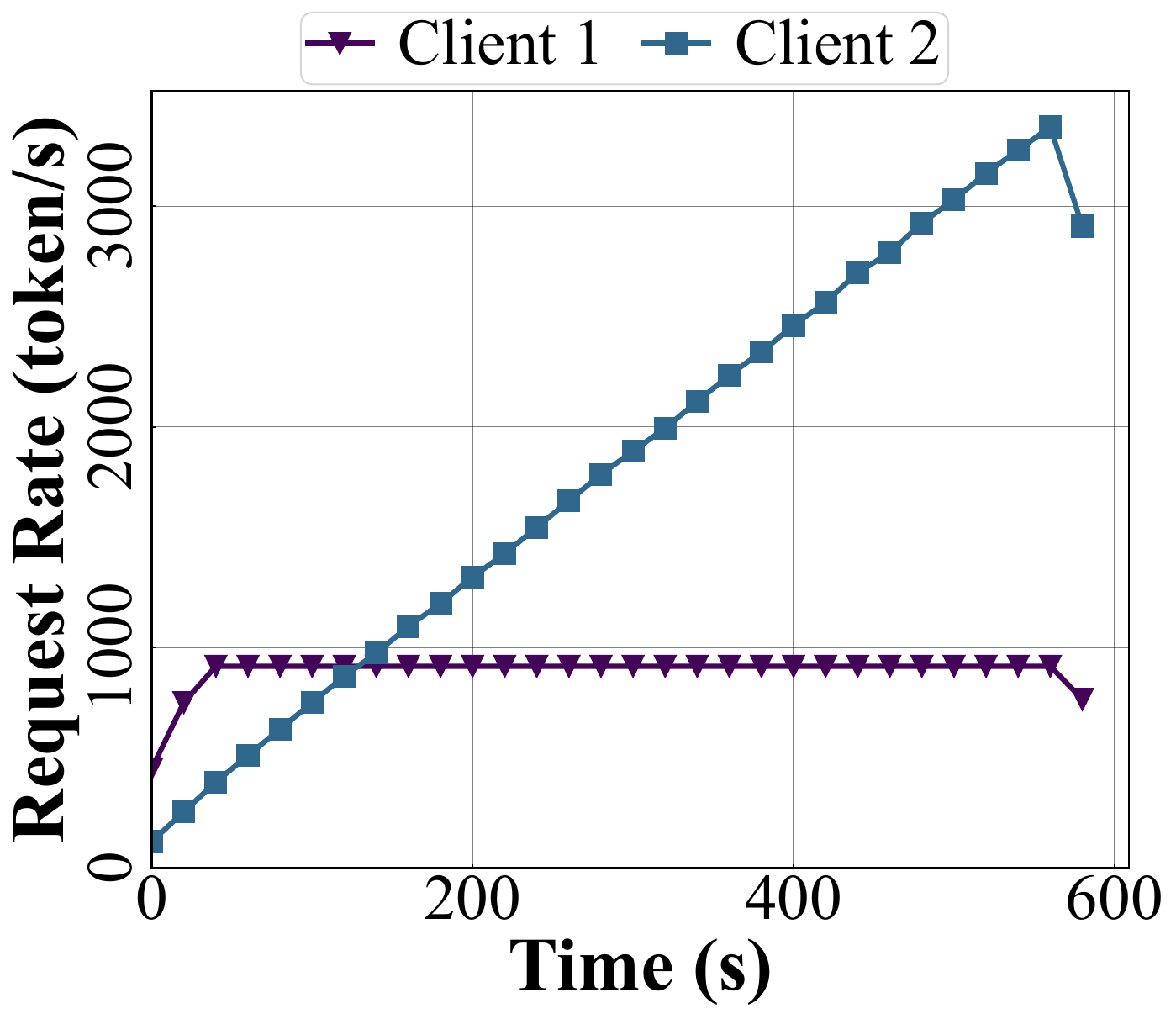}
        \caption{Client Request Rate}
        \label{fig:appendix_increase_req_rate}
    \end{subfigure}
    \hfill
    \begin{subfigure}[b]{0.24\linewidth}
        \includegraphics[width=\linewidth]{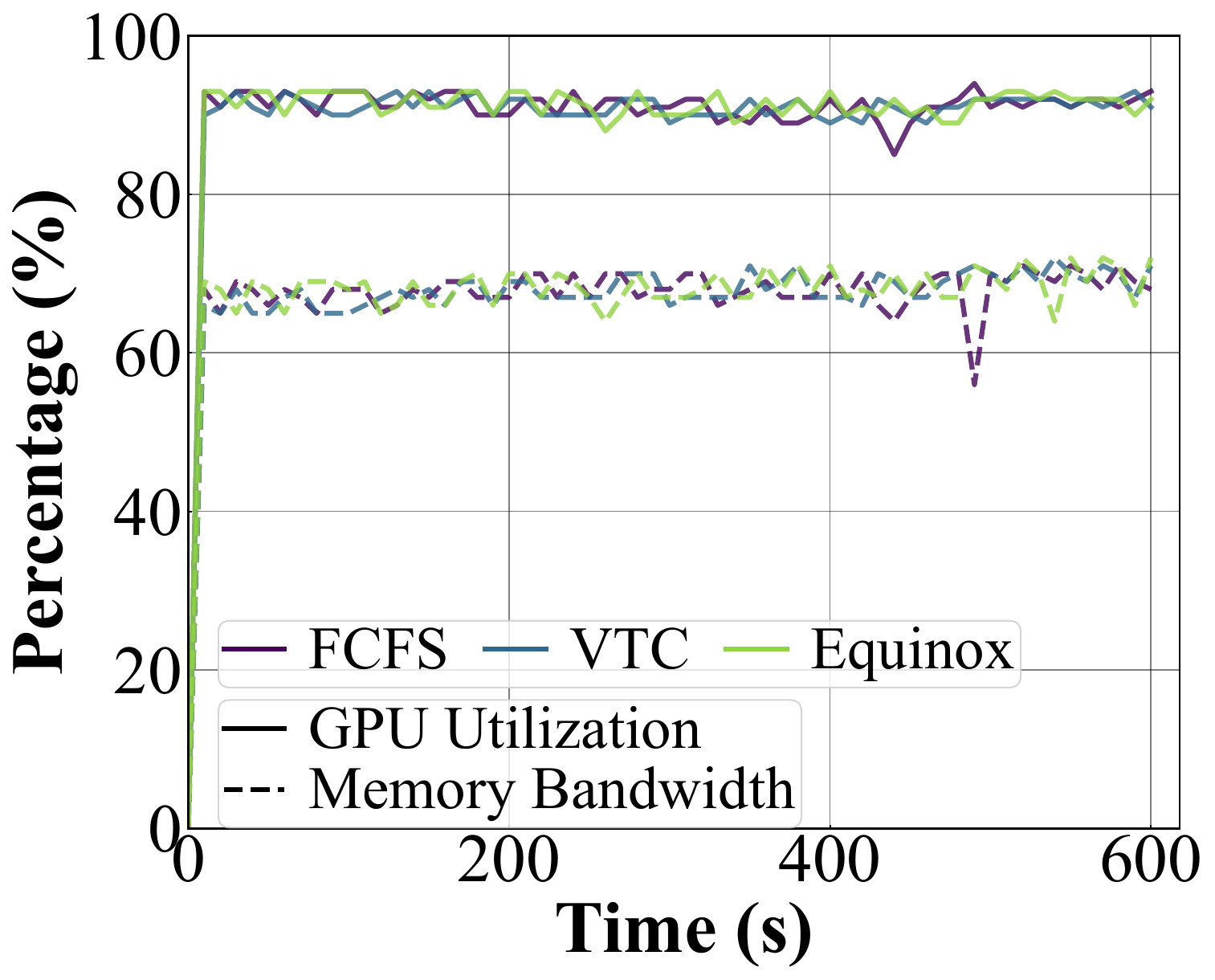}
        \caption{GPU Util. \& Bandwidth}
        \label{fig:appendix_increase_gpu}
    \end{subfigure}
    \hfill    
    \begin{subfigure}[b]{0.24\linewidth}
        \includegraphics[width=\linewidth]{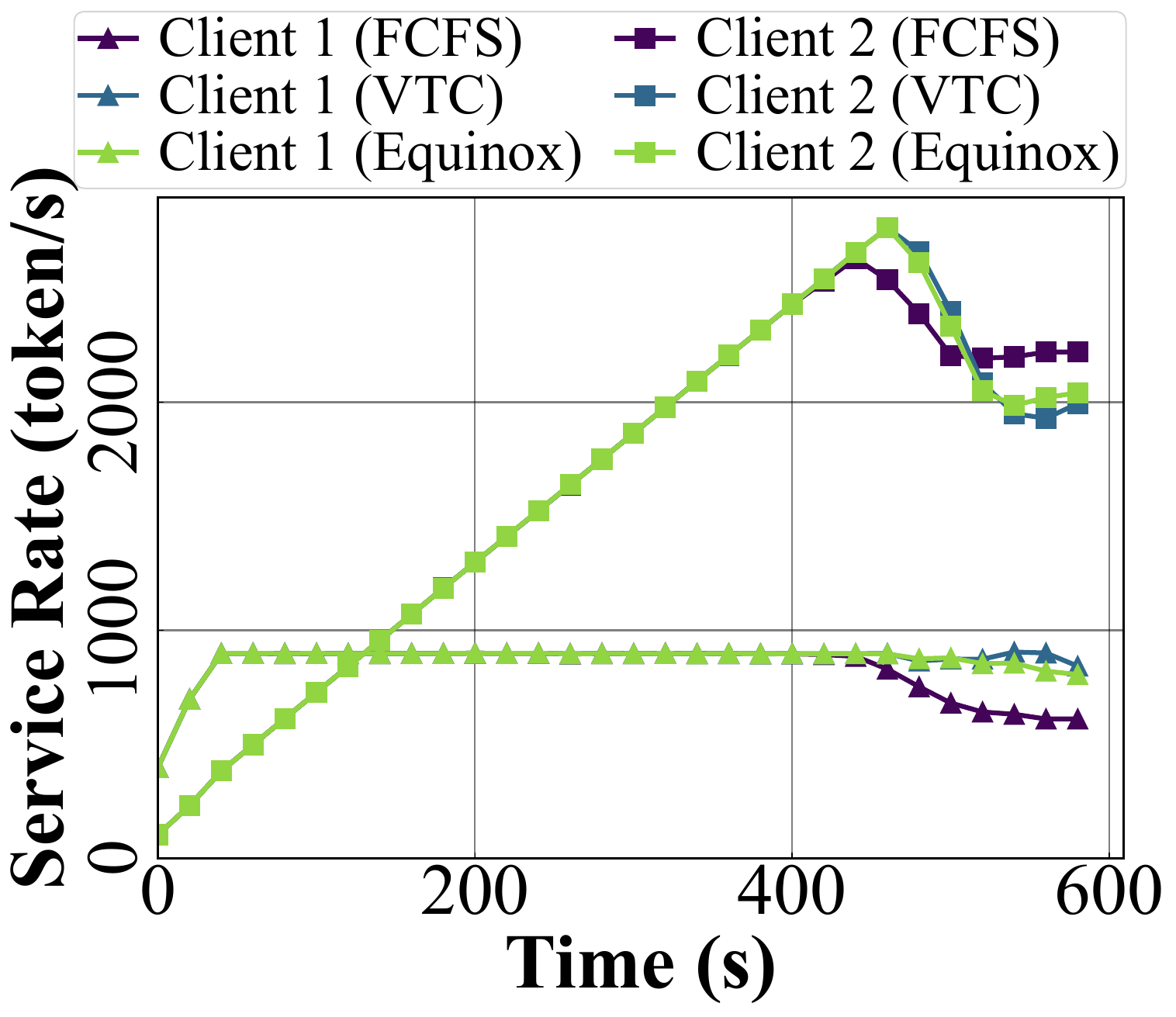}
        \caption{Per-client Service Rate}
        \label{fig:appendix_increase_service_rate}
    \end{subfigure}
    \hfill
    \begin{subfigure}[b]{0.24\linewidth}
        \includegraphics[width=\linewidth]{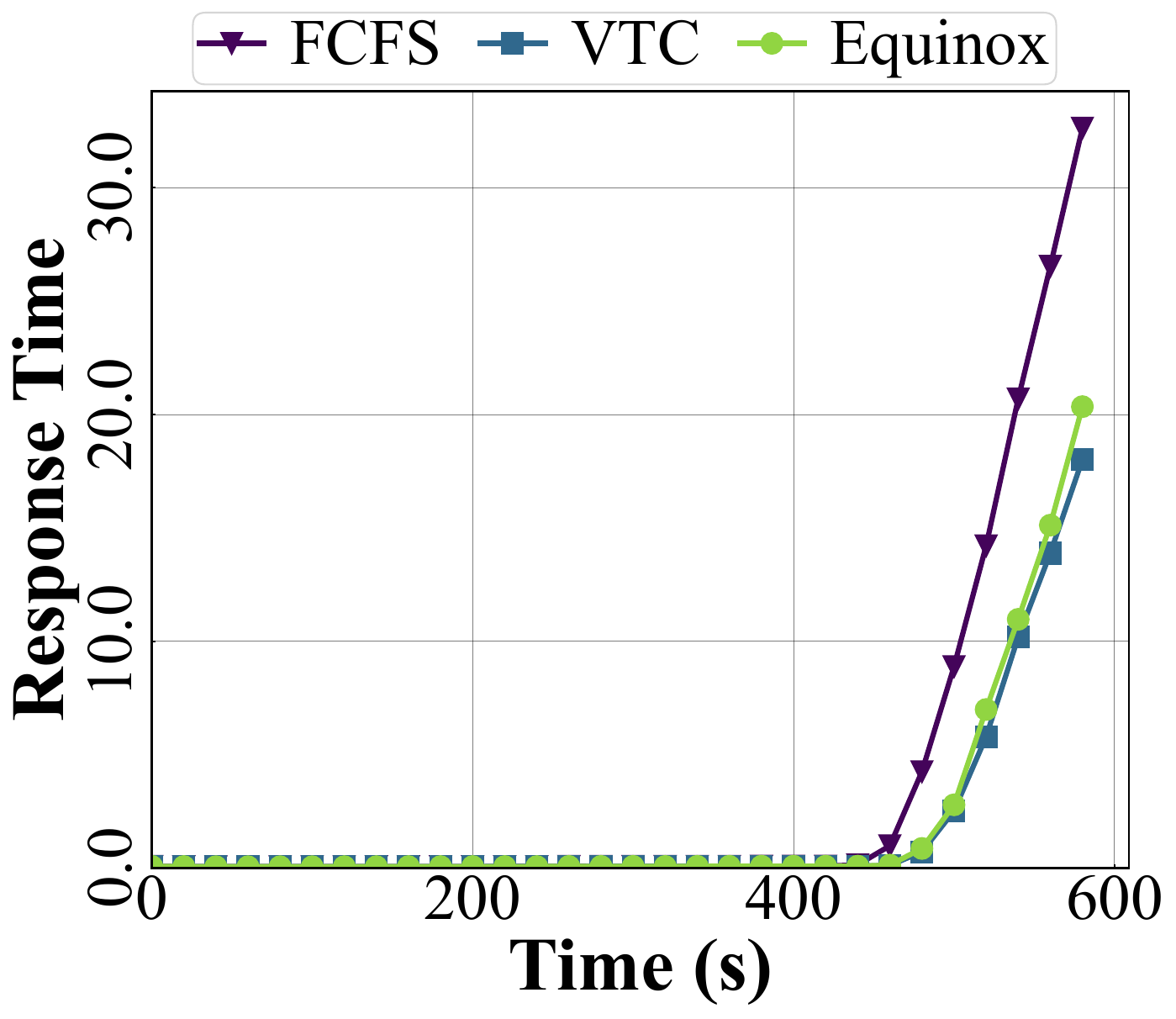}
        \caption{Total Response Time}
        \label{fig:appendix_increase_resp_time}
    \end{subfigure}
    \caption{Increasing load scenario.}
    \label{fig:appendix_increase}
\end{figure*}

\section{Supplementary Synthetic Workload Evaluations}
\label{sec:appendix_synthetic}

This appendix provides supplementary results from synthetic workload experiments, illustrating Equinox's behavior under specific conditions complementary to those presented in ~\autoref{sec:synthetic_evaluation}.

\begin{figure*}[htbp]
    \centering
    \begin{subfigure}[b]{0.33\linewidth}
        \includegraphics[width=\linewidth]{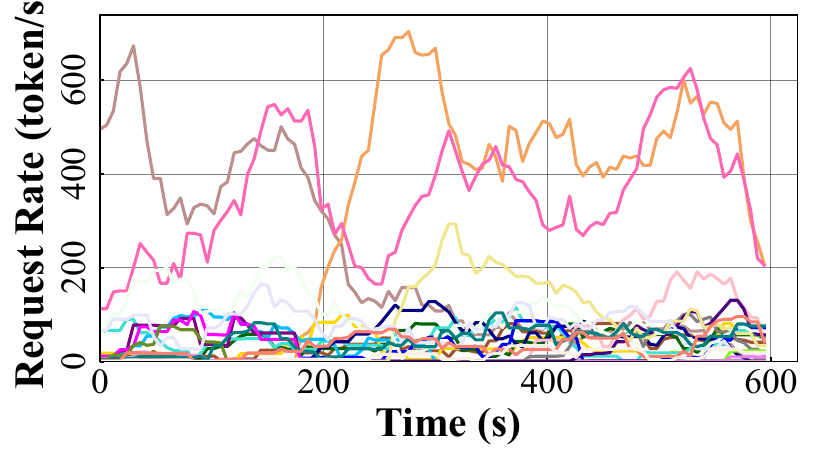}
        \caption{Per-client Request Rate}
        \label{fig:real_1}
    \end{subfigure}
    \hfill
    \begin{subfigure}[b]{0.33\linewidth}
        \includegraphics[width=\linewidth]{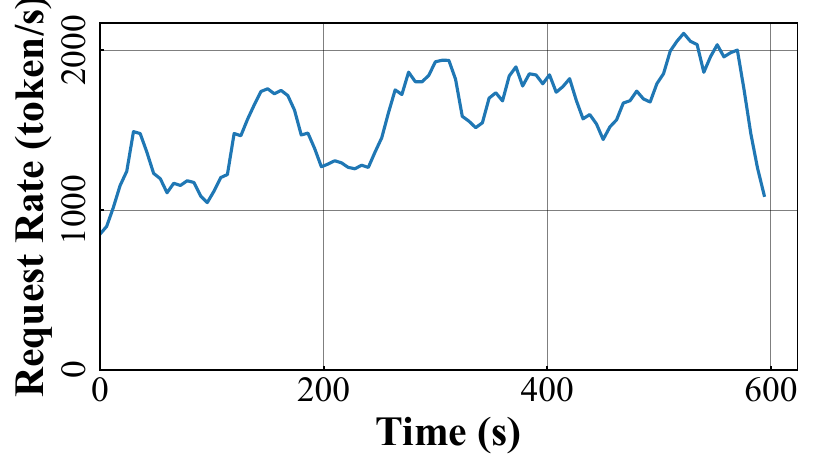}
        \caption{Total Request Rate}
        \label{fig:real_2}
    \end{subfigure}
    \hfill
    \begin{subfigure}[b]{0.33\linewidth}
        \includegraphics[width=\linewidth]{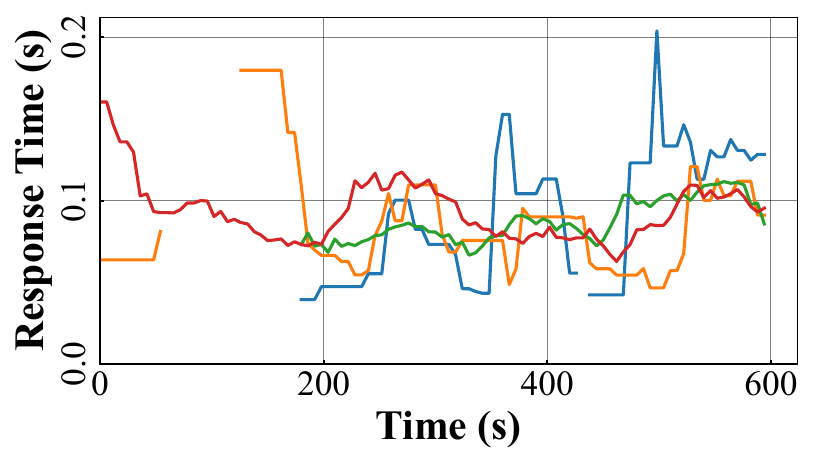}
        \caption{Selected-client Response Time}
        \label{fig:real_3}
    \end{subfigure}
    \caption{Real workload dynamics from LMSYS trace evaluated in S-LoRA, involving 27 clients. Shows variability in client request rates and resulting system response times.} 
    \label{fig:real}
\end{figure*}

\subsubsection*{Constant Overload Scenario}

We configure a scenario with constant, extreme overload to test fairness under high contention. Two clients send requests deterministically: Client 1 at a high rate of 20 req/s for short requests (input lengths of 20 and output lengths of 180), and Client 2 at a lower rate of 2 req/s but for very long requests (input lengths of 200 and output lengths of 1800). Both client demands exceed system capacity.

\noindent {\bf Overall performance.} ~\autoref{fig:appendix_overload} demonstrates that under constant overload, Equinox achieves dual advantages: it maintains the same fairness level as VTC while delivering enhanced service rates. Both schedulers successfully enforce fairness between clients with vastly different request rates and service demands, in contrast to FCFS which fails to provide equitable allocation in this high-contention regime. Notably, Equinox accomplishes this fairness guarantee while surpassing VTC in total throughput efficiency.

\noindent {\bf Breakdown.} As evidenced by ~\autoref{fig:appendix_overload_service_diff}, Equinox matches VTC's fairness characteristics through their equivalently small and bounded cumulative service differences. The service rate parity between clients (~\autoref{fig:appendix_overload_per_client_service}) further confirms that Equinox maintains VTC-caliber fairness while achieving service rate improvements - its per-client allocations remain balanced yet attain higher absolute values compared to VTC. Crucially, as shown in ~\autoref{fig:appendix_overload_service_rate}, Equinox elevates the total service rate beyond VTC's baseline while sustaining comparable GPU utilization (~\autoref{fig:appendix_overload_gpu}), demonstrating its unique ability to optimize throughput without compromising fairness.





\subsubsection*{Dynamic Load Increase Scenario}

This experiment evaluates Equinox's adaptability to dynamic changes in workload intensity. Client 1 maintains a constant request rate of 1 req/s for input lengths of 100 and output lengths of 400. Client 2 initially sends requests at the same input lengths and output lengths but increases its rate fourfold to 4 req/s midway through the experiment, significantly elevating the overall system load.

\noindent {\bf Overall performance.} Equinox demonstrates effective adaptation to the increasing load while upholding fairness principles, as shown in ~\autoref{fig:appendix_increase}. As Client 2 becomes more demanding, Equinox adjusts resource allocation dynamically without allowing this client to monopolize service.

\noindent {\bf Breakdown.} The sharp increase in Client 2's request rate is visible in ~\autoref{fig:appendix_increase_req_rate}. In response, Equinox dynamically recalibrates the per-client service rates (~\autoref{fig:appendix_increase_service_rate}), ensuring Client 1 continues to receive its fair share despite Client 2's increased demand. This adaptation is accompanied by an expected rise in overall response times (~\autoref{fig:appendix_increase_resp_time}) as system load intensifies. Correspondingly, GPU utilization increases (~\autoref{fig:appendix_increase_gpu}), reflecting the heightened demand.

\section{Supplementary Real-world Workload Evaluations}
\label{sec:appendix_real_world} 

This appendix contains supplementary results from real-world workload evaluations conducted using S-LoRA and vLLM systems, complementing the main results presented in ~\autoref{sec:real_world_evaluation}.

\subsubsection*{S-LoRA with LMSYS Dataset}
We conducted experiments using S-LoRA~\cite{slora} under a workload constructed from the trace logs of LMSYS Chatbot Arena. This scenario involved 27 distinct clients submitting requests based on patterns observed in these real-world interactions.

\noindent {\bf Workload Dynamics and System Response.} ~\autoref{fig:real} illustrates key characteristics of this dynamic workload and the system's response within S-LoRA. The per-client request rates (~\autoref{fig:real_1}) exhibit considerable variation over time, leading to fluctuations in the total instantaneous request rate offered to the system (~\autoref{fig:real_2}). Consequently, the observed per-client response times (following VTC, we select 13th, 14th and 26th, 27th clients based on the number of requests they send from minimum to maximum) (~\autoref{fig:real_3}) also vary, reflecting the interplay between the dynamic request arrivals, their resource requirements, and the scheduling decisions made by the system. While this setup is used for the cross-system fairness comparison (~\autoref{fig:fairness_comparison}), these specifically show the workload dynamics.

\end{document}